\documentclass[a4paper]{JHEP3}
\usepackage{amsmath,latexsym,amssymb}
\usepackage{epsfig}

\def\sqr#1#2{{\vcenter{\vbox{\hrule height.#2pt
        \hbox{\vrule width.#2pt height#1pt \kern#1pt
           \vrule width.#2pt}
        \hrule height.#2pt}}}}

\def\lsim{{\displaystyle
{{\raise-8pt\hbox{$ <$}}
\atop{\raise5pt\hbox{$\sim$}}}}}
\def\gsim{{\displaystyle
{{\raise-8pt\hbox{$ >$}}
\atop{\raise5pt\hbox{$\sim$}}}}}
\def\slsim{{\displaystyle
{{\raise-8pt\hbox{$\scriptstyle <$}}
\atop{\raise5pt\hbox{$\scriptstyle \sim$}}}}}
\def\sgsim{{\displaystyle
{{\raise-8pt\hbox{$\scriptstyle  >$}}
\atop{\raise5pt\hbox{$\scriptstyle \sim$}}}}}
\newskip\humongous \humongous=0pt plus 1000pt minus 1000pt

\newcommand{\oao}[2]{{#1\atopwithdelims[]#2}}
\newcommand{\oaop}[2]{{#1\atopwithdelims()#2}}

\newcommand{\sump}[0]{\sum_{(h,g)}\!{\raise 4pt \hbox{$'$}}\,}

\def\thefootnote{\fnsymbol{footnote}}
\def\be{\begin{equation}}
\def\ee{\end{equation}}
\def\ba{\begin{eqnarray}}
\def\ea{\end{eqnarray}}
\def\bs{\begin{subequations}}
\def\es{\end{subequations}}

\def\sp{\ , \ \ }

\newcommand{\uarrw}[0]{\mathrel{
{\raise.5ex\vbox{\hrule width 1cm}\hskip-6pt\rightarrow}}}
\newcommand{\underarrow}[1]{\mathop{\uarrw}_{#1}}

\boldmath
\title{Superstrings on NS5 backgrounds, deformed AdS$_3$ and
  holography\thanks{
Research partially supported by the EEC under the contracts
HPRN-CT-2000-00122, HPRN-CT-2000-00131, HPRN-CT-2000-00148.} }
\unboldmath
\author{Dan Isra\"el$^\Diamond$,
Costas Kounnas$^\Diamond$ and P. Marios Petropoulos$^\spadesuit$
\\
$\ $ \\
$\ $ \\
$^\Diamond$ Laboratoire de Physique Th{\'e}orique
de l'Ecole Normale Sup{\'e}rieure\footnote{Unit{\'e} mixte  du
CNRS et de l'Ecole Normale Sup{\'e}rieure,
UMR 8549.} \\
$\;\;\;$24  rue Lhomond, 75231 Paris Cedex 05, France\\
$\ $ \\
$^{\spadesuit}$ Centre de Physique Th\'eorique, Ecole Polytechnique\footnote{Unit\'e
mixte du CNRS et de l'Ecole Polytechnique, UMR 7644.} \\
$\;\;\;$ 91128 Palaiseau, France \\
$\ $ \\
E-mail:  \email{israel@lpt.ens.fr}, \email{kounnas@lpt.ens.fr},
\email{marios@cpht.polytechnique.fr} } \abstract{ We study a
non-standard decoupling limit of the D1/D5-brane system, which
interpolates between the near-horizon geometry of the D1/D5
background and the near-horizon limit of the pure D5-brane
geometry. The S-dual description of this background is actually an
exactly solvable two-dimensional (worldsheet) conformal field
theory: \{null-deformed $SL(2,\mathbb{R})\} \times SU(2) \times
T^4$ or $K3$. This model is free of strong-coupling singularities.
By a careful treatment of the $SL(2,\mathbb{R})$, based on the
better-understood $SL(2,\mathbb{R})/U(1)$ coset, we obtain the
full partition function for superstrings on $SL(2,\mathbb{R})
\times SU(2) \times T^4 /\mathbb{Z}_2$. This allows us to compute
the partition functions for the $J^3 \bar{J}^3$  and $J^2
\bar{J}^2$ deformations, as well as the full line of
supersymmetric null deformations, which links the
$SL(2,\mathbb{R})$ conformal field theory with linear-dilaton
theory. The holographic interpretation of this setup is a
renormalization-group flow between the decoupled NS5-brane
world-volume theory in the ultraviolet (little string theory), and
the low-energy dynamics of super Yang--Mills string-like
instantons in six dimensions. } \preprint{
LPTENS-03/07 \\
CPTH-S015.0403 \\
hep-th/0306053 \\}

\begin{document}

\setcounter{footnote}{0}
\renewcommand{\thefootnote}{\arabic{footnote}}
\setcounter{section}{0}

\section{Introduction}
So far only few exact, solvable string supersymmetric backgrounds
with a neat brane interpretation are known. The most popular is
certainly the near-horizon limit of the NS5-brane
background~\cite{Callan:1991ky}, which is an exact worldsheet
conformal field theory based on $SU(2)_k \times U(1)_Q$ (a
three-sphere plus a linear dilaton), and preserves 16 supercharges
thanks to the $N=4$ superconformal algebra on the
worldsheet~\cite{Antoniadis:1994sr}. This background includes a
strong-coupling region, that can be excised by distributing the
five-branes either on a circle~\cite{Sfetsos:1998xd}
\cite{Giveon:1999px} or on a spherical
shell~\cite{Kiritsis:2002xr}.

Another well-known exact string vacuum is the near-horizon
geometry of NS5-branes wrapped on a four-torus (or on a $K3$
manifold) and fundamental strings
~\cite{Antoniadis:1989mn}~\cite{Boonstra:1998yu}~\cite{Maldacena:1998bw}~\cite{Giveon:1998ns}.
In type IIB string theory, one can use S-duality to map this
solution to the D1/D5-brane system. The supersymmetry of this
background is enhanced from 8 to 16 supercharges in the
near-horizon limit. In this case the exact conformal field theory
is $SL(2,\mathbb{R}) \times SU(2) \times U(1)^4$. However, until
recently, the $SL(2,\mathbb{R})$
CFT~\cite{Balog:1988jb}~\cite{Petropoulos:1989fc} was poorly
understood (see~\cite{Petropoulos:1999nc} and references therein).
Substantial progress in the determination of the correct Hilbert
space of this theory was made
in~\cite{Maldacena:2000hw},~\cite{Maldacena:2000kv} and
\cite{Maldacena:2001km}. The key ingredient, first used
in~\cite{Henningson:1991jc}, was the observation that one must add
all the representations obtained by the spectral flow of the
affine algebra $\widehat{SL}(2,\mathbb{R})_L \times
\widehat{SL}(2,\mathbb{R})_R$. This allows to reconcile the
unitarity bound on the spin of the $SL(2,\mathbb{R})$
representations ($0<j\leq  k/2$) with the requirement that the
operator product algebra be closed.

A partition function for bosonic strings on thermal AdS$_3$ (i.e.
$H_{3}^+ /\mathbb{Z}$) was proposed in~\cite{Maldacena:2000kv}, by
using the older result by~\cite{Gawedzki:1991yu}.
In~\cite{Hanany:2002ev}, the partition function for the axial
coset $SL(2,\mathbb{R}) / U(1)_{\mathrm{A}}$ -- whose target-space
interpretation is a Euclidian two-dimensional black
hole~\cite{Witten:1991yr} -- was analyzed in the same spirit; it
allowed to extract the full spectrum in agreement with previous
semi-classical analysis. However, as a consequence of the
non-compact nature of the group, these partition functions are
plagued with a divergence, which should be handled with care in
order to obtain sensible results. Finding a modular-invariant
partition function for $SL(2,\mathbb{R})$ that reproduces the
spectrum found in~\cite{Maldacena:2000hw} is to our knowledge
still an open problem. One of the aims of the present paper is to
fill this gap, which is a first step towards the complete
understanding of superstrings on $SL(2,\mathbb{R}) \times SU(2)
\times T^4 \mathrm{\ or\ } K3$ as well as deformations of this
background. The structure of the partition function will be
understood from a different viewpoint, by using the orbifold
language, and the supersymmetrization will be discussed by
considering the extended superconformal algebra on the worldsheet.

The above two string backgrounds are in fact members of a family
of conformal field theories interpolating between them both in
space--time and in moduli space. These theories can be viewed as
exact marginal deformations of the $SL(2,\mathbb{R})$ WZW model,
driven by a left-right combination of null currents (i.e. currents
generating null subgroups)~\cite{Forste:1994wp}. The endpoint of
this deformation gives the linear dilaton and two light-cone free
coordinates~\cite{Klimcik:1994wp}. This geometry corresponds
actually to a near-horizon limit for the NS5-branes \emph{only}
(such a limit was also mentioned in~\cite{Giveon:1999zm}). We
explain in this paper how this background can be obtained by a
particular decoupling limit of the NS5/F1 background, 
actually the \emph{little-string-theory decoupling limit in the
presence of macroscopic fundamental strings}. In the D-brane
picture, this limit involves also the decompactification of the
torus. An important achievement of the present work is that this
model can also be viewed as a regularization of the
strong-coupling region of the NS5-brane theory, which thus
provides an alternative to~\cite{Kiritsis:2002xr}, with a
better-controlled worldsheet conformal field theory.

This class of backgrounds clearly preserves a fraction of
target-space supersymmetry, which is generically one quarter,
enhanced to one half on the endpoints of the deformation. An
interesting feature of this deformation is that it is completely
fixed by the requirement of $N=2$ superconformal invariance on the
worldsheet. We will show that it reduces to an orthogonal rotation
between the worldsheet bosons and fermions interpolating between
the $N=2$ superconformal algebras of $U(1)_Q \times
\mathbb{R}^{1,1} \times SU(2)$ and $ SL(2,\mathbb{R}) \times
SU(2)$.

There has been in the last years a considerable renewal of
interest for these theories because, besides their intrinsic
interest as exact string backgrounds, they enter in several
gauge/gravity dualities. The celebrated AdS/CFT
correspondence~\cite{Maldacena:1997re} is a conjectured
equivalence between the near-horizon geometry of the D3- or
D1/D5-brane background (respectively AdS$_5 \times S^5$ and AdS$_3
\times S^3 \times T^4$ ) and the extreme infra-red theory living
on their world-volume, a superconformal gauge theory with maximal
supersymmetry. In a similar fashion, a holographic duality between
the decoupled NS5-brane world-volume theory -- the so-called
little string theory (LST) -- and the linear-dilaton background
has been conjectured in~\cite{Boonstra:1998mp}~\cite{Aharony:1998ub}.

The holographic interpretation of our setup is clear. The
ultraviolet region of the holographic ``gauge'' dual corresponds
to the asymptotic geometry, and is therefore the decoupled
world-volume theory living on the NS5-branes. This theory is not a
field theory, since there is no ultraviolet field-theoretic fixed
point, and contains string-like excitations~\cite{Seiberg:1997zk},
hence the name ``little string theory''~\cite{Aharony:1998ub}. In
the type IIB case, this theory is described in the infra-red by a
gauge theory in six dimensions with $U(N_5)$ gauge group and
$N=(1,1)$ supersymmetry. The standard NS5 background corresponds
to a renormalization-group flow towards a free fixed point in the
infra-red, whose dual picture is a strong-coupling region in the
gravitational background.

In the case of the null deformation of $SL(2,\mathbb{R})$, the
addition of fundamental strings in the background corresponds in
the dual theory to a configuration of string-like instantons of
the low-energy gauge theory. Therefore the physics near the
infra-red fixed point is governed by the dynamics of the moduli
space of these instantons. The effective theory in 1+1 dimensions
is superconformal, hence the dilaton stops running. This CFT has
been studied intensively in the last years, in particular in the
context of black-hole quantum mechanics (see~\cite{David:2002wn} for references).

The paper is organized as follows. In Sec.~\ref{sect1} we present
the precise decoupling limit which leads to the background of
interest, and explain why this background is an exact conformal
theory. Section~\ref{sect2} is devoted to bosonic strings in
AdS$_3$, with special emphasis to the derivation of a partition
function for the $SL(2,\mathbb{R})$, where the spectrum is read
off unambiguously at finite or infinite radius. Then we introduce
the worldsheet fermions and discuss in Sec.~\ref{sect3} the
superstrings in the background AdS$_3 \times S^3 \times T^4
/\mathbb{Z}_2$ with particular attention to the construction of
extended worldsheet superconformal algebras. In Sec.~\ref{sect4} we
give the partition functions for the $J^3 \bar{J}^3$ and $J^2
\bar{J}^2$ deformations of $SL(2,\mathbb{R})$, and study in detail its
null deformation, which is the main motivation of this article.
Null deformations of the supersymmetric background
AdS$_3 \times S^3 \times T^4$ are extensively discussed in
Sec. \ref{sect4adss}, with a particular attention to the requirement
of preserving $N=2$ superconformal algebra. Section~\ref{sect5} gives
a brief outlook of the holographic interpretation of this superstring
vacuum.

\boldmath
\section{A new decoupling limit for the D1/D5-brane system}\label{sect1}
\unboldmath

In this section, we will present a decoupling limit for the
D1/D5-brane system or conversely the NS5/F1 dual configuration. In
this limit, we obtain a line of exact conformal theories, which
turn out to be connected by a marginal deformation. Supersymmetry
properties and spectra will be analyzed later.

\subsection{The supergravity solution and a partial near-horizon limit}

We consider the D1/D5-brane system in type IIB string theory. The
D5-branes extend over the coordinates $x, x^6, \ldots , x^9$,
whereas the D1-branes are smeared along the four-torus spanned by
$x^6, \ldots , x^9$. The volume of this torus is asymptotically
$V= (2\pi)^4 \alpha^{\prime 2} v$. With these conventions, in the
sigma-model frame, the supergravity solution at hand reads
(metric, dilaton and Ramond--Ramond field strength):
\begin{eqnarray}
 d\tilde s^2 & = & {1\over \sqrt{H_{1} H_{5}}}
\left(-dt^2 + dx^2 \right) + \sqrt{H_{1}H_{5}} \left(dr^2 + r^2
d\Omega_{3}^{2} \right) + \sqrt{H_{1}\over H_{5}} \sum_{i=6}^9
(dx^i)^2, \label{sugrsol} \\
\mathrm{e}^{2\tilde \phi} & = & g_{\rm s}^2 {H_1 \over H_5},  \label{sugrsold}\\
F_{[3]} & = &  - \frac{1}{g_{\rm s}} d H_{1}^{-1}  \wedge dt
\wedge dx + 2\alpha' N_5 \Omega_3\label{sugrsola}
\end{eqnarray}
($\Omega_3$ is the volume form of the three-sphere) with
\begin{equation}
H_1 = 1 + \frac{g_{\rm s} \alpha' N_1}{vr^2} \ , \ \  H_5 = 1 +
\frac{g_{\rm s} \alpha' N_5}{r^2}.\nonumber
\end{equation}
The four-torus can be replaced by a Calabi--Yau two-fold $K3$,
provided that the charges of the D1 and D5 branes are of the same
sign. The near-horizon ($r \to 0$) string coupling constant and
the ten-dimensional gravitational coupling constant are
\begin{equation}
g_{10}^2=g_{\rm s}^2 \frac{N_1}{vN_5} \  , \ \ 2 \kappa_{10}^{2} =
(2\pi )^7 \mathrm{e}^{2 \langle \tilde \phi \rangle }
\alpha^{\prime 4} .\nonumber
\end{equation}

The standard decoupling limit of Maldacena, which leads to the
AdS$_3$/CFT$_2$ correspondence~\cite{Maldacena:1997re}, is
\begin{equation}
\nonumber
\begin{array}{l}
\alpha'   \to   0, \\
U\equiv r/\alpha' \   \mathrm{fixed},\\
v   \   \mathrm{fixed}.
\end{array}
\end{equation}
In this limit, the holographic description is a two-dimensional
superconformal field theory living on the boundary of AdS$_3$ that
corresponds to the world-volume theory of the D1/D5 system
compactified on a $T^4$ whose volume is held fixed in Planck
units~\cite{Giveon:1998ns}~\cite{deBoer:1998ip}.

In order to reach a decoupling limit that corresponds to the
near-horizon geometry for the D5-branes only, one has to consider
the limit:
\begin{equation}
\label{decoupl}
\begin{array}{l}
\alpha' \to 0, \\
U= r / \alpha' \ \mathrm{fixed}, \\
g_{\rm s} \alpha' \ \mathrm{fixed}, \\
\alpha^{\prime  2} v \ \mathrm{fixed}.
\end{array}
\end{equation}
The last condition is equivalent to keeping fixed the
six-dimensional string coupling constant:
\begin{equation}
g_{6}^2 = {g_{\rm s}^2 \over v}.\nonumber
\end{equation}
Since the gravitational coupling constant vanishes in this limit,
the world-volume theory decouples from the bulk. The geometrical
picture of the setup is the following: as $v \to \infty$, the
torus decompactifies and the density of D-strings diluted in the
world-volume of the D5-branes goes to zero.

The string coupling remains finite in this near-horizon limit,
while the asymptotic region is strongly coupled. A perturbative
description, valid everywhere is obtained by S-duality. The
supergravity solution (\ref{sugrsol}), (\ref{sugrsold}) in the
S-dual frame reads:
\begin{eqnarray}
ds^2 & = & \mathrm{e}^{-\tilde \phi} d\tilde s^2 = \frac{1}{g_{\rm
s}} \left\{ {1\over H_{1}} \left(-dt^2 + dx^2 \right)
 + \alpha^{\prime 2} H_5 \left(dU^2 + U^2 d\Omega_3^{ 2} \right)
 +  \sum_{i=6}^9 (dx^{i})^2\right\}, \label{sugrsolS}\\
\mathrm{e}^{2\phi} & = & \frac{1}{g_{\rm s}^2}
\frac{H_5}{H_1}\label{sugrsolSd}
\end{eqnarray}
with (in the limit (\ref{decoupl}) under consideration)
\begin{equation}
H_1 = 1 + \frac{g_{\rm s} N_1}{\alpha' v U^2} \ , \ \ H_5 =
\frac{g_{\rm s} N_5}{\alpha' U^2}. \label{newH15}
\end{equation}
The expression (\ref{sugrsola}) for the antisymmetric tensor
remains unchanged but it stands now for a NS flux and we will
trade $F_{[3]}$ for $H_{[3]}$.

We now introduce the new variables:
\begin{equation} u= {1\over U}\ , \ \  X^\pm =
X\pm T = \frac{x \pm t} {g_{6}\sqrt{N_1 N_5}},\nonumber
\end{equation}
and the following mass scale:
\begin{equation}
M^2 = \frac{g_{\rm s} N_1}{\alpha' v}. \label{defparam}
\end{equation}
In these coordinates, the solution (\ref{sugrsolS}),
(\ref{sugrsolSd}) and (\ref{sugrsola}), with (\ref{newH15}),
reads:
\begin{eqnarray}
\frac{ds^2}{\alpha'}  &=&  N_5 \left\{ \frac{du^2}{u^2}+
\frac{dX^2-dT^2}{u^2 + 1/M^2} + d\Omega_{3}^{ 2}
\right\} + \frac{1}{\alpha' g_{\rm s}} \sum_{i=6}^9(dx^i)^2, \nonumber \\
\mathrm{e}^{2\phi}  &=&  \frac{1}{g_{10}^2} \frac{u^2}{u^2 + 1/M^2}, \label{solns} \\
{H_{[3]}\over  \alpha '} &=&   N_5 \left\{   \frac{2u}{\left(u^2 +
1/M^2\right)^2} du\wedge dT \wedge dX + 2\Omega_3
\right\}.\nonumber
\end{eqnarray}
This is the geometry of a deformed AdS$_3$ times an $S^3\times
T^4$. Asymptotically ($u \to 0$), it describes the near-horizon
geometry of the NS5-brane background, $U(1)_Q \times
\mathbb{R}^{1,1} \times SU(2) \times T^4$ {\it in its weakly
coupled region}. In the $u\to \infty$ limit, the background
becomes that of the NS5/F1 near-horizon: $SL(2,\mathbb{R})\times
SU(2) \times T^4$, with a finite constant dilaton. In some sense,
we are regulating the strong-coupling region of the NS5-brane
background by adding an appropriate condensate of fundamental
strings. As we already advertised, this regularization is an
alternative to the one proposed in~\cite{Kiritsis:2002xr}; it
avoids the spherical target-space wall of the latter, and replaces
it by a smooth transition, driven by a marginal worldsheet
deformation, as will become clear in Sec. \ref{sigmamod}.

Before going into these issues, we would like to address the
question of supersymmetry. The configuration displayed in Eqs.
(\ref{solns}) preserves by construction one quarter of
supersymmetry. Consider indeed IIB supergravity.
The unbroken supersymmetries correspond to the covariantly constant spinors for
which the supersymmetry variations of the dilatino and gravitino
vanish:
\begin{eqnarray}
\delta \lambda & = & \left[\gamma^\mu \partial_\mu \phi  \ \sigma^3 -
\frac{1}{6} H_{\mu \nu \rho} \gamma^{\mu \nu \rho} \right]
\oaop{\eta_1}{\eta_2}  = 0,
\label{vardil} \\
\delta \psi_{\mu}^{\vphantom m} & = &
\left[\partial_{\mu}^{\vphantom m} + \frac{1}{4} \left( w_{\mu}^{\
ab} - H_{\mu}^{\ ab} \sigma^3 \right) \Gamma_{ab}^{\vphantom m} \right]
\oaop{\eta_1}{\eta_2}  =   0. \label{vargrav}
\end{eqnarray}
where $\sigma^3$ is the third Pauli matrix. The two supersymmetry
generators $\eta_1$ and $\eta_2$ have the same chirality:
$\Gamma^{11} \eta_{1,2} = \eta_{1,2}$.

Let us for example concentrate on the dilatino variation, Eq.
(\ref{vardil}):
\begin{eqnarray}
\delta \lambda & = & \left[ \Gamma^2_{\vphantom m}\,  e_{2}^{u}\,
\partial_{u}^{\vphantom m} \phi \ \sigma^3 - H_{uxt}^{\vphantom m} \, \mathrm{e}^{u}_{1} \, \mathrm{e}^{x}_2 \,
\mathrm{e}^{t}_{0}\,  \Gamma^{1}_{\vphantom m} \Gamma^2_{\vphantom
m} \Gamma^0_{\vphantom m}  + H_{\theta \varphi \chi}^{\vphantom
m}\, \mathrm{e}^{\theta}_{3}\,  \mathrm{e}^{\varphi}_4 \,
\mathrm{e}^{\chi}_{5}\, \Gamma^{3}_{\vphantom m}
\Gamma^4_{\vphantom m} \Gamma^5_{\vphantom m}
\right] \oaop{\eta_1}{\eta_2}  \nonumber \\
& = &\left[ \frac{1/M^2}{u^2 + 1/M^2} \Gamma^2 \ \sigma^3  - \frac{u^2}{u^2 +
1/M^2} \Gamma^{1} \Gamma^2 \Gamma^0  +  \Gamma^{3} \Gamma^4
\Gamma^5  \right] \oaop{\eta_1}{\eta_2}  , \nonumber
\end{eqnarray}
where Latin indices ${a,b,\ldots}$ refer to the tangent-space
orthonormal bases, with $\{0,1,2\}$ and $\{3,4,5\}$ corresponding
to the AdS$_3$ and $S^3$ submanifolds. The two $SO(9,1)$ spinors
are decomposed into $SO(1,1) \times SO(4) \times
SO(4)_T$:$$\mathbf{16} \to (+,\mathbf{2},\mathbf{2}) +
(+,\mathbf{2'},\mathbf{2'}) + (-,\mathbf{2'},\mathbf{2}) +
(-,\mathbf{2},\mathbf{2'}).$$ The first $SO(4)$ is the isometry
group of the transverse space (coordinates $x^{2, \ldots, 5}$) and
$SO(4)_T$ corresponds to the four-torus.\\
In the infinite-deformation limit, $M^2 \to 0$, this equation
projects out the $SO(4)$ spinors of one chirality:
$$\left[\sigma^3 + \Gamma^{2} \Gamma^{3} \Gamma^{4} \Gamma^5
\right] \oaop{\eta_1}{\eta_2} = 0. $$ The surviving supersymmetry
generators, are, for $\eta_1$, $(+,\mathbf{2},\mathbf{2})$ and
$(-,\mathbf{2},\mathbf{2'})$, and for $\eta_2$,
$(+,\mathbf{2'},\mathbf{2'})$ and $(-,\mathbf{2'},\mathbf{2})$.

In the opposite limit of undeformed AdS$_3$, $M^2 \to \infty$, we
have instead: $$\left[ 1 - \Gamma^{0} \Gamma^{1} \Gamma^{2}
\Gamma^{3}\Gamma^{4} \Gamma^{5} \right] \oaop{\eta_1}{\eta_2}
=0,$$ which projects out the $SO(5,1)$ spinor (coordinates
$x^{0,1,6,\ldots,9}$ of the five-brane world-volume) of left
chirality, i.e. keeps the representations
$(+,\mathbf{2'},\mathbf{2'})$ and $(-,\mathbf{2},\mathbf{2'})$ for
both supersymmetry generators. For any finite value of the
deformation,  both projections must be imposed:
\begin{equation}
\left[ \frac{1/M^2}{u^2 + 1/M^2} \left( \sigma^3 - \Gamma^2 \Gamma^{3}
\Gamma^{4} \Gamma^{5} \right)
 - \frac{u^2}{u^2 + 1/M^2} \left( \Gamma^{0} \Gamma^{1}  +
 \Gamma^{2} \Gamma^{3} \Gamma^{4} \Gamma^{5}
\right) \right]  \oaop{\eta_1}{\eta_2}  = 0, \nonumber
\end{equation}
which breaks an additional half supersymmetry. The remaining
supersymmetries are $(-,\mathbf{2},\mathbf{2'})$ for $\eta_1$ and
$(+,\mathbf{2'},\mathbf{2'})$ for $\eta_2$. The gravitino equation
gives no further restrictions as it should (it reduces to the
Killing-spinor equation on $S^3$ and deformed AdS$_3$), and we are
eventually left with one quarter of supersymmetry. Supersymmetry
enhancement occurs only for the limiting backgrounds -- AdS$_3
\times S^3$ or three-sphere plus linear dilaton, which preserve
one half of the original supersymmetry.

\boldmath
\subsection{Exact conformal-field-theory description: a null deformation of $SL(2,\mathbb{R})$}
\label{sigmamod} \unboldmath

We will now show that the deformed-AdS$_3$ factor in the
background (\ref{solns}) is the target space of an exactly
conformal sigma-model.

The action for a WZW model is in general
\begin{equation}
S = \frac{k}{16\pi} \int_{\partial \mathcal{B}} \mathrm{Tr} \left(
g^{-1} dg \wedge * g^{-1} dg \right) + \frac{ik}{24\pi}
\int_{\mathcal{B}} \mathrm{Tr} \left( g^{-1} dg \right)^3.
\label{wzw}
\end{equation}
In the case of $SL(2,\mathbb{R})$, one can use the Gauss
decomposition for the group elements:
\begin{equation}
g=g_- g_0 g_+ = \left( \begin{array}{cc} 1 & 0 \\ x^- & 1
\end{array} \right) \left( \begin{array}{cc} 1/u & 0 \\ 0 & u
\end{array} \right) \left( \begin{array}{cc} 1 & x^+ \\ 0 & 1
\end{array} \right),
\label{gauss}
\end{equation}
which provides the Poincar\'e coordinate system (see Appendix A).
With this choice, the sigma-model action reads:
\begin{equation}
S = \frac{k}{2\pi} \int d^2 z  \left( \frac{\partial u
\bar{\partial} u}{u^2} + \frac{ \partial x^+ \bar{\partial}
x^-}{u^2}\right). \label{sigma}
\end{equation}

As usual, the affine symmetry $\widehat{SL}(2,\mathbb{R})_L \times
\widehat{SL}(2,\mathbb{R})_R$ is generated by weight-one currents.
Since the group is non-compact, there are null directions, easily
identified in the Poincar\'e coordinates. The corresponding
$\widehat{U}(1)_L \times \widehat{U}(1)_R$ symmetries are linearly
realized and generated by the following {\it null
currents}{\footnote{Strictly speaking, these are not Cartan
generators. See Appendix A.}{\it:}
\begin{equation}
\label{nullcurr} J = \frac{\partial x^+}{u^2} \  , \ \ \bar{J} =
\frac{ \bar{\partial} x^-}{u^2}.
\end{equation}
The $(1,1)$ operator $J (z) \bar{J} (\bar{z} )$, is truly marginal
and can be used to generate a line of CFT's. Along this line, i.e.
for finite values of the deformation parameter $1/M^2$, the
geometry back-reaction must be taken into account. Integrating the
corrections (much like in~\cite{Hassan:1992gi} and
\cite{Giveon:1993ph} for compact cases) one obtains the following
null-deformed $SL(2,\mathbb{R})$-sigma-model
action~\cite{Forste:1994wp}:
\begin{equation}
\label{nulldef} S = \frac{k}{2\pi} \int d^2 z  \left(
\frac{\partial u \bar{\partial} u}{u^2} + \frac{ \partial x^+
\bar{\partial} x^-}{u^2 + 1/M^2}\right).
\end{equation}
The affine symmetry $\widehat{SL}(2,\mathbb{R})_L \times
\widehat{SL}(2,\mathbb{R})_R$ is broken down to $\widehat{U}(1)_L
\times \widehat{U}(1)_R$ and only the two null currents survive, which now
read:
\begin{equation}
J =  \frac{\partial x^+}{u^{2}+ 1/M^2} \ , \ \ \bar{J} =
 \frac{\bar{\partial} x^-}{u^{2}+ 1/M^2}.
\label{nulldefcur}
\end{equation}

The deformed target-space geometry and antisymmetric tensor are
read off directly from Eq. (\ref{nulldef}), and turn out to
coincide with the deformed-AdS$_3$ factor in (\ref{solns}).

There is an alternative way to reach the same conclusion. As shown
in~\cite{Hassan:1992gi}, marginal deformations of a WZW model
correspond to $O(d,d,\mathbb{R})$ transformations acting on the
Abelian isometries. In order to implement the latter in the case at
hand, we rewrite the
$SL(2,\mathbb{R})$-WZW action as:
\begin{equation}
S = \frac{k}{2\pi} \int d^2 z \left(  \frac{\partial u
\bar{\partial} u}{u^2} + \partial \left(\begin {matrix}{x^+  x^-
}\end{matrix}\right) \cdot E \cdot \bar{\partial}
\left(\begin{matrix}x^+ \\ x^- \end{matrix}\right)
\right)\label{altS}
\end{equation}
with
\begin{equation}
E = \left( \begin{array}{cc} 0 & 1/u^2 \\ 0 & 0
\end{array} \right).
\nonumber
\end{equation}
Acting on the corresponding background with the following
$O(2,2,\mathbb{R})$ element\footnote{The simplest setup for
illustrating this transformation is flat background --
compactified bosons: two light-cone coordinates with a constant
$B$-field. The original Lagrangian, $4\pi \mathcal{L} =
\partial x^+ \bar{\partial} x^-$, transforms into
$4\pi \tilde{\mathcal{L}}= \frac{M^2}{1+M^2} \partial x^+
\bar{\partial} x^-$; this amounts to a shift of radii.} (see
e.g.~\cite{Giveon:1994fu} for a review):
\begin{equation}
g = \left( \begin{array}{cc} \mathbb{I} & 0 \\
- \Theta /M^2 & \mathbb{I} \end{array} \right) \ ,  \ \
\mathrm{with}\ \ \Theta = \left( \begin{array}{cc} 0 & 1 \\ -1 & 0
\end{array} \right),
\label{eldual}
\end{equation}
we recover (\ref{altS}) with
\begin{equation}
E' = g(E) = \left( \begin{array}{cc} 0 & \frac{1}{u^2+1/M^2} \\ 0
& 0
\end{array} \right),
\nonumber
\end{equation}
which is precisely the null-deformed $SL(2,\mathbb{R})$-WZW
action, Eq. (\ref{nulldef}).

Notice finally that the deformed $SL(2,\mathbb{R})$-WZW model
under consideration can also be obtained as a coset by gauging a
$U(1)$ subgroup of $SL(2,\mathbb{R}) \times
U(1)$~\cite{Horowitz:1994ei}. We will come back later to this
CFT to compute the one-loop partition function,
which demands a careful treatment of global properties of the fields.
The question of supersymmetry along these deformations needs also to
be recast in the present, exact CFT framework, and will be discussed
in Sec. \ref{sect4adss}.

\boldmath
\section{Bosonic strings on AdS$_3$}\label{sect2}
\unboldmath

This section is devoted to the bosonic part of the AdS$_3$ (i.e.
$SL(2,\mathbb{R})$) component of the previous backgrounds. Despite
many efforts and achievements (a short summary is given Appendix
A), our understanding is not completely satisfactory. We show here
how to reach information about $SL(2,\mathbb{R})$ starting from
the better-understood $SL(2,\mathbb{R})/U(1)_{\mathrm{A}}$ axial
gauging. This enables us to provide a partition function for the
$SL(2,\mathbb{R})$, which carries full information about the
spectrum. Under this form, the $SL(2,\mathbb{R})$ WZW model
resembles a $\mathbb{Z}_N$ freely acting orbifold of $T^2 \times
S^1 \times S^1$ over $U(1)$, at large $N$. Spectra and partition
functions of $SL(2,\mathbb{R})$ deformations will be addressed in
Sec. \ref{sect4}.

\boldmath
\subsection{$SL(2,\mathbb{R})$ from $SL(2,\mathbb{R})/U(1)_{\mathrm{A}}$}
\unboldmath

There is a tight link between the spectrum of string theory on
AdS$_3$ and the spectrum of the axial-gauged coset
$SL(2,\mathbb{R})/U(1)_{\mathrm{A}}$ -- non-compact parafermions.
The states in the coset are those of the $SL(2,\mathbb{R})$ CFT
with the restriction $J^{3}_{n} | \mathrm{state} \rangle =
\bar{J}^{3}_{n} | \mathrm{state} \rangle = 0$ for $n>0$, and the
conditions on the zero modes $J^{3}_0  + \bar{J}^{3}_0 = -wk$ and
$J^{3}_0  - \bar{J}^{3}_0 = n$. It is therefore possible to
reconstruct the $SL(2,\mathbb{R})$ starting from its axial
gauging, much like in the case of compact parafermions, where the
$SU(2)/U(1)$ gauging enables for reconstructing the $SU(2)$ WZW
model \cite{Gepner:1986}. In the non-compact case, however, the
coset was shown to be a unitary conformal field theory
\cite{Dixon:1989}, whereas this holds for the $SL(2,\mathbb{R})$
only if Virasoro conditions are imposed \cite{Petropoulos:1989fc}
\cite{Mohammedi:1990} \cite{Hwang:1990aq} \cite{Evans:1998qu}. The
physical states can be chosen, up to a spurious state, to be
annihilated by the positive modes of the time-like current
$J^{3}_{n>0}$,  $\bar{J}^{3}_{n>0}$. This is the same as for the
coset, except for the zero modes.

Our aim is here to show how a partition function for the
$SL(2,\mathbb{R})$ can be reached starting from the partition
function of the $SL(2,\mathbb{R})/U(1)_{\mathrm{A}}$ proposed in
\cite{Hanany:2002ev}. We start with the WZW action~(\ref{wzw}) for
$g\in SL(2,\mathbb{R})$ parameterized with Euler angles (see Eq.
(\ref{euler})). We will gauge the $U(1)$ axial subgroup $g \to
hgh$ with $h= \mathrm{e}^{i \lambda \sigma_2 /2}$. The action for
the gauged model is
\begin{equation}
S(g,A) = S(g) + \frac{k}{2\pi} \int d^2 z \mathrm{Tr} \left(
A\bar{\partial}gg^{-1} + \bar{A} g^{-1} \partial g - A g \bar{A}
g^{-1} - A\bar{A} \right). \nonumber
\end{equation}
The gauge field is Hodge-decomposed as:
\begin{equation}
A =  \partial (\tilde{\rho}+\rho ) + \frac{i}{\tau_2} ( u_1 \bar{\tau} -u_2 ) \ ,
\ \ \bar{A}  =  \bar{\partial} (\tilde{\rho}-\rho) - \frac{i}{\tau_2} ( u_1
\tau -u_2 ). \nonumber
\end{equation}
After field redefinitions, the gauged-fixed action is given by an
$SL(2,\mathbb{R})$ times a compact boson, with global constraints
and a $(b,c)$ ghost system. This theory is unitary and the
corresponding target space is Euclidean.

The partition function has been computed by using path-integral
techniques in~\cite{Gawedzki:1991yu} and \cite{Hanany:2002ev}. We
would like to summarize the method and remind the final result.
Following \cite{Hanany:2002ev} the model is transformed, for
technical convenience, into a $U(1)$-gauging of the -- non-unitary
-- $H_{3}^+ = SL(2,\mathbb{C}) / SU(2)$ CFT:
\begin{eqnarray}
S &=& \frac{k}{2\pi} \int d^2 z  \left( \partial \phi
\bar{\partial} \phi + \left(\partial  \bar{v} + \bar{v} \partial
\phi\right)
\left(\bar{\partial} v + v \bar{\partial} \phi \right) \right)\nonumber \\
&+&\frac{k}{2\pi} \int d^2 z \partial \rho \bar{\partial} \rho +
\frac{1}{\pi} \int d^2 z  \left(b\bar{\partial}c +
\tilde{b}\partial \tilde{c} \right). \nonumber
\end{eqnarray}
The first part of the action is indeed the $H_{3}^+ =
SL(2,\mathbb{C}) / SU(2)$ sigma-model. The various fields acquire
non-trivial holonomies from the gauge field, and can be decomposed
as:
\begin{eqnarray}
\phi & = & \hat{\phi} + \frac{1}{4\tau_2} \left[ (u_1 \bar{\tau}
-u_2 ) z
+(u_1 \tau -u_2 ) \bar{z} \right], \nonumber \\
\rho & = & \hat{\rho} + \frac{1}{4\tau_2} \left[ (u_1 \bar{\tau}
-u_2 ) z
+(u_1 \tau -u_2 ) \bar{z} \right], \nonumber \\
v & = & \hat{v} \exp - \frac{1}{4\tau_2} \left[ (u_1 \bar{\tau}
-u_2 ) z +(u_1 \tau -u_2 ) \bar{z} \right]. \nonumber
\end{eqnarray}
The fields $v$ and $\bar{v}$  give the following contribution to
the partition function:
\begin{eqnarray}
\det \left| \partial + \frac{1}{2\tau_2} (u_1 \bar{\tau} -u_2) +
\partial{\hat{\phi}} \right|^{-2} & = &
\det \left| \partial + \frac{1}{2\tau_2} (u_1 \bar{\tau} -u_2)
\right|^{-2} \exp {\frac{2}{\pi} \int d^2 z  \partial \hat{\phi}
\bar{\partial} \hat{\phi}} \nonumber \\
& = & 4 \eta \bar{\eta}\frac{ \mathrm{e}^{\frac{2\pi}{\tau_2}
(\mathrm{Im} (u_1 \tau -u_2))^2}}{\left|\vartheta_1 (u_1 \tau -u_2
|\tau ) \right|^2} \exp {\frac{2}{\pi} \int d^2 z \partial
\hat{\phi} \bar{\partial} \hat{\phi}},\nonumber
\end{eqnarray}
where $\vartheta_1 (\nu | \tau)$ is an elliptic theta function
(see Appendix D). The periodicity properties of this determinant
allows for breaking $u_1$ and $u_2$ into an integer and a compact
real: $u_1=s_1 + w$, $u_2 = s_2 + m$, $s_i \in [0,1)$. Taking
finally into account the contributions of the free bosons $\phi$
and $\rho$ and that of the ghosts, leads to the result
\cite{Hanany:2002ev}:
\begin{equation}
Z_{SL(2,\mathbb{R})/U(1)_{\mathrm{A}}} =  4 \sqrt{k(k-2)} \, \eta
\bar{\eta}
 \int d^2s\frac{
\mathrm{e}^{\frac{2\pi}{\tau_2} (\mathrm{Im} (s_1 \tau -s_2))^2 }
}{\left|\vartheta_1 (s_1 \tau -s_2 |\tau ) \right|^2}
\sum_{m,w=-\infty}^{+\infty} \mathrm{e}^{-\frac{k\pi}{\tau_2}
\left| (s_1 + w)\tau - (s_2 + m) \right|^2} . \label{cospart}
\end{equation}
We can recast the latter in terms of the free-boson conformal
blocks (\ref{coblofre}):
\begin{equation}
Z_{SL(2,\mathbb{R})/U(1)_{\mathrm{A}}} =  4 \sqrt{(k-2)\tau_2} \,
\eta \bar{\eta}
 \int d^2s\frac{
\mathrm{e}^{\frac{2\pi}{\tau_2} (\mathrm{Im} (s_1 \tau -s_2))^2 }
}{\left|\vartheta_1 (s_1 \tau -s_2 |\tau ) \right|^2} \sum_{m,w\in
\mathbb{Z}} \zeta \oao{w+s_1}{m+s_2}(k),
 \label{cospartcb}
\end{equation}
which meets our intuition that there are one compact and one
non-compact bosons in the cigar geometry.

A few remarks are in order here. The integration over  $s_1, s_2$
should be thought of as a constraint on the Hilbert space, which
defines the non-compact parafermionic $\mathbb{Z}$ charge. The
allowed parafermionic charges are $m = n/2$, $\bar{m} = -n/2$ for
the unflowed sector, and  $\tilde{m} = (n - wk)/2$ and
$\tilde{\bar{m}} = -(n+wk)/2$ for the $w$-flowed sector. Another
important issue is the logarithmic divergence originating from
$s_1=s_2=0$, and due to the non-compact nature of the group.
Baring this divergence, the partition function (\ref{cospart}) is
modular-invariant, as can be easily checked by using the modular
properties of Jacobi functions. Its content in terms of
non-compact-parafermion discrete and continuous series can be
further investigated (see \cite{Hanany:2002ev} for details).
Moreover, in the large-$k$ (flat-space) limit,
$Z_{SL(2,\mathbb{R})/U(1)_{\mathrm{A}}} \sim k\left(\pi
\sqrt{\tau_2} \eta\bar\eta\right)^{-2}$ up to an infinite-volume
factor\footnote{This factor comes as $ \int_{-\infty}^{+\infty}{dx
dy\over x^2 + y^2} \exp -\pi(x^2 + y^2)$.}: we recover two free,
uncompactified bosons.

We will now show that it is possible to recover a partition
function for the $SL(2,\mathbb{R})$ WZW model, starting from the
above result for the coset $SL(2,\mathbb{R})/U(1)_{\mathrm{A}}$.
In many respects this is similar to what happens in the compact
case: $SU(2)_k$ can be reconstructed as $\left(SU(2)_k/U(1) \times
U(1)_{\sqrt{2k}}\right)\big/ \mathbb{Z}_k$, where $\mathbb{Z}_k$ is
the compact-parafermionic symmetry of the coset $SU(2)_k/U(1)$,
and acts freely on the compact $U(1)_{\sqrt{2k}}$. To some extent,
however, manipulations involving divergent expressions such as
(\ref{cospart}) can be quite formal, and require to proceed with
care.

As we already pointed out, the states in
$SL(2,\mathbb{R})/U(1)_{\mathrm{A}}$ are those of
$SL(2,\mathbb{R})$ that are annihilated by the modes $J_n^3$ and
$\bar J_n^3$ $n>0$, and have $J_0^3$ and $\bar J_0^3$ eigenvalues
$(n-wk)/2$ and $-(n+wk)/2$. Therefore, in order to reconstruct the
$SL(2,\mathbb{R})$ partition function, we need to couple the coset
blocks with an appropriately chosen lattice for the Cartan
generators $J^3$ and $\bar J^3$ corresponding to a free time-like
boson. This coupling should mimic the $\mathbb{Z}_k$ free action
that appears in the $SU(2)_k$ (see the discussion for the $SU(2)$
in \cite{Kiritsis:1993ju}), in a non-compact parafermionic
version, though. Since the  non-compact parafermions have a
$\mathbb{Z}$ symmetry, we consider here a $\mathbb{Z}$ free
action. By using the conformal blocks for free bosons given in
Appendix B (see Eq. (\ref{coblofre})), we reach the following
partition function for the {\it universal cover} of
$SL(2,\mathbb{R})$, in the Lagrangian representation\footnote{To
find the partition function for the $N$-th cover, one has to
replace the $\mathbb{Z}$ orbifold by a $\mathbb{Z}_{Nk}$ orbifold:
$t_1 \to \frac{\gamma}{Nk}$, $t_2 \to \frac{\delta}{Nk}$. Then
  the left and right spectral flow are $w_{+}^{L,R} = w_{+} \pm N \ell$.}:
\begin{eqnarray}
Z_{SL(2,\mathbb{R})}&=& 4 \sqrt{\tau_2} (k-2)^{3/2} \int d^2 s \,
d^2 t \frac{ \mathrm{e}^{\frac{2\pi}{\tau_2} (\mathrm{Im} (s_1
\tau -s_2))^2 } }{\left|\vartheta_1 (s_1 \tau -s_2 |\tau )
\right|^2}\times  \nonumber \\
& &\times  \sum_{m,w,m',w'\in \mathbb{Z}} \zeta\oao{w + s_1 - t_1
}{m +s_2 - t_2 }\left(k \right)  \zeta\oao{w' + t_1 }{m' + t_2
}\left(-k \right). \label{fctpartsl}
\end{eqnarray}

Modular invariance is manifest in this expression, since it has
the structure of a freely acting orbifold (this can also be easily
checked by using formulas of Appendices B and D). The extra $k-2$
factor comes along with the $J^3$, $\bar{J}^3$ contribution; it
ensures the correct density scaling in the large-$k$ limit, as
explained in Appendix B about Eq. (\ref{coblofrecon}). We perform
a Poisson resummation\footnote{The resummation on $m'$ is of
course performed by means of analytic continuation, as usual when
dealing with a  time-like direction.} on $m$ and $m'$, which are
trade for $n$ and $n'$. We define $n^{\pm} = n \pm n'$ and
$w_{\pm} = w \pm w'$, and rewrite the partition function in the
Hamiltonian representation:
\begin{eqnarray}
Z_{SL(2,\mathbb{R})}&=& 4 \sqrt{\tau_2} (k-2)^{3/2} \int d^2 s \,
d^2 t \frac{ \mathrm{e}^{\frac{2\pi}{\tau_2} (\mathrm{Im} (s_1
\tau -s_2))^2 } }{\left|\vartheta_1 (s_1 \tau -s_2 |\tau )
\right|^2}\sum_{n^{\pm}, w_{\pm}\in \mathbb{Z}} \mathrm{e}^{-
i\pi \left(n^- ( s_{2}-2 t_{2}) + n^+ s_{2}  \right)}\times  \nonumber \\
& &\times \; \mathrm{e}^{-\frac{\pi \tau_2}{k} \left(n^+ n^- + k^2
(w_+ + s_{1})(w_-  + s_{1}- 2t_1) \right)+ i\pi \tau_1 \left( n^-
(w_- + s_{1} - 2t_1 ) + n^+ (w_{+} + s_{1} ) \right)} .
\label{partsl}
\end{eqnarray}

Expression (\ref{partsl}) becomes more transparent by introducing
light-cone directions with corresponding left and right lattice
momenta:
\begin{eqnarray}
P^+_\mathrm{L,R}&=& {n^+ \over \sqrt{2k}}\pm  \sqrt{k\over
2}w_{-},
\label{lcmom+} \\
P^-_\mathrm{L,R}&=& {n^- \over \sqrt{2k}}\pm  \sqrt{k\over
2}w_{+}. \label{lcmom-}
\end{eqnarray}
In terms of these unshifted\footnote{Due to the shifted-orbifold
structure of the partition function the relevant quantities are
actually the shifted momenta:
$$ P^{\mathrm{s}+}_\mathrm{L,R}= P^+_\mathrm{L,R}\pm \sqrt{k\over
2} (s_{1} -2t_1 )\ , \ \ P^{\mathrm{s}-}_\mathrm{L,R}=
P^-_\mathrm{L,R}\pm \sqrt{k\over 2}s_{1}.$$} momenta the partition
function at hand reads:
\begin{eqnarray}
Z_{SL(2,\mathbb{R})}&=& 4 \sqrt{\tau_2} (k-2)^{3/2} \int d^2 s \,
d^2 t \frac{ \mathrm{e}^{\frac{2\pi}{\tau_2} (\mathrm{Im} (s_1
\tau -s_2))^2 } }{\left|\vartheta_1 (s_1 \tau -s_2 |\tau )
\right|^2}\times  \nonumber \\
& &\times \; \sum_{n^{\pm}, w_{\pm}\in \mathbb{Z}} \mathrm{e}^{-
i\pi \sqrt{k\over 2} \left(\left(P^+_\mathrm{L} +
P^+_\mathrm{R}\right) s_{2}  + \left(P^-_\mathrm{L} +
P^-_\mathrm{R}\right)( s_{2}-2 t_{2})\right)}\times  \nonumber \\
& &\times \; q^{\frac{1}{2}  \left(P^+_\mathrm{L} + \sqrt{k\over
2}(s_{1}- 2t_1)\right)\left(P^-_\mathrm{L} + \sqrt{k\over 2}
s_{1}\right)}{\bar q}^{\frac{1}{2} \left(P^+_\mathrm{R} -
\sqrt{k\over 2}(s_{1}- 2t_1)\right)\left(P^-_\mathrm{R} -
\sqrt{k\over 2} s_{1}\right)}. \label{partslmom}
\end{eqnarray}

\subsection{Uncovering the spectrum}
\label{uncover}

The latter expression is formally divergent. It is in fact a
generalized function, which contains all the information about the
spectrum. The integral over $t_2$ leads to the constraint
$\delta_{n^-, 0}$ and, due to the shift $t_1$, the ``winding" $w_-
-2t_1$ is continuous. The momenta $P^{-}_{L,R}$ correspond
therefore to a boson $X^+$ ``compactified" at zero radius.
Conversely, the other light-cone degree of freedom is compact.
This \emph{light-cone compactification} (for a related discussion,
see~\cite{Hikida:2000ry}) is not so surprising. Indeed, the
quantum numbers of a given state in the $SL(2,\mathbb{R})$ CFT,
$j=\bar j$, $m+\bar{m}$, $ m-\bar{m}$, $w$ are respectively those
of a Liouville field, a non-compact coordinate, and a compact one.
Since we consider the universal cover of $SL(2,\mathbb{R})$, such
that the time is non-compact, the only possibility is that one of
the light-cone directions is compactified at radius $\sqrt{2k}$.
Note also a particular feature of a two-dimensional lattice for
two light-cone coordinates : if the radius of one light-cone
coordinate shrinks to zero, the momenta and windings of the other
light-cone coordinate are exchanged. This fact explains why the
energy, as it appears in the partition function of
$SL(2,\mathbb{R})$, is actually a (shifted) winding mode.

We will now proceed and analyze further the partition function
given in Eq. (\ref{partslmom}). After integrating out $t_2$ we
obtain:
\begin{eqnarray}
Z_{SL(2,\mathbb{R})}&=& 4 \sqrt{\tau_2} (k-2)^{3/2} \int d^2 s
\frac{ \mathrm{e}^{\frac{2\pi}{\tau_2} (\mathrm{Im} (s_1 \tau
-s_2))^2 } }{\left|\vartheta_1 (s_1 \tau -s_2 |\tau )
\right|^2}\sum_{n^{+}, w_{+}\in \mathbb{Z}} \mathrm{e}^{-i\pi
n^+\left(s_2 -\tau_1(w_{+} + s_{1})\right)}\times  \nonumber \\ &
&\times \int_0^1 dt_1 \sum_{w_{-}\in \mathbb{Z}}\mathrm{e}^{-\pi
\tau_2 k (w_+ + s_{1})(w_-  + s_{1}- 2t_1)} . \label{partslint}
\end{eqnarray}
It is straightforward to show that the large-$k$ limit of this
partition function is, up to the usual infinite-volume factor, $Z
\sim k^{3/2} \pi\left(\pi \sqrt{\tau_2} \eta\bar\eta\right)^{-3}$.
This was somehow built-in when writing (\ref{fctpartsl}) out of
(\ref{cospartcb}); it meets the expectations for ordinary
flat-space spectrum.

It is possible to extract the spectrum of the theory at any finite
$k$, and trace back its origin in terms of $SL(2,\mathbb{R})$
representations. We proceed along the lines of
~\cite{Maldacena:2000kv} and~\cite{Hanany:2002ev}. The precise
derivation of the spectrum is given in Appendix C. Here we only
collect the results.

\noindent\underline{Discrete representations.} The discrete
representations appear in the range $\frac{1}{2} < j <
\frac{k-1}{2}$. Their conformal weights are the following:
\begin{eqnarray}
L_0 = - \frac{j(j-1)}{k-2} + w_+ \left( -\tilde{m} - \frac{k}{4} w_+ \right) +N , \nonumber \\
\bar{L}_0 = - \frac{j(j-1)}{k-2} + w_+ \left( -\tilde{\bar{m}}-
\frac{k}{4} w_+ \right) +\bar{N}, \nonumber
\end{eqnarray}
with $\tilde{m}+\tilde{\bar{m}} = -k(w-t_1)$ and $\tilde{m}-\tilde{\bar{m}} = n$.

\noindent\underline{Continuous representations.} The continuous
spectrum appears with the density of states:
\begin{equation}
\rho (s) = \frac{1}{\pi}\log \epsilon + \frac{1}{4\pi i}
\frac{d}{ds} \log \frac{\Gamma \left(\frac{1}{2}-is-\tilde{m}
\right)\Gamma \left(\frac{1}{2}-is+\tilde{\bar{m}}\right) } {\Gamma
\left(\frac{1}{2}+ is-\tilde{m}\right) \Gamma
\left(\frac{1}{2}+is+\tilde{\bar{m}}\right) }\nonumber
\end{equation}
The weights of the continuous spectrum are
\begin{eqnarray}
L_0 = \frac{s^2 +1/4}{k-2} + w_+ \left( -\tilde{m} - \frac{k}{4} w_+ \right) +N,  \nonumber \\
\bar{L}_0 = \frac{s^2+1/4}{k-2} + w_+ \left( -\tilde{\bar{m}}- \frac{k}{4}
w_+ \right) +\bar{N},\nonumber
\end{eqnarray}
with $\tilde{m},\tilde{\bar{m}}$ as previously.

These results are in agreement with the unitary spectrum proposed
in~\cite{Maldacena:2000hw}. Here this spectrum was extracted
straightforwardly from a modular-invariant partition function,
constructed in the Lorentzian AdS$_3$ .

Our aim is now to better understand the coupling between the
oscillators and the zero modes of the light-cone coordinates, as
appearing in the partition function. To this end, we write the
algebra $\widehat{SL}(2,\mathbb{R})_L$ by using the free-field
representation of non-compact
parafermions~\cite{Griffin:1990zt}~\cite{Satoh:1997xe}.
The currents read:
\begin{eqnarray}
\label{ffrep}
J^{\pm}  &=&   -\left( \sqrt{\frac{k}{2}} \partial X \pm i
\sqrt{\frac{k-2}{2}}
\partial \rho \right) \mathrm{e}^{\pm i \sqrt{\frac{2}{k}}  (X-T)},\label{jpmfree}\\
J^{3}  &=&  i   \sqrt{\frac{k}{2}} \partial T ,\label{j3free}
\end{eqnarray}
with the following stress tensor:
\begin {equation}
T = \frac{1}{2} (\partial T)^2 -\frac{1}{2} (\partial X)^2 -
\frac{1}{2} (\partial \rho)^2 + \frac{1}{\sqrt{2(k-2)}} \partial^2
\rho.\label{Tfree}
\end{equation}
There is a linear dilaton with background charge $Q=
\sqrt{2/(k-2)}$ along the coordinate $\rho$. It contributes the
central charge, which adds up to $c=3+ 6/(k-2)$. The spectral flow
symmetry can be realized by adding $w_+$
units of momentum along $T$:
\begin{equation}
J^3  \to J^3 - \frac{k}{2z} w_+ \ , \ \ J^{\pm} (z) \to z^{\mp
w_+} J^{\pm} (z). \nonumber
\end{equation}

The primary operators are those of a free-field theory with a
peculiar zero-mode structure though, which is read off directly
from the lattice component of the partition function, Eq.
(\ref{partsl})\footnote{The fact that the roles of $X^+$ and $X^-$ are reversed
between the left-moving and the right-moving sectors will be
explained in the fifth section. For the moment note that
the partition function is invariant under:
$\bar{X}^{\pm} \to \bar{X}^{\mp}$.}:
\begin{equation}
\exp \Bigg\{ \sqrt{\frac{2}{k-2}} j \rho + i\sqrt{\frac{2}{k}}
\left[ \frac{k}{4} w_+ X^+ + \left(\tilde{m}-\frac{k}{4} w_+
\right)  X^- + \frac{k}{4} w_+ \bar{X}^- + \left(\tilde{\bar{m}}
-\frac{k}{4} w_+ \right) \bar{X}^+ \right] \Bigg\}. \nonumber
\end{equation}
One should stress, however, that even if the theory can be
represented with free fields, the descendants are constructed by
acting with the modes of the affine currents: the oscillator
number and the zero modes are shifted simultaneously. We are
therefore lead\footnote{This is close to the construction of
gravitational waves, see ~\cite{Kiritsis:1994ij}.} to use the
the Lagrange multipliers $s_1$ and $s_2$ in the partition
function~(\ref{partsl}) to enforce this twisting.

\subsection{About the structure of the partition function}

Our approach has been to build a modular-invariant partition
function for $SL(2,\mathbb{R})$ starting from that of the coset
model $SL(2,\mathbb{R})/U(1)_{\mathrm{A}}$. We have reached
expression (\ref{fctpartsl}) or equivalently (\ref{partslmom}).
These expressions are generalized functions which are formally
divergent, as was originally the partition function for the coset,
Eq. (\ref{cospartcb}). However, the presence of a divergence is
not an obstruction for uncovering the spectrum encoded in the
partition functions, as shown in \cite{Hanany:2002ev} for the
coset and here for the AdS$_3$. In this section, we would like to
make contact with the -- not fully satisfactory -- expressions
found in \cite{Petropoulos:1999nc} and \cite{Maldacena:2000hw},
explain why the methods used previously failed, and clarify the
underlying freely acting orbifold structure.

We start with Eq. (\ref{partslint}) that we regulate by shifting
$s_1 \tau - s_2 \to s_1 \tau - s_2 + \theta$ in the elliptic
theta function. This breaks modular
invariance unless, together with $\tau\to -1/\tau$, $\theta$
transforms into $- \theta/\tau$. Then, summation over $w_-$ and
integration over $t_1$ can be merged into an integration over the
{\it light-cone energy} $E= k (w_- - 2t_1)$, which is performed
after analytic continuation. A Poisson resummation over $n^+$
finally leads to the following result:
\begin{eqnarray}
Z_{SL(2,\mathbb{R})} &=& 4 \frac{(k-2)^{3/2}}{k\sqrt{\tau_2}} \int
d^2s \sum_{m_+ ,w_+ \in \mathbb{Z}} \delta \left( w_+ + s_{1}
\right) \delta \left( m_+ + s_{2} \right) \frac{
\mathrm{e}^{\frac{2\pi}{\tau_2} (\mathrm{Im} (s_1 \tau -s_2 +
\theta))^2 } }{\left|\vartheta_1 (s_1 \tau -s_2 + \theta |\tau )
\right|^2}\nonumber \\
&=& 4 \frac{(k-2)^{3/2}}{k\sqrt{\tau_2}}
\frac{\mathrm{e}^{\frac{2\pi}{\tau_2} (\mathrm{Im} \theta)^2}}
{\left|\vartheta_1 \left( \theta |\tau \right) \right|^2}.
\label{intsl}
\end{eqnarray}
This result is precisely that of~\cite{Petropoulos:1999nc} and
\cite{Maldacena:2000hw}, up to an overall normalization.

In unitary conformal field theories, the partition function is
usually decomposed in characters of the chiral holomorphic and
anti-holomorphic algebras:
\begin{equation}
Z_{\mathrm{genus-one}}(\tau,\bar \tau)=
\sum_{\mathrm{L,R}}\mathcal{N}_{\mathrm{L,R}}
\chi_{\mathrm{L}}(\tau) \bar \chi_{\mathrm{R}}(\bar \tau),
\nonumber
\end{equation}
where the summation is performed over all left-right
representations present in the spectrum with multiplicities
$\mathcal{N}_{\mathrm{L,R}}$, and $\chi(\tau) $ are the
corresponding characters:
\begin{equation}
\chi(\tau) = \mathrm{Tr}_{\mathrm{rep}} q^{L_0 -{c\over 24}}.
\nonumber
\end{equation}
This decomposition is very powerful for classifying models (i.e.
multiplicities $\mathcal{N}_{\mathrm{L,R}}$) by following the
requirements of modular invariance. From the path-integral point
of view, different modular-invariant combinations correspond to
different choices for boundary conditions on the fields. However,
this decomposition relies on the very existence of the characters.
This holds for WZW models on compact groups. It does not apply to
the case of non-compact groups, unless the group is Abelian --
free bosons. Then the zero-mode representations are
one-dimensional, the characters of the affine algebra are
well-defined, and the infinite-volume divergence decouples into an
overall factor. For non-Abelian groups, unitary\footnote{If we
give up unitarity, finite-dimensional zero-mode representations do
exist, but Virasoro conditions do not eliminate all spurious
states.} representations of the zero-modes are
infinite-dimensional, and the characters of the affine algebra
diverge. This degeneracy can be lifted by switching on a source
coupled to some Cartan generator:
\begin{equation}
\chi (\tau, \theta) = \mathrm{Tr}_{\mathrm{rep}}q^{L_0 - {c\over
24}} \mathrm{e}^{2i\pi \theta J_0}.\nonumber
\end{equation}
Such a
regularization is not fully satisfactory because it alters
modular-covariance and does not allow to cure the characters of
the continuous part of the spectrum. Moreover, the best these
generalized characters can do, is to lead (after formal
manipulations) to expressions like (\ref{intsl}). The information
carried by the latter is quite poor: it diverges at $\theta \to 0$
and the divergence cannot be isolated as a volume factor; the
large-$k$ limit is obscure; only the discrete part of the spectrum
seems to contribute. These caveats are avoided in the integral
representation we have presented here (Eqs. (\ref{fctpartsl}) or
(\ref{partslmom})), which is closer in spirit with the work of
\cite{Maldacena:2000kv} for the thermal AdS$_3$. Although
divergent, it is modular-invariant, contains a nice spectral
decomposition and has a well-defined large-$k$ limit in agreement
with our expectations.

To close this chapter, we would like to comment on the freely
acting orbifold structure implemented in Eq. (\ref{fctpartsl}). We
consider for illustration the $\mathbb{Z}_N$ orbifold of a compact
boson of radius $R$ times a two-torus, presented in Appendix B.
The $\mathbb{Z}_N$ acts as a twist on $T^2$ and as a shift on the
orthogonal $S^1$. The partition function for this model is given
by Eq. (\ref{ZN}). Although it is not compatible with the
symmetries of a two-dimensional lattice, we take formally the
large-$N$ limit of this expression. The first term of (\ref{ZN})
vanishes while the sum over $h,g$ in the second term becomes an
integral over $s_1,s_2 \in [0,1]$. We also drop out the
geometrical factor $\sin^2 \pi \frac{\Lambda (h,g)}{N}$ which is
meaningless here, and find:
\begin{equation}
\lim_{N\to \infty}\frac{Z_{\mathbb{Z}_N}}{N} =  4 \int d^2s \frac{
\mathrm{e}^{\frac{2\pi}{\tau_2} (\mathrm{Im} (s_1 \tau -s_2))^2 }
}{\left|\vartheta_1 (s_1 \tau -s_2 |\tau ) \right|^2} \sum_{m,w
\in \mathbb{Z}}\zeta\oao{w+s_1}{m+s_2}\left(R^2\right) .
\label{larZN}
\end{equation}
In order to make final contact with the partition function of the
$SL(2,\mathbb{R})/U(1)_{\mathrm{A}}$ (Eq. (\ref{cospartcb})) we
must identify $R^2$  with $k$, and mod out a non-compact free
boson, i.e. multiply (\ref{larZN}) by $\sqrt{\tau_2} \, \eta
\bar{\eta}$. We insist that we do not claim that the theory
$SL(2,\mathbb{R})/U(1)_{\mathrm{A}}$ is the same as an freely
acting orbifold of flat space, but only that the structure is very
similar. An important difference is that the oscillators of the
field $X$ of the free-field representation (Eq.~(\ref{ffrep})) are
twisted and its zero modes shifted simultaneously. This is not
possible in flat space.

One can similarly understand the orbifold structure underlying the
full $SL(2,\mathbb{R})$ model. To this end we consider the
$\mathbb{Z}_N\times \mathbb{Z}_N$ model given in Appendix B. In
the formal large-$N$ limit, all but the last term vanish in the
partition function (\ref{ZNZN}); the sums over $h_1,g_1$ and
$h_2,g_2$ are trade for integrals over $s_1,s_2$ and $t_1,t_2 \in
[0,1] \times [0,1]$:
\begin{eqnarray}
\lim_{N\to \infty}\frac{Z_{\mathbb{Z}_N\times
\mathbb{Z}_N}}{N^2}&=& 4 \frac{1}{\eta \bar\eta}\int d^2 s \, d^2
t \frac{ \mathrm{e}^{\frac{2\pi}{\tau_2} (\mathrm{Im} (s_1 \tau
-s_2))^2 } }{\left|\vartheta_1 (s_1 \tau -s_2 |\tau )
\right|^2}\times  \nonumber \\
& &\times  \sum_{m_1,w_1,m_2,w_2\in \mathbb{Z}} \zeta\oao{w_1 +
s_1 - t_1 }{m_1 +s_2 - t_2 }\left(R_1^2\right)\,  \zeta\oao{w_2 +
t_1 }{m_2 + t_2 }\left(R_2^2 \right).\label{larZNZN}
\end{eqnarray}
Comparison with the partition function of the $SL(2,\mathbb{R})$
(Eq. (\ref{fctpartsl})) is possible provided we identify in
(\ref{ZNZN}), $R_1^2$ with $k$, $R_2^2$ with $-k$, and mod out a
non-compact free boson, i.e. multiply Eq. (\ref{larZNZN}) by
$\sqrt{\tau_2} \, \eta \bar{\eta}$.

The function $\vartheta_1 (0,\tau ) $, which is
identically zero, never appears in the orbifold since its
corresponds to the untwisted, unprojected sector and
is replaced by the usual toroidal partition sum:
$$\frac{\Gamma_{2,2} (T,U)}{\eta^4 \bar{\eta}^4}.$$
In the case of  $SL(2,\mathbb{R})$ (the $N \to \infty$ limit), the
sum on the sectors is replaced by an integral over $s_1$ and
$s_2$. The integration over the energy picks up precisely the
untwisted, unprojected sector, as it gives the constraint
$\delta^{(2)} (s_1 \tau -s_2 )$. We can rewrite formally the
integrated partition function in terms of the functional
determinant for the twisted bosons:
$$Z = \frac{(k-2)^{3/2}}{k\sqrt{\tau_2}\eta \bar{\eta}} \int d^2 s \
\delta^{(2)} (s_1 \tau - s_2 ) \ \mathrm{det} \left| \partial +
\frac{1}{2\tau_2} \left( s_1 \bar{\tau} - s_2 \right) \right|^{-2} =
\frac{(k-2)^{3/2}}{k\sqrt{\tau_2}\eta \bar{\eta}} \ \mathrm{det}
\left|
\partial \right|^{-2}.$$
Thus we find that the partition function of $SL(2,\mathbb{R})$ is
the same as a linear dilaton and two light-cone free coordinates
\footnote{As already stressed, there is a central charge deficit
coming for the other CFT's defining the string theory which
corresponds to the lowest-weight of Liouville continuous
representations~\cite{Antoniadis:1994sr}.}.

\boldmath
\section{Superstrings on AdS$_3 \times S^3 \times T^4$}\label{sect3}
\unboldmath

We develop in this chapter the supersymmetry tools, which are
needed for studying superstrings on NS5- or NS5/F1-brane
backgrounds, as well as on the deformed AdS$_3$ geometries
interpolating between them. This includes explicit realizations of
extended $N=2$ and $N=4$ supersymmetry algebras. In principle it
is possible to construct a space--time supersymmetric string
background with $SL(2,\mathbb{R}) \times SU(2) \times
\mathcal{M}$, where $\mathcal{M}$ is any $N=2$ superconformal
field theory with the correct central charge $\hat{c} = 4$. To
make contact with the NS5/F1 background, we can choose either
$T^4$ or a CFT realization of $K3$. We finally present the
partition function of the model on AdS$_3 \times S^3 \times K3$.

\subsection{Extended superconformal algebras}

Since the AdS$_3 \times S^3 \times T^4$ background preserves one
half of the supersymmetry, the worldsheet theory should factorize
into an $N=4$ superconformal theory with $\hat{c} =4$ and an $N=2$
free theory with $\hat{c}=2$~\cite{Banks:1988yz}. An explicit
realization of the relevant extended algebras is necessary in
order to prove that supersymmetry survives at the string level; it
is also important for the determination of the couplings between
the bosonic and the fermionic degrees of freedom. However, as we
will see, the straightforward application of the rules of $N=2$
constructions is not the correct way to implement space--time
supersymmetry in $SL(2,\mathbb{R}) \times SU(2)$.

\subsubsection{NS5 background}\label{chs}
We would like here to remind the construction of the ``small'' $N=4$
superconformal algebra for the wormhole
background~\cite{Kounnas:1990ud}, which is the
simplest case beyond flat space~\cite{Ademollo:1976wv};
it is also relevant for discussing the supersymmetric null deformation
of $SL(2,\mathbb{R})$.

The near-horizon geometry of the NS5-brane background is the
target space of an exactly conformal sigma-model based on $SU(2)_k
\times U(1)_Q\times U(1)^6$~\cite{Callan:1991ky}. Let us
concentrate on the $SU(2)_k \times U(1)_Q$ factor. The full
algebra of this four-dimensional internal subspace consists of the
bosonic $SU(2)_k$, the Liouville coordinate, and four free
fermions:
\begin{eqnarray}
J^i (z) J^j (0) & \sim & \frac{k}{2}\frac{\delta^{ij}}{z^2} +  i
\sum_{\ell = 1}^3 \frac{\epsilon^{ij\ell} J^\ell(0)}{z}
\sp i,j, \ell=1,2,3, \nonumber \\
\partial \rho (z) \partial \rho (0) & \sim & - \frac{1}{z^2},  \nonumber \\
\psi^a (z) \psi^b (0) & \sim & \frac{\delta^{ab}}{z} \sp
a,b=1,\ldots,4.\nonumber
\end{eqnarray}
The ``small'' $N=4$ algebra is generated by twisting the ``large''
$N=4$ algebra~\cite{Sevrin:1988ew} based on the affine symmetry
$SU(2)_{k+1} \times SU(2)_1 \times U(1)$. The generators of that
algebra read:
\begin{eqnarray}
T & = & \frac{1}{k+2} \sum_{i = 1}^3 J^i J^i - \frac{1}{2}
\partial \rho \partial \rho
- \frac{1}{2} \sum_{a = 1}^4\psi^a \partial \psi^a , \nonumber \\
G^{i} & = &  \sqrt{\frac{2}{k+2}} \left[-J^i \psi^4 +\sum_{j, \ell
= 1}^3  \epsilon^{ij\ell} \left(J^j -
\psi^4 \psi^j\right) \psi^\ell \right] + i \psi^i \partial \rho  ,\nonumber \\
G^{4} & = &  \sqrt{\frac{2}{k+2}} \sum_{i = 1}^3\left[J^i \psi^i +
\frac{1}{3} \sum_{j, \ell = 1}^3\epsilon^{ij\ell} \psi^i \psi^j
\psi^\ell
\right] + i \psi^4 \partial \rho  ,\label{extN4} \\
S^{i}  & = & \frac{1}{2} \left( \psi^4 \psi^i +
\frac{1}{2}\sum_{j, \ell = 1}^3 \epsilon^{ij\ell} \psi^j \psi^\ell
\right) , \nonumber \\
\tilde{S}^{i} & = &  \frac{1}{2} \left( \psi^4 \psi^i -
\frac{1}{2} \sum_{j, \ell = 1}^3\epsilon^{ij\ell} \psi^j \psi^\ell
\right) + J^i ,  \nonumber
\end{eqnarray}
where $\rho$ is an ordinary bosonic coordinate.

The large $N=4$ algebra is contracted to the required small $N=4$,
provided the $\rho$ coordinate is promoted to a Liouville field by
adding a background charge $Q$ (i.e. a linear dilaton in the
corresponding direction). The effect on the algebra (\ref{extN4})
is the following:
\begin{equation}
T \to  T - Q \partial^2 \rho \sp  G^a \to G^a - iQ
\partial \psi^a.\nonumber
\end{equation}
The background charge $Q$ is such that we obtain a $\hat{c}=4$
theory: $Q = \sqrt{2/(k+2)}$. The linear dilaton background
compensates the central charge deficit of the $SU(2)_k$.\\
We bosonize the self-dual combination of fermions:
\begin{equation}
i \sqrt{2} \partial H^{+} =  \psi^1 \psi^2 + \psi^4 \psi ^3.
\nonumber
\end{equation}
It defines the $R$-symmetry $SU(2)$ algebra {\it at level
one} generated by the
currents:
\begin{equation}
(S^3,S^\pm) = \left(\frac{i}{\sqrt{2}} \partial H^+ ,
\mathrm{e}^{\pm i \sqrt{2} H^+} \right).\label{RN4}
\end{equation}

The resulting ``small'' ($N=4$)-algebra generators
are combined into two conjugate $SU(2)$ $R$-symmetry doublets:
\begin{eqnarray}
 G^+ , \tilde{G}^- & = & \left[ Q J^-
\mathrm{e}^{\frac{i}{\sqrt{2}} H^-} + i\left(
  Q\left(J^3 + i \sqrt{2} \partial H^-\right) -i\partial \rho \right)
\mathrm{e}^{\frac{-i}{\sqrt{2}} H^-} \right] \mathrm{e}^{\frac{\pm i}{\sqrt{2}} H^+}, \label{NS5alg1} \\
\tilde{G}^+, G^- & = & \left[ Q J^+
\mathrm{e}^{\frac{-i}{\sqrt{2}} H^-} + i\left(
  Q\left(J^3 + i\sqrt{2} \partial H^-\right) + i\partial \rho  \right)
\mathrm{e}^{\frac{i}{\sqrt{2}} H^-} \right] \mathrm{e}^{\frac{\pm
i}{\sqrt{2}} H^+} \label{NS5alg2}.
\end{eqnarray}

\subsubsection{NS5/F1 background}

We now move to the AdS$_3 \times S^3 \times T^4$ background, which
describes the near-horizon geometry of the NS5/F1-brane system.
Our focus is the six-dimensional $SL(2,\mathbb{R}) \times SU(2)$
subspace that we want to split into one $\hat{c} = 4$ system with
$N=4$ superconformal symmetry and one with $\hat{c}=2$ with $N=2$.
The total central charge of this factor is given by:
\begin{equation} \hat{c} = \frac{3k_{SL(2,\mathbb{R})}}{k_{SL(2,\mathbb{R})}-2} +
\frac{3k_{SU(2)}}{k_{SU(2)}+2}. \label{chat}
\end{equation}
Therefore to obtain a critical string background ($\hat{c}=6$) for any
level, we must choose
\begin{equation}
k_{SL(2,\mathbb{R})} -4 = k_{SU(2)} = k.\label{kshift}
\end{equation}
The $N=1$ algebra of the theory is generated by
\begin{eqnarray}
T & = & \frac{1}{k+2} I_i I_j \delta^{ij} + \frac{1}{k-2}
J_{\alpha} J_{\beta} \eta^{\alpha \beta} -\frac{1}{2} \psi_i
\partial \psi^i  -\frac{1}{2} \chi_{\alpha} \partial
\chi^{\alpha},\label{enmo}
\\
G & = & \sqrt{\frac{2}{k+2}} \left[ \psi_i I^i - \frac{i}{3}
\epsilon_{ij\ell} \psi^i \psi^j \psi^\ell + \chi_{\alpha}
J^{\alpha}- \frac{i}{3} \epsilon_{\alpha \beta \gamma}
\chi^{\alpha} \chi^{\beta} \chi^{\gamma} \right],\label{supercurr}
\end{eqnarray}
where $I^i$ and $J^{\alpha}$ denote respectively the bosonic
currents of $SU(2)$ and $SL(2,\mathbb{R})$, $\psi^i$ and
$\chi^{\alpha}$ the corresponding fermions, and $\eta^{\alpha
\beta} = (+,+,-)$\footnote{Indices $i,j,\ldots$ and $\alpha,
\beta, \ldots$ run both over $1,2,3$, and we raise them with
$\delta^{ij}$ and $\eta^{\alpha \beta}$.}.
\boldmath
\paragraph{The $N=2$ algebras of
$SL(2,\mathbb{R}) \times SU(2)$.} \unboldmath

The various currents provided by the $SU(2)$ and
$SL(2,\mathbb{R})$ algebras and the associated fermions allow for
extracting one $N=2$, $\hat{c}=2$ algebra generated by:
\begin{eqnarray}
G_{2}^{\pm} & = & \frac{1}{\sqrt{2(k+2)}} \left[ \left(I^3  +
\psi^+ \psi^- \right)
\mp \left(J^3 + \chi^+ \chi^- \right) \right]\left(\psi^3 \pm \chi^3\right), \label{N2G} \\
J_{2} & = & \psi^3 \chi^3. \label{N2J}
\end{eqnarray}
We have combined the currents and the fermions as follows:
$$J^{\pm} = J^1 \pm i J^2 \ , \ \ I^{\pm} = I^1 \pm i I^2 \ , \ \
\psi^{\pm} = \frac{\psi^1 \pm i \psi^2}{\sqrt{2}} \ , \ \
\chi^{\pm} = \frac{\chi^1 \pm i \chi^2}{\sqrt{2}}.$$ The remaining
generators form another $N=2$ algebra decoupled from the first
one~\cite{Ivanov:ec}:
\begin{eqnarray}
G_{4}^{\pm} & = & \frac{1}{\sqrt{k+2}} \left[ I^{\mp} \psi^{\pm}
-i J^{\mp} \chi^{\pm} \right], \label{extsca1} \\
S^3 & = & \frac{1}{2(k+2)} \left( 2J^3 + (k+4) \chi^+ \chi^- -
2I^3  + k \psi^+ \psi^- \right). \label{extsca2}
\end{eqnarray}
The various coefficients in $S^3$ are such that: (\romannumeral1)
$S^3$ is regular with respect to $G_{2}^{\pm}$ in order to obtain
two independent algebras, and (\romannumeral2) $S^3(z)
G_{4}^{\pm}(0) \sim \pm \ G_{4}^{\pm}(0) /2z$. The normalization
of $S^3$ follows from $S^3(z) S^3(0) \sim 1/2z^2$. Therefore we
rewrite it in terms of a free boson as follows:
\begin{eqnarray}
\frac{i}{\sqrt{2}} \partial H^+ & = & \frac{1}{2} \left( \psi^+
\psi^- + \chi^+ \chi^- \right) + \frac{1}{k+2} \left( J^3 + \chi^+
\chi^- - I^3  - \psi^+ \psi^- \right)
\nonumber \\
& = & \frac{1}{2} \left( \psi^+ \psi^- + \chi^+ \chi^- \right) +
\frac{1}{k+2} \left( \mathcal{J}^3 - \mathcal{I}^3 \right).
\label{Rsymm}
\end{eqnarray}
We have introduced the total currents, including the fermionic part:
\begin{equation}
\mathcal{J}^3 = J^3 +  \chi^+ \chi^- \ \ \mathrm{and} \ \
\mathcal{I}^3 = I^3 + \psi^+ \psi^- ,\label{calJI}
\end{equation}
respectively for $SL(2,\mathbb{R})$ and $SU(2)$, both at level
$k+2$.

We would like to extend the superconformal symmetry to $N=4$, as
we expect from the target-space supersymmetry. This is possible
provided the $R$-symmetry current $i\partial H^+$ corresponds to a
free boson compactified at the self-dual radius, which seems
indeed the case here since the total currents $\mathcal{J}^3$ and
$\mathcal{I}^3$ are correctly normalized in Eqs.~(\ref{Rsymm}),
and can both be bosonized at radius $\sqrt{2(k+2)}$. However, in
order to form the superconformal characters of $SU(2)$ and
$SL(2,\mathbb{R})$, the characters of these currents are coupled
to the $N=2$ coset theories:
$$\frac{SU(2)}{U(1)} \times U(1)_{R=\sqrt{\frac{k}{k+2}}} \ \ \mathrm{and} \ \
\frac{SL(2,\mathbb{R})}{U(1)} \times U(1)_{R=\sqrt{\frac{k+2}{k}}}.$$
The coupling acts as a $\mathbb{Z}_{k+2}$ shift in the lattice of
$\mathcal{J}^3$ and $\mathcal{I}^3$; it is similar to the discussion
about the bosonic  $SL(2,\mathbb{R})$. Therefore, the charges of the
current $S^3$ are not those of a boson at self-dual radius.

\paragraph{Space--time supersymmetry.}

For a worldsheet superconformal theory with {\it accidental} $N=2$
superconformal symmetry, the space--time supersymmetry charges are
obtained by spectral flow of the $R$-symmetry
current~\cite{Gepner:1987qi}. However, this requires that the
charges of all the physical states with respect to this $U(1)$
current are {\it integral}. As we have seen above, the
$R$-symmetry current of the $\hat{c}=4$ block, whose expression is
given in~(\ref{Rsymm}) does not fulfill this requirement, because
its charges depend on the eigenvalues of $\mathcal{I}^3$ and
$\mathcal{J}^3$. Moreover, space--time supercharges based on the
above $N=2$ current lead to incompatibilities with target-space
symmetries~\cite{Berenstein:1999gj}. These problems arise because,
even in flat space, the space--time supercharges are constructed
with the fermion vertex operators at zero momentum. In the present
case, the space--time momentum enters directly in the $N=2$
charges and the $SU(2)$ charges, though compact, cannot be
considered as internal.

We will proceed as in~\cite{Giveon:1998ns}: we construct directly
the space--time supercharges with the spin fields of the free
fermions, which are BRST invariant and mutually local. This seems
sensible, since the fermions in a critical string theory based on
WZW models are free. The $N=2$ current of the theory:
\begin{equation}
J = J_2 + 2S^3 = \frac{2}{k+2} \left( \mathcal{J}^3 - \mathcal{I}^3 \right) +
\psi^+ \psi^- + \chi^+ \chi^-  + \psi^3 \chi^3, \nonumber
\end{equation}
differs from the free-fermionic one by the ``null" contribution
$\left(\mathcal{J}^3 - \mathcal{I}^3 \right) / (k+2)$. We can of
course wonder whether a more appropriate choice of $N=2$ structure
exists. Another choice of complex structure does exist, and is
provided by decomposing the $SL(2,\mathbb{R})$ as
$SL(2,\mathbb{R})/O(1,1) \times U(1)$. It suffers, however, from
the same problem.

\boldmath
\paragraph{The required projections in $SL(2,\mathbb{R})
\times SU(2) \times  U(1)^4$.} \unboldmath

Here we come to the full background AdS$_3 \times S^3 \times T^4$.
The worldsheet fermions of the $T^4$ are bosonized as:
\begin{equation}
\lambda^1 \lambda^2 = \partial H_3 \ , \ \ \lambda^3 \lambda^4 =
\partial H_4, \nonumber
\end{equation}
and those of $SU(2) \times SL(2,\mathbb{R})$ as:
\begin{equation}
\psi^+ \psi^-  =  i \partial H_2 \sp \chi^+ \chi^-  =   i
\partial H_1 \sp \psi^3 \chi^3  =  i \partial H_0.
\nonumber
\end{equation}
In the $-1/2$ picture, the spin fields are
\begin{equation}
\Theta_{\mathbf{\varepsilon}} (z) = \exp \left\{ -  {\varphi \over
2} + \frac{i}{2} \sum_{\ell=0}^4 \varepsilon_\ell H_\ell \right\},
\label{spfi}
\end{equation}
where $\mathrm{e}^{-\varphi /2}$ is the bosonized superghost
ground state in the Ramond sector~\cite{Friedan:1985ge}.

The standard GSO projection keeps all spin fields satisfying
\begin{equation}
\varepsilon_0 \varepsilon_1 \varepsilon_2 \varepsilon_3 \varepsilon_4 = 1, \label{GSO}
\end{equation}
which is required by BRST invariance. In type IIB superstrings,
the GSO projection is the same on the spin fields from the right-moving sector, while for
type IIA it is the opposite.
In the AdS$_3 \times S^3
\times T^4$ background, the fields (\ref{spfi}) must obey the
additional relation:
\begin{equation}
\varepsilon_0 \varepsilon_1 \varepsilon_2 = 1. \label{projadd}
\end{equation}
Equivalently, for type IIB, by using  the GSO projection~(\ref{GSO}), the restriction
on the allowed spin fields can be imposed on the fermions of the four-torus:
\begin{equation}
\varepsilon_3 \varepsilon_4 = 1. \label{projad}
\end{equation}
Relations (\ref{projadd}) or (\ref{projad}) ensure the absence of
$1/z^{3 \over 2}$ poles in the OPE of the spin fields with the
$N=1$ supercurrent~(\ref{supercurr}) that would otherwise appear
as a consequence of the torsion terms. Note also that in the
superconformal algebra for the right-moving sector, the torsion
terms come with a negative sign in the supercurrent, but the
projection remains the same.

In order to preserve the $N=1$ supercurrent, we must implement the
projection (\ref{projad}) as a $\mathbb{Z}_2$ orbifold on the
coordinates of the four-torus. This is why we are effectively
dealing with the background AdS$_3 \times S^3 \times K3$. Then,
from the S-dual viewpoint, the model we are describing consists of
D5-branes wrapped on a $K3$ manifold and D-strings. It is
known~\cite{Bershadsky:1995qy} that in this case there is an {\it
induced} D-string charge which is the opposite of the total
D5-brane charge. Indeed, including gravitational corrections to
the Wess--Zumino term of the D5-brane action, generates the
following D1-brane charge:
\begin{equation}
Q_{1}^\mathrm{ind} = N_5 \int \frac{c_2 (K3)}{24} = -N_5.
\nonumber
\end{equation}
Therefore the net number of D1-branes accompanying the D5-branes
is $N_1 + N_5$. This has little effect on the S-dual model under
consideration here, because the number of D-strings affects the
value of the string coupling but not the worldsheet CFT.

In conclusion, we have seen that, in order to obtain a
supersymmetric spectrum consistent with the BRST symmetry, we have
to project out half of the space--time spinors from the Ramond
sector. By modular invariance, this projection must act as an
orbifold on the fermionic characters and also on the bosonic part
to be consistent with superconformal invariance. For simplicity we
have chosen to act on the fermions associated with the four-torus;
therefore the background is now AdS$_3 \times S^3 \times T^4 /
\mathbb{Z}_2$. The $T^4 / \mathbb{Z}_2$ orbifold can be replaced
by another realization of $K3$ in CFT, such as a Gepner
model~\cite{Gepner:1987qi}. Another way to realize the projection
is to twist the fermionic characters associated with AdS$_3 \times
S^3$. In that case, we have to act nontrivially on the
$SL(2,\mathbb{R})$  and $ SU(2)$ bosonic characters. Then the
toroidal part of the background remains untwisted: the NS5-branes
are wrapped on a $T^4$ rather than on a $K3$.

\boldmath
\subsection{The partition function for superstrings on AdS$_3 \times S^3 \times K3$}
\unboldmath

We are now in position to write the partition function for type
IIB superstrings on AdS$^3 \times S^3 \times K3$. We must combine
the various conformal blocks in a modular-invariant way, and
impose the left and right GSO projections together with the
additional projections dictated by the presence of torsion.

The standard orbifold conformal blocks are given in Appendix B.
The $SU(2)_k$ partition function is chosen to be the diagonal
modular-invariant combination \cite{Cappelli:xt} and the
$SL(2,\mathbb{R})$ factor was discussed in Sec. \ref{sect2}.
Putting everything together, including the conformal and
superconformal ghosts, we obtain:
\begin{eqnarray}
Z_{\rm IIB} & = & {\mathrm{Im} \tau \over  \eta^2 \bar\eta^2 }
Z_{SU(2)} Z_{SL(2,\mathbb{R})} {1\over 2} \sum_{h,g=0}^1
Z_{T^4/{\mathbb{Z}_2}}^{\rm twisted} \oao{h}{g} \nonumber \\
&&\nonumber \\
&& \times {1 \over 2} \sum_{a,b=0}^1 (-)^{a+b}\
\vartheta^2 \oao{a}{b} \vartheta \oao{a+h}{b+g}
\vartheta \oao{a-h}{b-g} \nonumber \\
&&\nonumber \\
&& \times {1 \over 2} \sum_{\bar{a},\bar{b}=0}^1
(-)^{\bar{a}+\bar{b}} \ \bar{\vartheta}^2 \oao{\bar{a}}{\bar{b}}
\bar{\vartheta} \oao{\bar{a}+h}{\bar{b}+g} \bar{\vartheta}
\oao{\bar{a}-h}{\bar{b}-g} . \label{zII}
\end{eqnarray}

We can read from the latter expression the spectrum of chiral
primaries with respect to the space--time superconformal
algebra~\cite{Giveon:1998ns}. The vertex operators for such
left-moving states in the NS sector are given in the ($-1$) ghost
picture by (see~\cite{Kutasov:1998zh}):
\begin{eqnarray}
\mathcal{V}^{I}_j & = & \mathrm{e}^{- \varphi} \ \lambda^I \
\Phi^{SL(2,\mathbb{R})}_{j,\ m} \ \Phi^{SU(2)}_{j-1, \ m'}
\nonumber \\
\mathcal{W}^{\pm}_{j} & = & \mathrm{e}^{-\varphi} \ \left[\chi
\Phi^{SL(2,\mathbb{R})} \right]_{j \pm 1, \ m}
\ \Phi^{SU(2)}_{j-1, \ m'} \nonumber \\
\mathcal{X}^{\pm}_j & = & \mathrm{e}^{-\varphi} \
\Phi^{SL(2,\mathbb{R})}_{j, \ m} \ \left[ \psi
\Phi^{SU(2)}\right]_{j-1 \pm 1, \ m'}\nonumber
\end{eqnarray}
where $\Phi^{SU(2)}_{j',m'}$ and $\Phi^{SL(2,\mathbb{R})}_{j,m}$
are respectively the bosonic primary fields of the holomorphic
current algebras $SU(2)_{k}$ and $SL(2,\mathbb{R})_{k+4}$. They
are combined with the worldsheet fermions into representations of
$SU(2)_{k+2}$ and $SL(2,\mathbb{R})_{k+2}$.

The above states live in the five-plus-one dimensional
world-volume of the NS5-branes. In order to obtain the closed
string spectrum, we tensorize this left-moving spectrum with the
right-moving one, and impose the $\mathbb{Z}_2$ projection on the
torus. Additional states localized in one-plus-one dimensions are
constructed with the twisted sectors of the orbifold:
\begin{eqnarray}
\tilde{\mathcal{W}}^{\pm}_{j} & = & \mathrm{e}^{- \mathcal
\varphi} \ \mathcal{V}^{\ tw} \left[\psi \Phi^{SU(2)} \right]_{j
\pm 1, \ m} \ \Phi^{SL(2,\mathbb{R})}_{j-1, \ m'},
\nonumber \\
\tilde{\mathcal{X}}^{\pm}_{j} & = & \mathrm{e}^{- \mathcal
\varphi} \ \mathcal{V}^{\ tw} \ \Phi^{SU(2)}_{j, \ m} \ \left[
\chi \Phi^{SL(2,\mathbb{R})}\right]_{j-1 \pm 1, \ m'},\nonumber
\end{eqnarray}
where $\mathcal{V}^{\ tw}$ are the twist fields of the NS sector.

\boldmath
\section{Marginal deformations of $SL(2,\mathbb{R})$}\label{sect4}
\unboldmath

The $SL(2,\mathbb{R})$ geometry in Euler (global) coordinates
reads\footnote{We systematically set the AdS$_3$ radius to one in
the expressions for the background fields. One has to keep in
mind, however, that a factor $k$ is missing in the metric. This
plays a role when performing T-dualities by applying the Buscher
rules \cite{Buscher:sk} because it can affect the periodicity
properties of some coordinates.}:
\begin{eqnarray}
ds^2 & = & dr^2 -\cosh^2 r dt^2 + \sinh^2 r d\phi^2, \nonumber \\
B & = & \cosh^2 r d\phi \wedge dt \nonumber
\end{eqnarray}
and there is no dilaton. Strictly speaking the time coordinate $t$
is $2\pi$-periodic for the $SL(2,\mathbb{R})$, but non-compact for
its universal covering (AdS$_3$); $\phi$ is $2\pi$-periodic and
$r>0$.

Conformal deformations of this background are generated by truly
marginal operators i.e. dimension-$(1,1)$ operators that survive
their own perturbation. In the presence of holomorphic and
anti-holomorphic current algebras, marginal operators are
constructed as products $J^\alpha \bar J^\beta$, not all being
necessarily truly marginal. In the $SU(2)$ WZW model, nine
marginal operators do exist. However, they are related by
$SU(2)\times SU(2)$ symmetry to one of them, say $J^3 \bar J^3$.
Hence, only one line of continuous deformations appears.

The situation is different for $SL(2,\mathbb{R})$, because here
one cannot connect any two vectors by an $SL(2,\mathbb{R})$
transformation. This is intimately related to the existence of
several families of conjugacy classes: the elliptic and
hyperbolic, which correspond to the two different choices of {\it
Cartan} subalgebra, and the parabolic corresponding to the {\it
null} subalgebra. Three different truly marginal
left-right-symmetric deformations are possible,
leading therefore to three families of continuously
connected conformal sigma models. Each of them preserves
a different $U(1)_L \times U(1)_R$ subalgebra of the
undeformed WZW model.

The marginal deformations of $SU(2)$ have been thoroughly
investigated \cite{Hassan:1992gi} \cite{Giveon:1993ph}
\cite{Forste:2003km} with respect to: (\romannumeral1) the
identification with the $\left(SU(2)_k/U(1) \times
U(1)_{\sqrt{2k\alpha}}\right)\big/ \mathbb{Z}_k$, where
$SU(2)_k/U(1)$ is the gauging of the $SU(2)_k$ WZW model, and
$\alpha$ is the deformation parameter; (\romannumeral2) the
geometrical (sigma-model) interpretation in terms of metric,
torsion and dilaton backgrounds; (\romannumeral3) the
determination of the toroidal partition function and the spectrum
as functions of $R$. For the $SL(2,\mathbb{R})$ deformations, the
available results are less exhaustive, especially concerning the
spectrum and the partition function \cite{Forste:1994wp}
\cite{Giveon:1993ph} \cite{Elitzur:cb}. Our aim is to understand
the spectra -- partition functions -- as well as the issue of
supersymmetry, when these deformations appear in a more general set up
like the NS5/F1.

\boldmath
\subsection{The $J^3 \bar{J}^3$ deformation}
\unboldmath

The $J^3 \bar{J}^3$ deformation of $SL(2,\mathbb{R})$ is the one
that naturally appears by analytically continuing the deformed
$SU(2)$. It was analyzed in \cite{Giveon:1993ph}. Much like the
latter, the metric, antisymmetric tensor and dilaton can be
obtained by considering the $SL(2,\mathbb{R}) \times U(1) / U(1)$
coset, where the $U(1)$ in the product is compact with a radius
related to the deformation parameter, and the gauged one is a
diagonal combination of the extra $U(1)$ with the ``time-like"
$U(1)$ in the $SL(2,\mathbb{R})$, defined by $h=\exp
i\frac{\lambda}{2} \sigma^2$. The geometry corresponding to the
$J^3 \bar{J}^3$-deformed $SL(2,\mathbb{R})$, with deformation
parameter $\alpha -1>0$, is thus found to be
\begin{eqnarray}
ds^2 & =  & dr^2 + \frac{- \cosh^2 r dt^2 +
\alpha \sinh^2 r d\phi^2}{\alpha \cosh^2 r - \sinh^2 r}  \label{j3defm} \\
B & = & \frac{\alpha \cosh^2 r d\phi \wedge dt}{\alpha \cosh^2 r -
\sinh^2 r} ,\label{j3defa}
\end{eqnarray}
with dilaton
\begin{equation}
\mathrm{e}^{2\Phi}=\mathrm{e}^{2\Phi_0}\frac{\alpha - 1}{\alpha
\cosh^2 r-\sinh^2 r}.\label{j3defd}
\end{equation}
The scalar curvature of this geometry is
\begin{equation}
R=  2 \frac{\left(1-\tanh^2 r\right) \left( 2 \left(\alpha^2 -
\tanh^2 r \right) - 5 \alpha \left(1 - \tanh^2 r\right)
\right)}{\left(\alpha -\tanh^2 r\right)^2}.\nonumber
\end{equation}
Notice that the background fields are usually expected to receive
$1/k$ corrections; hence, they are valid semi-classically only,
except when they are protected by symmetries as in the unperturbed
WZW models.

At $\alpha=1$ we recover the AdS$_3$ metric and antisymmetric
tensor. For $\alpha \geq 1$, the geometry is everywhere regular
whereas for $\alpha < 1$ the curvature diverges at
$r=\mathrm{arctanh} \sqrt{\alpha}$. Similarly, the string coupling
$g_{\mathrm{s}} = \exp \Phi$ is finite everywhere for $\alpha \geq
1$ and blows up at $r=\mathrm{arctanh} \sqrt{\alpha}$ for $\alpha
< 1$. This means that the semi-classical approximation fails for
$\alpha < 1$. The string theory is however well defined. It is
actually related by T-duality to the range $\alpha > 1$ as will be
discussed later.

The two endpoints of the deformed background are remarkable:

\noindent\underline{The $\alpha \to \infty$ limit.} In this case
the background fields become:
\begin{eqnarray}
ds^2 & =  & dr^2 +  \tanh^2 r d\phi^2 -
\frac{dt^2}{\alpha}\label{cigmet} \\
\mathrm{e}^{-\Phi}&=&\mathrm{e}^{-\Phi_0}\cosh r\label{cigdil}
\end{eqnarray}
with no antisymmetric tensor. This is the {\it cigar} geometry
times a free time-like coordinate. The cigar -- Euclidean black
hole \cite{Witten:1991yr} -- is the axial gauging ($g\to hgh$) of
the $U(1)$ subgroup defined by $h=\exp i\frac{\lambda}{2}
\sigma^2$. It generates time translations (see Eq. (\ref{rotran}))
and acts without fixed points. The corresponding geometry is
regular and the $2\pi$-periodicity of $\phi$ inherited from the
$SL(2,\mathbb{R})$ ensures the absence of conical singularity.

\noindent\underline{The $\alpha \to 0$ limit.} Now we recover the
{\it trumpet} plus a free time-like coordinate:
\begin{eqnarray}
ds^2 & =  & dr^2 +  \coth^2 r dt^2 -
\alpha d\phi^2 \label{trumet} \\
\mathrm{e}^{-\Phi}&=&\mathrm{e}^{-\Phi_0}\sinh r. \label{trudil}
\end{eqnarray}
The trumpet is the vector gauging ($g\to hgh^{-1}$) of the same
$U(1)$, and generates now rotations around the center (see Eq.
(\ref{rotran})). Throughout this gauging the time coordinate
becomes space-like and vice-versa. The subgroup acts with fixed
points, and this accounts for the appearance of a singularity at
$r=0$, which is present no matter the choice for the periodicity
of $t$ (in fact this coordinate is not periodic if we start from
the universal covering of $SL(2,\mathbb{R})$). Notice also that
the time coordinate $\phi$ is compact, but infinitely rescaled
though.

A similar phenomenon occurs in the $J^3 \bar J^3$ deformation line
of the $SU(2)$ WZW model. The two endpoints are the axial and
vector gaugings of $SU(2)$ by a $U(1)$, times a free boson, at
zero or infinite radius. Axial and vector gaugings are identical
in the compact case \cite{Kiritsis:1993ju}. They are T-dual
descriptions of the same CFT, whose background geometry is the
{\it bell}. Therefore, in the compact case, there is a real
T-duality relating large and small deformation parameters. The
situation is quite different in the non-compact case.

We would like now to better understand the algebraic point of view
and determine the partition function of the deformed theory at any
$\alpha$. To all orders in the deformation parameter, the
deformation acts only on the charge lattice of the Cartan
subalgebra along which the WZW theory is deformed. Therefore, as
in the $SU(2)$ case, the deformation at hand corresponds to a
shift of the ``radius" of the $J^3, \bar J^3$ lattice:
$\sqrt{2k}\to \sqrt{2k \alpha}$. The form (\ref{fctpartsl}) of the
original $SL(2,\mathbb{R})$ partition function enables us to
implement this radius shift in the time-like lattice with the
modular-invariant result:
\begin{eqnarray}
Z_{3\bar3}(\alpha)&=& 4 \sqrt{\tau_2} (k-2)^{3/2} \int d^2 s \,
d^2 t \frac{ \mathrm{e}^{\frac{2\pi}{\tau_2} (\mathrm{Im} (s_1
\tau -s_2))^2 } }{\left|\vartheta_1 (s_1 \tau -s_2 |\tau )
\right|^2}\times  \nonumber \\
& &\times  \sum_{m,w,m',w'\in \mathbb{Z}} \zeta\oao{w + s_1 - t_1
}{m +s_2 - t_2 }\left(k \right)\,  \zeta\oao{w' + t_1 }{m' + t_2
}\left(- k \alpha \right).  \label{j3defpart}
\end{eqnarray}
We can first expand the spectrum around the symmetric
$SL(2,\mathbb{R})$ point (i.e. for $\alpha = 1+ \varepsilon$,
$|\varepsilon| \ll 1$). It allows to express the spectrum in terms
of the $SL(2,\mathbb{R})$ quantum numbers. Using the same
techniques as in Appendix C, we first perform a Poisson
resummation, and integrate over $t_2$. We find the exponential
factor:
\begin{eqnarray}
\label{cartdef}
\exp \left\{  -\pi \tau_2 \varepsilon
\left[ \frac{n^2}{k} - k(w_+ - (w -t_1))^2 \right] \right.
\nonumber \\
-2\pi \tau_2  w_+ \left(k(w - t_1) - \frac{k}{2}w_+ \right) -2\pi
\tau_2 s_1 (q+\bar{q}+1+k(w-t_1))
\nonumber \\
\left. + \vphantom{\frac{1}{2}} 2i\pi s_2 (q - \bar{q} -n) - (k-2)
\pi \tau_2 s_{1}^2 +2i\pi \tau_1 (w_+ + s_1)n \right\}. \nonumber
\end{eqnarray}
The second and third lines are exactly the same as the undeformed
$SL(2,\mathbb{R})$ and lead to the same analysis. For $\varepsilon
\ll 1$, the first line gives simply a shift on the weights of the
operators according to their $J_{0}^3$,$\bar{J}_{0}^3$
eigenvalues:
\begin{eqnarray}
L_{0}^{\varepsilon} = L_{0} - \frac{\varepsilon}{k}
\left(\tilde{m} + \frac{k}{2} w_+\right)\left(\tilde{\bar{m}} +
\frac{k}{2} w_+ \right),
\nonumber \\
\bar{L}_{0}^{\varepsilon} = \bar{L}_{0} - \frac{\varepsilon}{k}
\left(\tilde{m} + \frac{k}{2} w_+\right)\left(\tilde{\bar{m}} +
\frac{k}{2} w_+ \right).\nonumber
\end{eqnarray}
In terms of the unflowed eigenvalues: $m = \tilde{m} + kw_+ /2 \ ,
\ \ \bar{m}= \tilde{\bar{m}} + kw_+ /2$, the deformation term is
$$
\delta h = \delta \bar{h} = - \varepsilon m \bar{m} /k
$$
It breaks of course the $SL(2,\mathbb{R})_L \times
SL(2,\mathbb{R})_R$ symmetry of the CFT.

In the limits $\alpha \to 0$ or $\infty$, the $J^3, \bar J^3$
lattice decouples. We then recover a free time-like boson (of zero
or infinite radius) times an $SL(2,\mathbb{R})/U(1)$ coset. By
using the large/small-radius limits presented in Appendix B, we
can trace the effect of these limits at the level of the partition
function\footnote{The $\alpha \to \infty$ limit requires analytic
continuation because the lattice is time-like.}:
\begin{equation}
Z_{3\bar3}(\alpha) {\underarrow{\alpha\to
\infty}}\frac{k-2}{\sqrt{k \tau_2} \eta \bar{\eta}}
Z_{SL(2,\mathbb{R})/U(1)_{\mathrm{A}}} \label{Ali}
\end{equation}
with $Z_{SL(2,\mathbb{R})/U(1)_{\mathrm{A}}}$ given in
(\ref{cospartcb}). This is precisely the geometrical expectation
since the large-$\alpha$ limit of the background
(\ref{j3defm})--(\ref{j3defd}) is the cigar, Eqs. (\ref{cigmet}),
(\ref{cigdil}) describing the semiclassical geometry of the axial
$U(1)$ gauging of $SL(2,\mathbb{R})$. Similarly, we find
\begin{eqnarray}
Z_{3\bar3}(\alpha) {\underarrow{\alpha\to 0}} &  &4 \sqrt{k\alpha}
(k-2)^{3/2} \int d^2 s \, d^2 t \frac{
\mathrm{e}^{\frac{2\pi}{\tau_2} (\mathrm{Im} (s_1 \tau -s_2))^2 }
}{\left|\vartheta_1 (s_1 \tau -s_2 |\tau )
\right|^2}\times  \nonumber \\
& &\times  \sum_{m,w\in \mathbb{Z}} \zeta\oao{w + s_1 - t_1 }{m
+s_2 - t_2 }\left(k \right).  \label{Vli}
\end{eqnarray}
In terms of background geometry, this limit describes the trumpet
(Eqs. (\ref{trumet})and (\ref{trudil})) times a free time-like
coordinate. We can therefore read off from expression (\ref{Vli})
the partition function of the vector coset:
\begin{eqnarray}
Z_{SL(2,\mathbb{R})/U(1)_{\mathrm{V}}}&=& 4 (k-2)^{3/2}
\sqrt{\tau_2} {\eta\bar\eta}\int d^2 s \, d^2 t \frac{
\mathrm{e}^{\frac{2\pi}{\tau_2} (\mathrm{Im} (s_1 \tau -s_2))^2 }
}{\left|\vartheta_1 (s_1 \tau -s_2 |\tau )
\right|^2}\times  \nonumber \\
& &\times  \sum_{m,w\in \mathbb{Z}} \zeta\oao{w + s_1 - t_1 }{m
+s_2 - t_2 }\left(k \right). \label{cospartv}
\end{eqnarray}
The spectrum of primary fields of this coset can be computed
straightforwardly, and reads:
\begin{equation}
L_{0}^{\rm vector} = \bar{L}^{\rm vector}_0 = - \frac{j(j-1)}{k-2}
+ \frac{\mu^2}{k} \ , \ \ \mathrm{with} \ \ \ \mu = - \frac{k}{2}
(w-t_1) \in \mathbb{R}, \label{spectrump}
\end{equation}
both for continuous and discrete representations. The Gaussian
variable $w-t_1$ can be integrated out, and leads to the partition
function:
\begin{equation}
Z_{SL(2,\mathbb{R})/U(1)_{\mathrm{V}}} = 4
\frac{(k-2)^{3/2}}{\sqrt{k}} {\eta\bar\eta}\int d^2s
\frac{\mathrm{e}^{\frac{2\pi}{\tau_2} (\mathrm{Im} (s_1 \tau
-s_2))^2}}{\left|\vartheta_1 \left( s_1 \tau - s_2 |\tau \right)
\right|^2} .\label{cospartvint}
\end{equation}

The cigar and trumpet geometries are semi-classically T-dual: they
are related by Buscher duality~\cite{Buscher:sk}. Any two points
$(\alpha, 1/\alpha)$ in the above line of deformations are in fact
connected by an element of
$O(2,2,\mathbb{R})$~\cite{Giveon:1993ph}. However, recalling the
spectrum of primaries of the axial coset:
\begin{eqnarray}
L^{\rm axial}_0  = - \frac{j(j-1)}{k-2} + \frac{({n\over2} -
{kw\over 2})^2}{k}\ , \ \ \mathrm{with} \ \ \ \bar{L}^{\rm
axial}_0 = - \frac{j(j-1)}{k-2} + \frac{({n\over 2} + {kw\over
2})^2}{k},\label{cigspec}
\end{eqnarray}
we observe that they are different from the vector ones. Our
previous discussion gives the explanation for this apparent
``failure'' of T-duality. The coordinate $\phi$ in the cigar
metric~(\ref{cigmet}) inherits the $2\pi \sqrt{2k}$ periodicity
(in the asymptotic region $r\to \infty$) from the angular
coordinate of AdS$_3$, irrespectively of the cover of
$SL(2,\mathbb{R})$ under consideration; therefore the
corresponding lattice in the partition function is compact with
radius $\sqrt{2k}$. However, the $t$ coordinate of the trumpet
metric~(\ref{trumet}) is $2\pi N \sqrt{2k}$-periodic, for the
$N$-th cover of $SL(2,\mathbb{R})$ (see Fig.~\ref{cover}).
\FIGURE{ \centerline{\epsfxsize=105mm\epsfbox{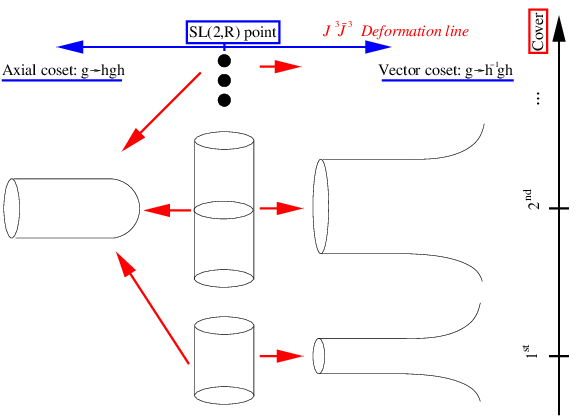}}
\caption{The T-duality for the different covers of
$SL(2,\mathbb{R})$.} \label{cover}} Consequently, since our
original choice was the universal cover of the algebra, the
spectrum we found -- Eq.~(\ref{spectrump}) -- corresponds to the
``universal cover'' of the trumpet, i.e. with a non-compact
transverse coordinate.

The spectrum of the above coset theories has been studied
in~\cite{Dijkgraaf:1991ba}. In that work, axial and vector cosets
were argued to be T-dual, but the definition used for the
T-duality amounts to exchange the momenta and winding modes of the
physical states. In the case of $SL(2,\mathbb{R})$, this is
equivalent to the more rigorous definition of T-duality
--~residual discrete symmetry of a broken gauge symmetry --
\emph{only} on the \emph{single cover} of the group manifold.
Indeed, if we consider the single cover of $SL(2,\mathbb{R})$, we
obtain in the limits $\alpha \to 0$ and $\alpha \to \infty$ the
same spectrum of primaries for the vector and axial cosets,
Eq.~(\ref{cigspec}), but with different constraints: $\tilde{m}
\pm \bar{\tilde{m}} = n$, $\tilde{m} \mp \bar{\tilde{m}} = -kw$,
where plus and minus refer to the vector and axial cosets
respectively.

Before closing this chapter, we would like to comment on the
unitarity of the undeformed model. To this end, we consider a
small deformation parameterized by $\varepsilon$:
\begin{equation}
\alpha = 1 - \varepsilon \ , \ \ \varepsilon > 0.\nonumber
\end{equation}
Then the quadratic term in Eq.~(\ref{cartdef}) is space-like and
enables us to perform the energy integration \emph{without} any
analytic continuation (in fact, we slightly rotate the energy
direction into the light-cone). In the limit $\varepsilon \to 0$ ,
we obtain a delta-function that gives the trivial partition
function previously discussed. It is possible to analyze this
result from a different perspective. We decompose the characters
of the discrete representations e.g., by using a non-compact
generalization of the Ka\u{c}-Peterson
formula~\cite{Bakas:1991fs}: $ \chi^{j,+}_{SL(2,R)} (\tau ) =
\sum_{m \in \mathbb{N}} c_{m}^{j} q^{-(j+m)^2 /k} $ with the
string functions $ c_{m}^j = q^{-j(j-1) / (k-2) + (j+m)^2 /k}
\times (\mathrm{oscillators}) $. Changing the Cartan radius
$\sqrt{2k} \to \alpha \sqrt{2k}$, $\alpha < 1$, all characters
become convergent.

\boldmath
\subsection{The $J^2 \bar{J}^2$ deformation}
\unboldmath

The operator $J^2 \bar{J}^2$ is also suitable for marginal
deformations of the theory. It is not equivalent to $J^3
\bar{J}^3$ because it corresponds to a choice of \emph{space-like}
Cartan subalgebra instead of a time-like one. Conversely the
corresponding deformation can be realized as  $SL(2,\mathbb{R})
\times U(1) / U(1)$ where the gauged $U(1)$ is now a diagonal
combination of the extra $U(1)$ factor with $SL(2,\mathbb{R})$
elements of the type $h=\exp -\frac{\lambda}{2} \sigma^3$.

Owing to the previous discussion, we easily determine the
partition function at any value of the deformation parameter. This
is realized by deforming the corresponding cycle in the Cartan
torus, which amounts in shifting the radius of the space-like
$J^2$, $\bar{J}^2$ lattice. In order to present this partition
function in a form closer to the one of the $J^3 \bar{J}^3$
deformation, we diagonalize $J^2$  instead of $J^3$. This is
achieved by redefining the lattice variables in
Eq.~(\ref{fctpartsl}) as:
\begin{eqnarray}
w+s_1 - t_1 \to w - t_1 & , & \  w'+ t_1 \to w' + t_1 +s_1,
 \nonumber \\
m+ s_2 - t_2 \to m-t_2 & , & \  m'+ t_2   \to m'+ t_2+s_2.
\nonumber
\end{eqnarray}
Then we write the following partition function for the
$J^2 \bar{J}^2$ deformation:
\begin{eqnarray}
Z_{2\bar2}(\alpha)&=& 4 \sqrt{\tau_2} (k-2)^{3/2} \int d^2 s \,
d^2 t \frac{ \mathrm{e}^{\frac{2\pi}{\tau_2} (\mathrm{Im} (s_1
\tau -s_2))^2 } }{\left|\vartheta_1 (s_1 \tau -s_2 |\tau )
\right|^2}\times  \nonumber \\
& &\times  \sum_{m,w,m',w'\in \mathbb{Z}}
\zeta\oao{w - t_1
}{m -  t_2 }\left(k  \alpha \right)\,
\zeta\oao{w' + t_1 +s_1}{m'+ t_2 +s_2}\left(- k\right).  \label{minkdef}
\end{eqnarray}

Getting the effective geometry is interesting {\it per sei} but
its systematic analysis goes beyond the scope of the present work.
For extreme deformations, a space-like coordinate is factorized,
and we are left with an $SL(2,\mathbb{R})/ U(1)$ coset with
Lorentzian target space: the Lorentzian two-dimensional black
hole.

Again the $\alpha \to 0$ and $\alpha \to \infty$ limits are
related by T-duality, which at the level of the semi-classical
geometry describe various space--time regions of the black hole
\cite{Dijkgraaf:1991ba}. For $\alpha \to \infty$ we obtain:
\begin{equation}
Z_{2\bar2}(\alpha) {\underarrow{\alpha\to \infty}}\frac{1}{\sqrt{k
\alpha\tau_2}{\eta \bar\eta}} Z_{\mathrm{BH}} \label{AliBH}
\end{equation}
with
\begin{equation}
Z_{\mathrm{BH}}=4 \sqrt{\tau_2} (k-2)^{3/2} \, \eta \bar{\eta}
\int d^2s\frac{ \mathrm{e}^{\frac{2\pi}{\tau_2} (\mathrm{Im} (s_1
\tau -s_2))^2 } }{\left|\vartheta_1 (s_1 \tau -s_2 |\tau )
\right|^2} \sum_{m',w'\in \mathbb{Z}} \zeta
\oao{w'+s_1}{m'+s_2}(-k). \label{ZBH}
\end{equation}

There is a subtlety in the latter expression compared to the
ordinary Euclidean axial black hole (\ref{cospartcb}). In the
path-integral calculation of the partition function for the
Euclidian coset, the oscillators were coupled to the full real
momentum of the free boson. We used the periodicity of the
determinant to break the zero modes into an integer part and a
real compact part. The integer part was interpreted as the lattice
of the zero modes of the compact boson and the real part as
Lagrange multipliers which impose constraints on the Hilbert
space. In the present case, we have to perform an analytic
continuation in order to move to the Hamiltonian representation of
the partition function,
\begin{equation}
Z_{\mathrm{BH}}=4 \sqrt{k} (k-2)^{3/2} \, \eta \bar{\eta}
\int_{\mathbb{R}^2} d^2v\, \frac{\mathrm{e}^{\frac{2\pi}{\tau_2}
[\mathrm{Im} (i(v_1 \tau -v_2))]^2}
\mathrm{e}^{-\frac{k\pi}{\tau_2} \left| v_1 \tau - v_2 \right|^2}
}{\left|\vartheta_1 \left( i(v_1 \tau - v_2)  |\tau \right)
\right|^2},\nonumber
\end{equation}
and read the string spectrum. Now, because the coupling is
imaginary, the determinant is no longer periodic; therefore we
have a \emph{non-compact} time-like coordinate coupled to the
oscillators.


\subsection{The null deformation}
\label{null}

As we already mentioned in Sec. \ref{sigmamod}, the
$SL(2,\mathbb{R})$ WZW model allows for extra, unconventional,
marginal deformations, which are not generated by Cartan
left-right bilinears such as $J^2 \bar{J}^2$ or $J^3 \bar{J}^3$.
Instead, the marginal operator we will consider is the following:
\begin{equation}
J\bar{J} \sim \left(J^1  + J^3\right) \left(\bar J^1 + \bar
J^3\right),\nonumber
\end{equation}
(see Eqs. (\ref{nullcurr}) and (\ref{curJ})--(\ref{curJ2b})).

In the holographic dual description, these null currents are the
translation generators of the conformal group acting on the
boundary in Poincar\'e coordinates~\cite{Balasubramanian:1998sn}.
We will here analyze their action from the sigma-model viewpoint
and determine the spectrum and partition function of the deformed
model. Supersymmetry issues will be addressed in the more complete
set up of $\{$ null-deformed $SL(2,\mathbb{R}) \ \} \times SU(2)
$, in Sec. \ref{sect4adss}.

We recall that the metric of the deformed background is (see Eqs.
(\ref{solns}))
\begin{equation}
ds^2 = \frac{du^2}{u^2}+ \frac{-dT^2+dX^2}{u^2+1/M^2}.\nonumber
\end{equation}
The scalar curvature reads:
\begin{equation}
R = -2 \frac{u^2 \left(3u^2 -4 /M^2 \right) }{\left(u^2 +
1/M^2\right)^2}.\nonumber
\end{equation}
This geometry is smooth everywhere for $M^2 > 0$. On the opposite,
$M^2 < 0$ gives a singular geometry, which seems however
interesting, and corresponds to the repulsion
solution~\cite{Johnson:1999qt}.

In order to study the null deformation of AdS$_3$, it is useful to
introduce a free-field representation of $SL(2,\mathbb{R})$, in
which operators $J$ and $\bar J$ have a simple expression. We
first introduce $\phi = -\log u$, $\bar \gamma = x^+$ and $\gamma
= x^-$. The worldsheet Lagrangian (see Eq.
(\ref{sigma}))\footnote{Note that the Euclidean rotation from
$SL(2,\mathbb{R})$ to $H_{3}^+$ is performed by just considering
$\gamma$ and $\bar \gamma$ as complex conjugate.} reads:
\begin{equation}
\frac{2\pi}{k} \mathcal{L} = \partial \phi \bar{\partial} \phi +
\mathrm{e}^{2\phi} \partial \bar{\gamma} \bar{\partial} \gamma.
\label{sigmaimp}
\end{equation}
It can be represented with a $(\beta ,\gamma)$ ghost system of
conformal dimensions $(1,0)$:
\begin{equation}
\frac{2\pi}{k} \mathcal{L} = \partial \phi \bar{\partial} \phi
+\beta \bar{\partial} \gamma+ \bar{\beta} \partial \bar{\gamma} -
\beta \bar{\beta} \mathrm{e}^{-2\phi}.\label{sigmacla}
\end{equation}
Classically, $\beta$ and $\bar{\beta}$ are Lagrange multipliers.
They can be eliminated by using their equations of motion, and
this gives back the action (\ref{sigmaimp}). At the quantum level,
however, we must integrate them out; taking into account the
change of the measure and the renormalization of the exponent, we
obtain, after rescaling the fields:
\begin{equation}
2\pi \mathcal{L} = \partial \phi \bar{\partial} \phi +\beta
\bar{\partial} \gamma+ \bar{\beta} \partial \bar{\gamma} - \beta
\bar{\beta} \mathrm{e}^{-\sqrt{\frac{2}{k-2}}
\phi}-\sqrt{\frac{2}{k-2}} R^{(2)} \phi .\label{sigmaqua}
\end{equation}
The last term is the screening charge necessary to compute
correlation functions in the presence of the background charge for
the field $\phi$ ($R^{(2)}$-term). The OPE of the free fields are
$\phi(z,\bar z) \phi(0) \sim - \ln (z\bar{z})$ and $\beta(z)
\gamma(0) \sim 1/z$. By using these free fields, the holomorphic
currents (\ref{curJ})--(\ref{curJ2}) are recast, at quantum level,
as:
\begin{eqnarray}
J^1 + J^3 & = &\beta, \nonumber \\
J^2 & = &- i\beta \gamma - i\sqrt{\frac{k-2}{2}} \partial \phi, \nonumber \\
J^1 - J^3 & = &  \beta \gamma^2 +  \sqrt{2(k-2)} \gamma \partial
\phi + k \partial \gamma.\nonumber
\end{eqnarray}
They satisfy the $\widehat{SL}(2,\mathbb{R})_{\mathrm{L}}$ OPA,
Eqs. (\ref{LOPA}). Notice finally that the holomorphic primaries
of the $SL(2,\mathbb{R})$ CFT read in this basis:
\begin{equation}
\Phi^{j}_m = \gamma^{j-m} \mathrm{e}^{\sqrt{\frac{2}{k-2}} j
\phi}. \label{prim}
\end{equation}
The conformal weight of this operator is entirely given by the
Liouville primary, whereas the $J^2$ eigenvalue corresponds to the
sum of the ``ghost number'' and the Liouville momentum.

We can use the above free-field representation to write the null
deformation of the AdS$_3$ sigma model, Eq. (\ref{nulldef}):
\begin{equation}
2\pi \mathcal{L} = \partial \phi \bar{\partial} \phi +\beta
\bar{\partial} \gamma+ \bar{\beta} \partial \bar{\gamma} - \beta
\bar{\beta} \left[ \frac{1}{M^2} +
\mathrm{e}^{-\sqrt{\frac{2}{k-2}}
\phi}\right]-\sqrt{\frac{2}{k-2}} R^{(2)} \phi.\nonumber
\end{equation}
For any non-zero value of the deformation parameter, the fields
$\beta$ and $\bar{\beta}$ can be eliminated, leading to two
light-cone free coordinates. The energy--momentum tensor reads:
\begin{equation}
T = - \frac{1}{2}\partial \phi \partial \phi -
\frac{1}{\sqrt{2(k-2)}}
\partial^2 \phi - M^2 \partial \bar{\gamma} \partial \gamma.
\nonumber
\end{equation}

We would like to determine the partition function of the model at
hand. To this end we will follow the procedure which has by now
become familiar: implement the deformation into the twist-shift
orbifold structure of the AdS$_3$. The ghost number is no longer
conserved because the ghost-number current $\beta \gamma$ (equal
to $\gamma \partial \bar{\gamma} \exp 2\phi$, on  shell) is not a
conformal field (this means in particular that the fields
(\ref{prim}) are now ill-defined). However, there is still a
global $U(1)$symmetry: $\gamma \to \mathrm{e}^{i\alpha} \gamma$,
$\bar{\gamma} \to \mathrm{e}^{-i\alpha} \bar{\gamma}$ that we can
orbifoldize (this is a twist on a complex boson). However, with
this representation it is not possible to implement spectral flow
without adding extra fields; therefore we have to find the action
of the deformation on the zero-mode structure.

Owing the above ingredients, we can construct the partition
function that fulfills the following requirements:
\begin{enumerate}
\item The endpoint of the deformation, $M^2 \to 0$, should give
the Liouville theory times a free light-cone. At $M^2 \to \infty$
one should recover the undeformed $SL(2,\mathbb{R})$.
\item The partition function must be modular invariant.
\end{enumerate}
This enables us to propose the following partition function for
the null-deformed model:
\begin{eqnarray}
Z_{SL(2,\mathbb{R})}^{\rm \ null}\left(M^2\right)&=& 4
\sqrt{\tau_2} (k-2)^{3/2} \int d^2 s \, d^2 t \frac{
\mathrm{e}^{\frac{2\pi}{\tau_2} (\mathrm{Im} (s_1 \tau -s_2))^2 }
}{\left|\vartheta_1 (s_1 \tau -s_2 |\tau ) \right|^2}\times
\label{partnulldef} \\
& \times & \sum_{m,w,m',w'\in \mathbb{Z}} \zeta\oao{w + s_1 - t_1
}{m +s_2 - t_2 }\left(k{ 1+M^2\over  M^2}\right)  \zeta\oao{w' +
t_1 }{m' + t_2 }\left(-k{1+ M^2\over M^2} \right). \nonumber
\end{eqnarray}
which is obtained by changing the radii of both space-like ($J^1$,
$\bar J^1$)\footnote{ or equivalently ($J^2$, $\bar J^2$) since
they are exchanged by a group transformation} and time-like
($J^3$, $\bar J^3$) lattices by the same amount. It is 
natural since these moduli correspond to the 
deformations along $J^1$ and $J^3$.
In the pure $SL(2,\mathbb{R})$ theory, the spectrum is constructed by acting
on the primaries with the modes of the affine currents. The
difference with the linear dilaton model is that the shift of the
oscillator number is linked to the shift of the zero modes. In
writing (\ref{partnulldef}), which interpolates between these two
models, we have implemented that this shift should vanish at
infinite deformation (i.e. $M^2=0$), which indeed happens provided
the radius of the lattice of light-cone zero-modes becomes large.
We will expand on that in the next section. In the limit of
infinite deformation $M^2\to 0$, by using the standard technology
developed so far, we find
\begin{equation}
Z_{SL(2,\mathbb{R})}^{\rm \ null}\left(M^2\right)\sim {(k-2)^{3/2}
M^2 \over \pi^2 \tau_2^{3/2} \left( \eta\bar\eta\right)^{3}}\sp
{\rm at} \ M^2 \sim 0. \nonumber
\end{equation}
This coincides with the partition function for $U(1)_Q  \times
\mathbb{R}^{1,1}$.

The derivation of the spectrum goes as previously (see Appendix
C). Again, we concentrate on the vicinity of the unbroken
$SL(2,\mathbb{R})$ (although the spectrum is known at any $M^2$):
\begin{equation}
{M^2\over 1+ M^2} =1-\varepsilon \ , \ \ \varepsilon \ll
1.\nonumber
\end{equation}
The deformed discrete spectrum is
\begin{eqnarray}
L_0 & = & - \frac{j(j-1)}{k-2} - w_+ \left( \tilde{m} +
\frac{\varepsilon}{2} (\tilde{m}+\tilde{\bar{m}}) \right) - \frac{k}{4} (1+
\varepsilon) w_{+}^{2} + N \nonumber \\
& & + \frac{\varepsilon (1-2j)}{2(k-2)} \left( \tilde{m} + \tilde{\bar{m}} +
\frac{k}{2(k-2)} (1-2j) \right)
, \nonumber \\
\bar{L}_0 & = & - \frac{j(j-1)}{k-2} - w_+ \left( \tilde{\bar{m}} +
\frac{\varepsilon}{2} (\tilde{m}+\tilde{\bar{m}}) \right) - \frac{k}{4} (1+
\varepsilon) w_{+}^{2} + \bar{N}\nonumber \\
& & + \frac{\varepsilon (1-2j)}{2(k-2)} \left( \tilde{m} +
\tilde{\bar{m}} + \frac{k}{2(k-2)} (1-2j) \right).
\end{eqnarray}
The last terms of both equations are due to the displacement of the
poles corresponding to the discrete representations. For the
continuous spectrum, we have a similar shift on the weights:
\begin{eqnarray}
L_0 & = &  \frac{s^2+1/4}{k-2} - w_+ \left( \tilde{m} +
\frac{\varepsilon}{2} (\tilde{m}+\tilde{\bar{m}}) \right) +
\frac{k}{4} (1+ \varepsilon) w_{+}^{2} + N, \nonumber \\
\bar{L}_0 & = & \frac{s^2 + 1/4}{k-2} - w_+ \left( \tilde{\bar{m}} +
\frac{\varepsilon}{2} (\tilde{m}+\tilde{\bar{m}}) \right) + \frac{k}{4} (1+
\varepsilon) w_{+}^{2} + \bar{N}\nonumber
\end{eqnarray}
with a deformed density of states. We observe that the deformation
term is linked to the spectral flow.

\boldmath
\section{The supersymmetric null deformation of $SL(2,\mathbb{R})
\times SU(2)$}\label{sect4adss}
\unboldmath

So far, we have been considering conformal deformations of the
AdS$_3$ background, realized by marginal deformations of the
corresponding $SL(2,\mathbb{R})$ WZW model. In supersymmetric
NS5-brane configurations, the $SL(2,\mathbb{R})$ appears usually
along with $SU(2)$. We will here analyze the issue of
supersymmetry in presence of null deformations of the
$SL(2,\mathbb{R})$ factor. We will in particular show that the
requirement for the worldsheet $N=2$ superconformal symmetry to be
preserved, gives very tight constraints on the allowed
deformations.

\boldmath
\subsection{The $N=2$ algebra of the deformed theory}
\unboldmath

We first rewrite the $N=2$ algebra for $SL(2,\mathbb{R}) \times
SU(2)$,  Eqs. (\ref{N2G}), (\ref{N2J}) and (\ref{calJI})
in the free-field representation (Eqs. (\ref{jpmfree}),
(\ref{j3free})). We recall that $I^i$ are the $SU(2)$ currents,
$\psi^i$ and $\chi^i$ the fermions of respectively  $SU(2)$ and
$SL(2,\mathbb{R})$:
\begin{eqnarray}
\sqrt{2} G^{\pm}  & = & i \sqrt{\frac{2}{k+2}} \left(
\sqrt{\frac{k+4}{2}} \partial X \mp i \sqrt{\frac{k+2}{2}}
\partial \rho \right)
\mathrm{e}^{\mp i \sqrt{\frac{2}{k+4}} (X-T)} \chi^{\pm} \nonumber \\
& & + \  \sqrt{\frac{2}{k+2}} \left[ \mathcal{I}^3 \mp \left(
\sqrt{\frac{k+4}{2}} i\partial T +  \chi^+ \chi^- \right) \right]
\frac{\psi^3 \pm \chi^3}{2} + \sqrt{\frac{2}{k+2}} I^{\mp}
\psi^{\pm},
\label{slalgG} \\
\nonumber \\
J &  = & \psi^3 \chi^3 + \chi^+ \chi^- + \psi^+ \psi^- +
\frac{2}{k+2} \left[  \sqrt{\frac{k+4}{2}} i \partial T + \chi^+
\chi^- - \mathcal{I}^3 \right].\label{slalgJ}
\end{eqnarray}
The fermions $\chi^+$, $\chi^-$ are bosonized as $\chi^+ \chi^- =
i \partial H_1$. Note also a shift $k \to k+4$ with respect to the
formulas for pure AdS$_3$; $k$ is the level of $\widehat{SU}(2)$,
and this shift ensures that the total bosonic central charge
equals six (see (\ref{chat}) and (\ref{kshift})).

In~\cite{Hikida:2000ry}, a map was given between the free-field
representation of the superconformal algebra for
$SL(2,\mathbb{R})$ and the algebra for $N=2$ Liouville times two
free coordinates. As a first step towards the supersymmetrization
of the null deformation studied previously, we will show that
there exists a \emph {one-parameter family of $N=2$ algebras}
interpolating between $SL(2,\mathbb{R}) \times SU(2)$ and $U(1)_Q
\times \mathbb{R}^{1,1} \times SU(2)$.

The $N=2$ generators (with a non-standard complex structure) are,
for $U(1)_Q \times \mathbb{R}^{1,1} \times SU(2) $,
\begin{eqnarray}
\sqrt{2} \hat{G}^{\pm}  & = &  \left( i \partial \hat{X}
-\sqrt{\frac{2}{k+2}} i \partial \hat{H}_1 \pm \partial \rho
\right) \mathrm{e}^{\pm i \hat{H}_1}
\nonumber \\
& + & \left( \sqrt{\frac{2}{k+2}} \mathcal{I}^3 \mp i \partial
\hat{T} \right) \frac{\psi^3 \pm \chi^3}{\sqrt{2}} +
\sqrt{\frac{2}{k+2}} I^{\mp} \psi^{\pm}, \nonumber \\
\hat{J} & = & i\partial \hat{H}_1 + \psi^3 \chi^3 +  \chi^+ \chi^- +
 \frac{2}{k+2} \left[  \sqrt{\frac{k+2}{2}} i\partial \hat{X} -
   \mathcal{I}^3 \right],\nonumber
\end{eqnarray}
where $\hat{T}$, $\hat{X}$, are the light-cone coordinates. This
is not the usual $N=2$ subalgebra of the $N=4$ superconformal
algebra (Eqs.~(\ref{NS5alg1}) and~(\ref{NS5alg2})) of the  $U(1)_Q
\times SU(2) $ SCFT.

We now perform the following $SO(2,1)$ rotation, which leaves
unchanged the OPE's:
\begin{eqnarray}
\hat{H}_1 & = & H_1 - t X^-, \nonumber \\
\hat{T} & = & c T + s H_1,
\nonumber \\
\hat{X} & = & c X + s (H_1  - t X^- ),\nonumber
\end{eqnarray}
where $X^- = X-T$, and we have introduced:
$$ c=\cosh \sigma \ , \ \ t = \tanh \sigma \ , \ \
s = \sinh \sigma =  s_0 / p \ ,  \ \ \mathrm{with} \ s_0 =
\sqrt{\frac{2}{k+2}}\ \ \mathrm{and} \  p \geq 1.
$$
For $p=1$, the rotated $N=2$ algebra corresponds exactly to the
superconformal algebra of $SL(2,\mathbb{R}) \times SU(2)$,
Eqs.~(\ref{slalgG}) and (\ref{slalgJ}). Furthermore, one can check
that the $N=2$ superconformal structure is preserved for any $p$.

Coming back to the null deformation of $SL(2,\mathbb{R}) \times
SU(2)$, we conclude that the above one-parameter family of
supercurrents can be implemented along the line of deformation.
The $N=2$ $R$-symmetry current of the deformed theory takes the
form:
\begin{eqnarray}
J  & = &  \psi^3 \chi^3 + \chi^+ \chi^- + i\partial H_1 + p
s^2 \left( \frac{1}{t} i \partial T + i \partial H_1 \right) -
s_0^2 \ \mathcal{I}^3 + (p-1) t  \ i \partial X^-, \nonumber
\end{eqnarray}
and the deformed supercurrents read:
\begin{eqnarray}
\sqrt{2} G_2^{\pm}   & = &     \left(  s_0 \mathcal{I}^3 \mp i ( c
\
\partial T + s \ \partial H_1 ) \right) \frac{\psi^3 \pm
\chi^3}{\sqrt{2}} +  s_0 I^{\mp} \psi^{\pm}
\nonumber \\
 &   & +  \ i \left[ c \ \partial X
 - ( s_0 -s ) \left(\partial H_1 - t \ \partial X^- \right)
\mp i \partial \rho \right] \mathrm{e}^{\pm i \left(H_1 - t
X^-\right). } \label{defscur}
\end{eqnarray}
Clearly, the background contains both torsion (in the first line
of (\ref{defscur})) and a background charge (in the second line).
The field $H_1$ still corresponds to a free complex fermion,
provided we change the radius of the compact light-cone direction:
\begin{equation}
\hat{R}_+ = \frac{2}{t} = \sqrt{2\left(p^2 (k+2)+2\right)}.
\nonumber
\end{equation}
The latter statement holds because the deformation is null. It
therefore confirms our assumption (Sec. \ref{null}) that the null
deformation corresponds to a change of the radius of the compact
light-cone coordinate. The relation between this parameterization
and the mass scale introduced in Sec.~\ref{sect1} is, up to $1/k$
corrections:
\begin{equation}
\frac{1}{M^2} = \frac{k+2}{k+4} \left( p^2 -1\right).
\label{relparam}
\end{equation}
The holomorphic primary fields for the deformed $SL(2,\mathbb{R})$
are recast as:
\begin{equation}
\Phi^{w_+ ,\ \rm def}_{j\ m} = \exp \left\{ \sqrt{\frac{2}{k+2}}
j \phi + i t \left[ \left( \tilde{m} - \frac{1}{2t^2} w_+ \right) X^- +
\frac{1}{2t^2} w_+ (X+T) \right] \right\},\nonumber
\end{equation}
where, as in the undeformed theory, $\tilde{m}-\tilde{\bar{m}} \in
\mathbb{Z}$ 
and $\tilde{m}+\tilde{\bar{m}} -2w_+ /t^2$ is the energy. This
gives the following $N=2$ charges:
\begin{equation}
\mathcal{Q}_R
 =  q_1 + q_2 + q_0  - w_+
 +  \frac{2}{p(k+2)} \left[ \tilde{m} + \ q_1 \ - w_+ \right]
- \frac{2}{k+2} \left[ m_{SU(2)} \ + \ q_2 \right]\nonumber
\end{equation}
with $q_i$ the fermionic charges.

The worldsheet supersymmetry of the deformed theory works
similarly to the undeformed one ($p=1$). In fact, the
spectral-flow charge $w_+$ in the first bracket has to be
compensated by a shift of $q_1$, because the spectral-flow
symmetry must act on the total current~\cite{Pakman:2003cu}. We
are left with well-normalized charges for $\mathcal{I}^3$ and
deformed $\mathcal{J}^3$.

\subsection{Space--time supersymmetry}

The supersymmetry generators of the original $SL(2,\mathbb{R})
\times SU(2)$ model are given in Eq. (\ref{spfi}) with a
restriction on the allowed charges captured in (\ref{projad}), on
top of the usual GSO projection.

In the deformed theory, these operators are no longer physical
with respect to the supercurrent $G=G^+ + G^-$ given in
Eq.~(\ref{defscur}). The physical spin fields are instead
\begin{equation}
\label{defspfield} \Theta^{\rm def}_{\mathbf{\varepsilon}} (z) =
\exp \left\{ -  {\varphi \over 2} + \frac{i}{2}\sum_{\ell =0}^{4}
\varepsilon_\ell H_\ell + \frac{i}{2} \varepsilon_1  (p-1) t \,
X^- \right\}.
\end{equation}
Since the only modification resides in the $X^-$ term, it changes
neither the conformal dimensions of these fields nor their mutual
locality. By using the same projection as before, we obtain a set
of well-defined physical spin operators. Acting with one of these
operators on a left-moving vertex operator of the NS sector
($q_\ell \in \mathbb{Z}$),
\begin{equation}
V (z') \sim {\mathrm{e}}^{-\varphi} \Phi^{w_+ ,\ \rm def}_{j \, m }
\exp{i \sum_{\ell=0}^4 q_\ell \,H_\ell} (z'), \nonumber
\end{equation}
gives a leading term behaving like
\begin{equation}(z-z')^{\frac{1}{2} \left\{ -1+\sum_{\ell=0}^4 \varepsilon_\ell q_\ell
+ \varepsilon_1 (p-1) w^{+} \right\} }.\nonumber
\end{equation}
We conclude that the locality condition of space--time
supercharges with respect to the states requires {\it the
deformation parameter be an odd integer: $p \in 2\mathbb{Z}+1$}.
Note that since this quantization condition originates from the
massive states (those with $w_+ \neq 0$), it is not visible in the
supergravity analysis. In the limit of infinite deformation, this
choice generates modified space--time supercharges, constructed
with $H_1 - s_{0} X^-$. This choice is not the one obtained from
the $N=4$ algebra (see Eqs.~(\ref{NS5alg1}) and (\ref{NS5alg2})),
because the complex structure for the fermions is different. An
appropriate choice, though, avoids infinite shifts in the spin
fields, and allows for reaching a well-defined theory at the limit
of infinite deformation.

At this point we want to discuss the issue of supersymmetry
breaking. In type IIB, the first projection performed in the
undeformed theory keeps the spinors $(+,\mathbf{2'},\mathbf{2'})$
and $(-,\mathbf{2},\mathbf{2'})$ of $SO(1,1) \times SO(4) \times
SO(4)_T$ for both supersymmetry generators. When the AdS$_3$
factor of the background is deformed in the null direction, the
gravitino must be right-moving in space--time (hence, it depends
on $X^-$), which picks up the spinor $(-,\mathbf{2},\mathbf{2'})$.
So, although the number of covariantly constant spinors is reduced
by a factor of two in the deformed background, the number of
\emph{transverse} fermionic degrees of freedom appearing in the
spectrum is the same. A subtlety comes, however, while dealing
with the right-moving sector of the theory. As already mentioned,
in WZW models, the right superconformal algebra is written with a
torsion term of opposite sign. The correct torsion for the right
movers in the $SL(2,\mathbb{R})$ factor demands to rotate the
fields of $SU(2)_R \times U(1)_Q \times \mathbb{R}^{1,1}$ as:
\begin{eqnarray}
\hat{\bar{H}}_1 & = & \bar{H}_1 - t \bar{X}^+, \nonumber \\
\hat{\bar{T}} & = & c \bar{T} - s \bar{H}_1,
\nonumber \\
\hat{\bar{X}} & = & c \bar{X} + s (H_1  + t \bar{X}^+ ). \nonumber
\end{eqnarray}
Therefore, the right algebra is written with $\bar{X}^+$ rather
than with $\bar{X}^-$. For the undeformed model this is irrelevant
since the two free-field representations are isomorphic. However,
for non-zero deformation, it makes a difference because the
right-moving spin fields will be corrected with $\bar{X}^+$ rather
than $\bar{X}^-$:
\begin{equation}
\bar{\Theta}^{\rm def}_{\mathbf{\bar{\varepsilon}}} (z) = \exp
\left\{ - {\varphi \over 2} + \frac{i}{2}
\sum_{\ell=0}^{4}\bar{\varepsilon}_\ell \bar{H}_\ell +
\bar{\varepsilon}_1 (p-1) t \, \bar{X}^+ \right\}.\nonumber
\end{equation}
As a consequence, gravitinos from the right- and left-moving
sectors of the worldsheet CFT propagate in opposite light-cone
directions. They give space--time transverse supercharges
$(\mathbf{2},\mathbf{2'})$ from the left and
$(\mathbf{2'},\mathbf{2'})$ from the right. In type IIA, the same
reasoning leads to the  representations $(\mathbf{2},\mathbf{2'})$
for both generators. The conclusion is that the deformation {\it
flips the chirality of space--time fermions from the right
sector}.

\boldmath
\subsection{The partition function for superstrings on deformed
$SL(2,\mathbb{R}) \times SU(2)$} \unboldmath

As we have seen in the previous analysis, the only necessary
modification on the fermionic part is the flip of chirality for
the right-moving fermions. It is implemented by inserting
$(-)^{F_{\rm R}}$, where $F_{\rm R}$ is the space--time fermion
number for right-movers. This is an orbifold, that projects out
the Ramond states from the untwisted sector, while the twisted
sector restores the Ramond states, with opposite chirality though.
The fermionic vertex operators are thus constructed with $\exp \pm
i \left(H_1 + (p-1) t \ X^- \right)/2$. However, since this
modification has no effect on the conformal weights of the spin
fields, it does not alter the fermionic characters in the
partition function. The remaining parts of the partition function
have been discussed in previous sections. Putting everything
together, we find for the $\{$null-deformed $SL(2,\mathbb{R})\}
\times SU(2) \times T^4 / \mathbb{Z}_2$:
\begin{eqnarray}
Z_{\rm IIB}(p) & = & {\mathrm{Im} \tau \over  \eta^2 \bar\eta^2 }
Z_{SU(2)} Z_{SL(2,\mathbb{R})}^{\rm \ null}(p) {1\over 2}
\sum_{h,g=0}^1
Z_{T^4/{\mathbb{Z}_2}}^{\rm twisted} \oao{h}{g} \nonumber \\
&&\nonumber \\
&& \times {1 \over 2} \sum_{a,b=0}^1 (-)^{a+b}\ \vartheta^2
\oao{a}{b} \vartheta \oao{a+h}{b+g}
\vartheta \oao{a-h}{b-g} \nonumber \\
&&\nonumber \\
&& \times {1 \over 2} \sum_{\bar{a},\bar{b}=0}^1
(-)^{\bar{a}+\bar{b}} \ \frac{1}{2} \sum\limits_{h',g' =0}^{1} \
(-)^{\left( 1- \delta_{p,1}
\right)\left[\bar{a}g'+\bar{b}h'+h'g'\right]} \ \bar{\vartheta}^2
\oao{\bar{a}}{\bar{b}} \bar{\vartheta} \oao{\bar{a}+h}{\bar{b}+g}
\bar{\vartheta} \oao{\bar{a}-h}{\bar{b}-g} ,\nonumber
\end{eqnarray}
where $p\in 2{\mathbb{Z}} + 1$ and $Z_{SL(2,\mathbb{R})}^{\rm \
null}(p)$ is given in (\ref{partnulldef}) with $k\to k+4$; the
relation between $p$ and $M^2$ has been given by
Eq.~(\ref{relparam}). The sum over $h' $ and $g'$ flips
the chirality of the right-moving fermionic representation for any
$p \neq 1$, according to the left-right asymmetry discussed in the
text.

This is the {\it simplest} modular invariant combination of the
various ingredients, with the correct projections dictated by the
superconformal invariance. It should, by no means, be considered
as unique, and many other models do exist, which are equally
acceptable.

\boldmath
\section{Some comments about holography}\label{sect5}
\unboldmath

In this last section of the paper, we will give some remarks about
the holographic dual of string theory in the
background~(\ref{solns}). Since this string theory is exactly
solvable and perturbative everywhere, we can use the gauge/gravity
correspondence beyond the supergravity approximation. The purpose
here is only to identify the non-gravitational dual of the setup
and to explain its relevance to study little string theory.
Although the space--time studied in this paper is constructed as a
deformation of AdS$_3$, the holographic interpretation is
different from the undeformed case.

\subsection{The D1/D5 setup and AdS/CFT}

Let first review briefly the usual holographic dual of the theory
of D1/D5-branes~\cite{Maldacena:1997re}. Starting with the
supergravity solution of Eqs. (\ref{sugrsol}), (\ref{sugrsold})
and (\ref{sugrsola}),
we would like to take a limit where the theory on the branes is
decoupled from the bulk modes. This is obtained by the low-energy
limit:
$$
\alpha'   \to   0 \ \ , \ \ \ U= r/\alpha' \ \  \mathrm{fixed}
$$
so that the gravitational coupling constant goes to zero. All
open-string modes become infinitely massive and only the zero
modes survive; the dual theory is a field theory.

In the above limit, the dimensionless volume of the four-torus has
a fixed, finite value:
\begin{equation}
\hat{v} (T^4 ) = N_1 / N_5.\nonumber
\end{equation}
Therefore, the dual conformal theory is $(1+1)$-dimensional. In
the low-energy limit, the D1-branes are trapped inside the
D5-brane world-volume and can be considered as string-like
instantons of the six-dimensional $U(N_5)$ gauge theory of the
D5-branes. The $\mathcal{N} = (4,4)$ superconformal field theory
dual to the near-horizon limit of the background is the Higgs
branch of a sigma-model on the moduli space of these instantons,
of central charge $c= 6(N_1 N_5+1)$ \cite{Maldacena:1997re}.

In regions of the moduli space where the string coupling is large,
the theory is appropriately described in the S-dual frame. The
string background is then given by the WZW theory: $SU(2)_k \times
SL(2,\mathbb{R})_{k+4}$, where $k=N_5$. Now the constant value of
the six-dimensional string coupling is $g_6 = N_5 / N_1$. Note
that, although the number of fundamental strings do not appear in
the worldsheet CFT description, the string theory is weakly
coupled only for large values of $N_1$.

An interesting feature of this theory is that it is possible to
construct directly out of the worldsheet currents the generators
of the space--time Virasoro algebra~\cite{Giveon:1998ns}, which
act on the boundary of AdS$_3$.

\boldmath
\subsection{The null deformation of AdS$_3$}
\unboldmath

The null deformation of AdS$_3$ cannot be small; regardless the
value of the deformation parameter, the causal structure of the
space--time is completely changed. There is no longer conformal
boundary, but rather an asymptotic flat geometry with a linear
dilaton. Anyway, the deformation parameter can always been scaled
to one (if positive) by rescaling the non-compact coordinates of
the light-cone.

From the holographic point of view, the null deformation
corresponds to adding an (infra-red) irrelevant operator in the
Lagrangian. It is therefore more appropriate for holography to
start with the holographic dual of the linear dilaton background
and perturb it in the infra-red by a relevant operator.

Our decoupling limit~(\ref{decoupl}) is quite different from the
standard one. We send the ten-dimensional string coupling to
infinity. Therefore in order to study this part of the moduli
space of the brane theory we have to perform an S-duality. In the
dual variables, the limit under consideration is
\begin{equation}
\tilde{g}_s \to 0 \ \ , \ \ \ \tilde{\alpha}' \ \mathrm{fixed} \ \
, \ \ \ r/\tilde{g}_s \ \mathrm{fixed}. \label{lstlimit}
\end{equation}
This looks the same as the little-string-theory limit. However, in
the case at hand, the background contains fundamental strings,
which affect the geometry in the vicinity of the branes. In the
limit (\ref{lstlimit}) the energy of the D1-branes stretched
between the NS5-branes is kept fixed: $E_{D1} \sim
r/(\tilde{\alpha}' \tilde{g}_s )$. If instead we keep the energy
of the fundamental strings fixed, $E_{F1} \sim r/\tilde{\alpha}'$,
the contribution of the fundamental strings disappears and we are
left with the ``pure'' NS5-brane background
(see~\cite{Giveon:1998ns}). Another important difference between
the standard near-horizon limit and the partial near-horizon limit
is that the  later involves the decompactification of the torus,
in the D-brane picture  (we send the asymptotic value of the
dimensionless volume $v$ to infinity). Therefore the dual
``gauge'' theory should be six-dimensional; this is again the same
as little string theory.

\subsection{A little review of little string theory}

The little string theory is the decoupled theory living on the
world-volume of the NS5-branes (for a review,
see~\cite{Kutasov:2001uf}). The world-volume theory is decoupled
from the bulk by taking the limit:
\begin{equation}
\label{LSTlim} \tilde{g}_s \to 0 \ \ , \ \ \ \tilde{\alpha} ' \
\mathrm{fixed.}\nonumber
\end{equation}
This is not a low-energy limit, unlike in the AdS/CFT case. The
resulting theory is interacting and non-local, since it exhibits
the T-duality symmetry. In the case of type IIB string theory, the
low-energy limit of LST is a $U(N_5)$ gauge theory with $N =
(1,1)$ supersymmetry and a bare gauge coupling:
\begin{equation}
g_{YM}^2 = \tilde{\alpha} '.\nonumber
\end{equation}
The coupling grows at large energy, and additional degrees of
freedom, which are identified as string-like instantons in the
low- energy theory, appear at energies of order
$\tilde{\alpha}^{\prime - 1/2}$. These instantons are identified
with fundamental strings attached to the NS5-branes. In the
infra-red this theory flows to a free fixed point. In the case of
type IIA string theory, the infra-red limit is an $N=(2,0)$
interacting superconformal theory. It contains tensionless
strings.

The conjectured holography of~\cite{Aharony:1998ub} states that
this theory is dual to string theory in the NS5-brane background
in the near-horizon limit~\cite{Callan:1991ky}:
\begin{eqnarray}
ds^2 & = &  dx^{\mu} dx^{\nu} \eta_{\mu \nu} + \tilde{\alpha} '
N_5 \left(d\rho^2 + d\Omega_{3}^{\ 2} \right), \nonumber \\
\Phi & = & \Phi_0 - \rho, \nonumber\\
H & = & 2 \tilde{\alpha} ' N_5 \ \epsilon (\Omega_3 ).\nonumber
\end{eqnarray}
In this limit, $r/\tilde{g}_s$ -- the energy of the D1-branes
stretched between the NS5-branes -- is kept fixed. This geometry
corresponds to the exact conformal field theory $SU(2)_{k} \times
U(1)_{Q}$, with $k+2=N_5$ that was discussed in Sec.~(\ref{chs}).
This holographic description breaks down in the region $\Phi \to -
\infty$, near the branes, because the string coupling blows up. In
the type IIB case, this is related to the fact that the infra-red
fixed point is free. In the type IIA theory, this strong-coupling
region is resolved by lifting the background to
M-theory~\cite{Aharony:1998ub}; we obtain the background AdS$_7
\times S^4$ of eleven-dimensional supergravity in the vicinity of
the M5-branes (distributed on a circle in the eleventh dimension).

As in the AdS/CFT correspondence, on-shell correlators of
non-normalizable states in string theory corresponds to off-shell
Green functions of observables in LST. The non-normalizable states
of the NS5-brane  background are constructed with the discrete representations
of the linear dilaton:
\begin{equation}
\label{screening} V = \left( \psi \bar{\psi} \Phi_{j}
\bar{\Phi}_j\right)_{j+1} \mathrm{e}^{\frac{2j}{\sqrt{2(k+2)}}
\rho},
\end{equation}
where the primary operator of spin $j$ of $SU(2)$ and the fermions
are combined into an operator of spin $j+1$. From the worldsheet
point of view, the operators~(\ref{screening}) are necessary to
balance the background charge in the correlation functions. The
theory contains also delta-function-normalizable states, from the
continuous representations:
\begin{equation}
V \sim \mathrm{e}^{\left(- \frac{1}{\sqrt{2(k+2)}} + is\right)
\rho}\nonumber ,
\end{equation}
but their holographic interpretation is less clear.
In the bulk they are propagating fields in the linear
dilaton background. There is a third class of operators from the
discrete representations, normalizable in the ultraviolet but not
in the infra-red:
\begin{equation}
V \sim \mathrm{e}^{\frac{-2(j+1)}{\sqrt{2(k+2)}} \rho} \Phi_{j}
\bar{\Phi}_j.\nonumber
\end{equation}
These operators correspond to states localized on the five-branes.
When we deform the theory towards $SL(2,\mathbb{R})$, they become
the discrete representations of the unitary spectrum of
$SL(2,\mathbb{R})$.

\subsection{Low energy limit, strong coupling and fundamental strings}

The decoupling limit of the LST~(\ref{LSTlim}) does not really
make sense for type IIB superstrings in the deep infra-red region.
In fact, as the string coupling blows up here, sending its
asymptotic value to zero does not ensure that the theory living on
the branes decouples from the bulk at very low energies
(see~\cite{Maldacena:1997cg} for a related discussion). Since the
bulk theory is non-perturbative in this region, the resolution of
this puzzle is quite conjectural.

Our background, viewed from the LST side, provides a possible
mechanism to better describe the situation. We first take the
worldsheet description of the bulk physics; the bosonic Lagrangian
of the theory is
\begin{equation}
2\pi \mathcal{L} = \partial \rho \bar{\partial} \rho -
\sqrt{\frac{2}{k+2}} R^{(2)} \rho + \partial X^+ \bar{\partial}
X^- + 2\pi \mathcal{L}^{WZW}_{SU(2)_k}\nonumber
+ \sum_{i=6}^9
\partial X^i \bar{\partial} X^i.
\end{equation}
Starting from this extremity of the line of marginal deformations,
we add at first order the following $(1,1)$ operator to the
Lagrangian:
\begin{equation}
\delta \mathcal{L} \sim M^2 \mathrm{e}^{- \sqrt{\frac{2}{k+2}}
\rho} \ \partial X^+ \bar{\partial} X^-.\nonumber
\end{equation}
It provides a Liouville potential for the linear dilaton and thus
regulate the strong-coupling region. This potential adds a
non-trivial ``electric'' NSNS flux in the background. We know from
our previous analysis that deforming the theory in this way
corresponds to adding fundamental strings in the infra-red region
of the background. This operator is a singlet of $SU(2)_L \times
SU(2)_R$. The brane picture is that, as we go down into the throat
(the strong-coupling region) macroscopic fundamental strings
condense in the world-volume of the NS five-branes.

Now consider the dual, non-gravitational theory living on the
D5-branes. The low-energy bosonic Euclidean action is
\begin{equation}
S_{\rm gauge} = \int \ \frac{1}{\alpha' g_{\rm s}} \mathrm{Tr} \ F
\wedge * F + \frac{i}{\alpha'} C_0 \ dt \wedge dx \wedge
\mathrm{Tr} \ F \wedge F + \frac{1}{\alpha' g_{\rm s}} \left\{
DX^i \wedge
* DX^i + [X^i, X^j]^2 * \mathbb{I} \right\}.\nonumber
\end{equation}
Where $C_0$ is the flux generated by the D1-branes. It is well
known that for gauge theories in dimensions higher than four,
field configurations of non-zero energy (instanton-like) have an
infinite action and therefore do not contribute to the path
integral. The conclusion is different in the presence of the
D1-brane flux. Starting with any instanton solution in four
dimensions,
$$*_{4} F =  F,$$
we can lift it to a field configuration in six dimensions which
obey a generalized self-duality condition:
\begin{equation}
*_6 F = F \wedge dt \wedge dx,\nonumber
\end{equation}
Such a solution is not really an instanton, since it is invariant
under time translations. This configuration is one-half BPS,
because it imposes the following fermionic projection on supersymmetry
generators:
\begin{equation}
\frac{1}{2} \left( 1- \gamma^{6789} \right) \eta = 0,\nonumber
\end{equation}
where $6, 7 , 8, 9$ are the coordinates of the 4-torus in our
conventions. The gauge action for such a solution is
\begin{equation}
S_{\rm gauge} = \frac{1}{\alpha'} \left[ \frac{1}{g_{\rm s}} + i
C_0 \right] \int dt \wedge dx \wedge \mathrm{Tr}  F \wedge
F.\nonumber
\end{equation}
In the infra-red, the RR 2-form behaves as:
$$C_0 = - \frac{i}{g_{\rm s}} \left[ \frac{\alpha' v U^2}{g_{\rm s} N_1} -1 \right]
\mathop{\rightarrow}\limits_{U \to 0} \frac{i}{g_{\rm s}}.$$
Therefore, the classical action for such a configuration vanish.
The conclusion is that, in presence of the D1-brane flux, an
imaginary ``theta-like'' topological term is added to the SYM
action. Now stringy instanton solutions of the gauge theory are
minima of the action and contribute to the path integral; in the
infra-red the theory is not the free SYM fixed point, but rather
the $(1+1)$-dimensional dynamics of these objects.

At this point one can wonder if there are other natural infra-red
completions of the perturbative string background. Another
proposal, called ``double-scaled little string theory'' has been
made~\cite{Giveon:1999px}. The idea is to describe the Higgs phase
of the little string theory, where the NS5-branes are distributed
on a circle of radius $r_0$. The double scaling limit is defined
by: $\tilde{g}_s \to 0$, $r_0 / \sqrt{\tilde{\alpha}'} \to 0$. In
this limit, the worldsheet dual background is described by an
orbifold of an $N=2$ Liouville theory with a potential tensorized
with the coset $SU(2)/U(1)$. By using a duality discussed recently
in~\cite{Hori:2001ax}, this theory is argued to be equivalent to
$$\left( \frac{SU(2)}{U(1)} \times \frac{SL(2,\mathbb{R})}{U(1)}
\right) \bigg/ \mathbb{Z}_{N_5}.$$ In this model the
strong-coupling throat region of the NS5-branes background is
replaced by the tip of the cigar geometry, so the dilaton value is
bounded from above. However, because of the duality, this model is
\emph{not} continuously connected to the pure NS5-brane
background by a marginal deformation. The holographic interpretation of this mirror symmetry
needs further investigation.

\section{Discussion and conclusions}\label{concl}

Let us summarize the main conclusions of this paper. We have shown
that a new, interesting decoupling limit for the D1/D5-brane
theory (or NS5/F1 in the S-dual description) exists. It captures
not only the infra-red dynamics, but also the full
renormalization-group flow. Furthermore, this theory is free of
strong-coupling problems in the bulk, in contrast with the
little-string-theory limit in the moduli space. We have studied
this background mainly from the string worldsheet point of view,
since the bulk background in the NS5/F1 picture is an exactly
solvable worldsheet conformal field theory: \{null-deformed
$SL(2,\mathbb{R})\} \times SU(2) \times $($T^4 { \rm \ or \ }
K3$).

We have first analyzed the undeformed $SL(2,\mathbb{R})$ theory,
for which our achievement is the construction of the partition
function for the (Lorentzian) AdS$_3$, including both discrete and
continuous representations, in all the sectors of spectral flow.
Our procedure is to start from the coset theory $SL(2,\mathbb{R})/
U(1)_A$, and to reconstruct the $SL(2,\mathbb{R})$ partition
function by coupling the former with a lattice corresponding to
the time direction. The partition function is thus obtained in a
linearized form, where the energy integration is manifest.
Although formally divergent, the expression of the partition
function contains all the information about the full spectrum, and
behaves as expected in the large-$k$ limit. Upon integrating the
energy, we recover the partition function of
\cite{Petropoulos:1999nc}
\cite{Maldacena:2000hw}~\cite{Gawedzki:1991yu}. An important
feature of our partition function is that one light-cone direction
is compact, whereas the other is non-compact. This allows for a
natural definition of the light-cone Hamiltonian.

The $SL(2,\mathbb{R})$ WZW model is a building block for
physically interesting backgrounds, such as AdS$_3 \times S^3
\times T^4$, which preserve supersymmetry and have a brane origin.
We have written the extended worldsheet superconformal algebra for
this theory, although, as discussed
in~\cite{Giveon:1998ns}~\cite{Berenstein:1999gj}, it is somehow
problematic to define the space--time supersymmetry in that way.
In order to implement the projection that leaves unbroken one half
of flat-space supersymmetry, we have chosen to consider
superstrings on the $T^4 /\mathbb{Z}_2$ orbifold point of $K3$.
This implies only minor modifications in the brane interpretation
of the background: the topological sector of the orbifold
corresponds to additional D1-branes with negative charge (this is
clear from the D-brane gravitational couplings), while the
microscopic degrees of freedom living on the branes are different
from the $T^4$ case. We could instead have insisted on keeping the
$T^4$ as the internal manifold, but then the orbifold action
acting on $SU(2) \times SL(2,\mathbb{R})$ -- necessary for cutting
half of the supersymmetries -- would have been more complicated.
The lattice interpretation of the $SL(2,\mathbb{R})$ partition
function is a very powerful tool in the process of understanding
marginal deformations in AdS$_3$ backgrounds. We have considered
left-right symmetric bilinears in the currents. For
$SL(2,\mathbb{R})$ this amounts to three different deformations:
two marginal deformations corresponding to the two different
choices of Cartan subalgebra, and one marginal deformation along a
``null" direction.

The deformation with respect to the time-like generator $J^3$
relates the theory to the Euclidean black hole
$SL(2,\mathbb{R})/U(1)_A$, on one side, and to its T-dual
$SL(2,\mathbb{R})/U(1)_V$ on the other side. We have found that
the spectra of the two cosets do not match exactly. This is
related to the fact that we have started with a theory on the
universal cover of $SL(2,\mathbb{R})$. The
conclusion is different if we take the single cover
instead, but then string theory
does not make sense except at the coset points, because of
the presence of closed time-like curves. The  $J^3$ deformation
gives a geometry that also interpolates between the AdS$_3$
geometry (in global coordinates) and the linear dilaton
background. In the case at hand, however, the brane picture -- if
any -- remains to be understood.

Similar considerations hold for marginal deformations driven by
the space-like choice of Cartan generator, namely $J^2$. In the
limit of infinite deformation, we obtain now the Lorentzian
two-dimensional black hole.

The case of null deformation of $SL(2,\mathbb{R})$ has attracted
most of our attention, because of its brane interpretation and the
underlying decoupling limit. We have reached a modular-invariant
partition function for the purely bosonic case, and extended the
whole set up to the supersymmetric background. We have in
particular shown that the physical spin fields, which give the
space--time supercharges, are modified asymmetrically by the
background, and are restricted by the same projection as in the
absence of any deformation. The locality condition for this
charges with respect to the string states, however, gives an extra
quantization condition, on the deformation parameter. Therefore,
as a \emph{superconformal worldsheet theory}, the line of
deformation is \emph{continuous}, but \emph{space--time
supersymmetry} further selects a \emph{discrete} subset of
deformation points. We observe that, at least for $k$ large, these
special points are such that the $O(2,2,\mathbb{R})$
transformations of Eq.~(\ref{eldual}) that give the null-deformed
model belong to $O(2,2,\mathbb{\mathbb{Z}})$, i.e. become a
discrete line of \emph{dualities}.

The decoupling limit of the D1/D5-brane configuration that we have
presented here, calls for further holography investigation. In the
present paper, the analysis of the holographic picture of this
gravitational background has been very superficial: it provides a
natural infra-red regularization of little string theory, by
imposing an upper bound on the string coupling constant, without
changing the asymptotic ultraviolet geometry. More work is needed
to understand it.

There are many other issues that remain open, as for example:
\begin{itemize}
\item Study in more detail other realizations of the worldsheet supersymmetry
that do not involve the orbifold of the four-torus.
\item Give a complete picture of the AdS$_3 \times S^3$ landscape
by means of a systematic analysis of other
supersymmetry-preserving marginal deformations e.g. cosets or
limiting gravitational-wave backgrounds.
\item Interpret these backgrounds in terms of brane set-ups.
\item Put holography at work; the explicit calculation of
correlation functions in the deformed theory seems at first sight
rather difficult, and needs further investigation.
\end{itemize}

Finally it is worth stressing that connecting the near-horizon
geometry of NS5/F1-branes ($SL(2,\mathbb{R}) \times SU(2) \times
U(1)^4$) with the near-horizon limit for the NS5-branes alone
($\mathbb{R}^{1,1} \times U(1)_Q \times SU(2)\times U(1)^4$) is a
step towards the search of an exact CFT description of a
background which is $SL(2,\mathbb{R})$ in some region of
space--time and asymptotically flat in another one.

\acknowledgments
We have enjoyed very useful discussions with C. Bachas,
A. Fotopoulos, E. Kiritsis, B. Pioline and S. Ribault. We are also
grateful to E. Kiritsis for a careful reading of the manuscript.

\appendix
\boldmath
\section{The $SL(2, \mathbb{R})$ WZW model: a reminder}
\unboldmath

We collect in this appendix some well-known facts about the
$SL(2,\mathbb{R})$ WZW model, within a consistent set of
conventions. The commutation relations for the generators of the
$SL(2,\mathbb{R})$ algebra are
\begin{equation}
\left[ J^1 , J^2 \right]  = - i J^3 \sp \left[ J^2 , J^3 \right] =
i J^1\sp \left[ J^3 , J^1 \right]  =   i J^2. \label{comm}
\end{equation}
The sign in the first relation is the only difference with respect
to the $SU(2)$. Introducing
\begin{equation}
J^\pm = i J^1 \mp J^2, \nonumber
\end{equation}
yields\footnote{In some conventions $J^\pm =  J^1 \pm i J^2$, as
for $SU(2)$.}
\begin{equation}
\left[ J^3 , J^\pm \right]  = \pm J^\pm \sp \left[ J^+ , J^-
\right] = 2 J^3, \label{commpm}
\end{equation}
which are also valid for $SU(2)$. This is the $s\ell(2)$ algebra.
Its representations are the same for both $SL(2,\mathbb{R})$ and
$SU(2)$; only their unitarity properties are different (see e.g.
\cite{groups}).

The quadratic Casimir for $SL(2,\mathbb{R})$ is defined as:
\begin{equation}
C_2 = \left(J^1\right)^2 + \left(J^2\right)^2 - \left(
J^3\right)^2 = - \frac{1}{2}\left( J^+ J^- +J^- J^+ \right) -
\left( J^3\right)^2 , \label{casimir}
\end{equation}
and its eigenvalues parametrized by\footnote{There is an
arbitrariness in the sign of $C_2$, as well as on that of $j$. The
ones we consider here are the most popular in the community.
However, the most efficient are the opposite ones both for $C_2$
and $j$, since they allow for a unified presentation of $SU(2)$
and $SL(2,\mathbb{R})$ representations.} $ C_2 = j(1-j)$.

Irreducible representations of the above algebra are essentially
of two kinds: discrete $\mathcal{D}^{\mp}(j)$ or continuous
principal $\mathcal{C}_{\rm p}(b,a)$ and continuous supplementary
$\mathcal{C}_{\rm s}(j,a)$. The discrete ones have highest
($\mathcal{D}^-$) or lowest ($\mathcal{D}^+$) weight, whereas the
continuous ones do not. The spin $j$ of the discrete
representations is real\footnote{In order to avoid closed
time-like curves, we are considering the universal covering of
$SL(2,\mathbb{R})$. Therefore, $j$ is not quantized.}, and their
states are labelled by $\vert j m\rangle$, $m=\mp j, \mp j\mp 1,
\mp j \mp 2, \ldots $ For the principal continuous ones,
$j={1\over 2}+i b,\ b>0$, and the magnetic number is $m=a, a\pm 1,
a\pm 2, \ldots ,\ -{1\over 2} \leq a< {1\over 2}$, $a,b \in
\mathbb {R}$; for the supplementary continuous ones, $0 < j \leq
{1\over 2}$ and $-{1\over 2} \leq a< {1\over 2}$, with the
constraint $\left\vert j - {1\over 2}\right\vert < {1\over
2}-\vert  a \vert $, $a,j \in \mathbb{R}$. These representations
are unitary and infinite-dimensional; $\mathcal{D}^{\pm}(j)$
become finite-dimensional when $j$ is  a negative integer or
half-integer, and are non-unitary for any negative $j$. Notice
finally that the quadratic Casimir $C_2$ is positive for both
continuous series; for the discrete ones it is positive or
negative when $0<j<1$ or $1<j$, respectively.

The three-dimensional anti-de Sitter space is the universal
covering of the $SL(2,\mathbb{R})$ group manifold. The latter can
be embedded in Minkowski space with signature $(-,+,+,-)$ and
coordinates $(x^0,x^1,x^2,x^3)$ -- we set the radius to one:
\begin{equation}
g  =  \left( \begin{array}{cc}x^0 + x^2 & x^1 + x^3 \\ x^1 - x^3 &
x^0 - x^2 \end{array} \right). \label{4emb}
\end{equation}
The Poincar\'e patch introduced in the Gauss decomposition,
$(u,x^\pm) \in \mathbb{R}^3$ covers exactly once the
$SL(2,\mathbb{R})$. Comparing Eqs. (\ref{gauss}) and (\ref{4emb})
yields
\begin{equation}
x^0+x^2=\frac{1}{u}\sp x^1 \pm x^3 =\frac{x^\pm}{u} \sp x^0-x^2=u
+ \frac{x^+x^-}{u}.\nonumber
\end{equation}
The metric and antisymmetric tensor read:
\begin{equation}
ds^2 = \frac{du^2 + dx^+ dx^-}{u^2}\sp H = dB =\frac{du \wedge
dx^+\wedge dx^-}{u^3} \label{poincare}.
\end{equation}

The isometry group of the $SL(2,\mathbb{R})$ group manifold is
generated by left or right actions on $g$: $g\to hg$ or $g\to gh$
$\forall h \in SL(2,\mathbb{R})$. From the four-dimensional point
of view, it is generated by the Lorentz boosts or rotations
$\zeta_{ab}= i\left( x_a\partial_b -  x_b \partial_a\right)$ with
$x_a=\eta_{ab}x^b$. We list here explicitly the six generators in
the Poincar\'e coordinates,  as well as the action they correspond
to:
\begin{eqnarray}
L_1 &=& \frac{1}{2}\left(\zeta_{32} - \zeta_{01}\right) =-i\left(
\frac{x^-}{2}u\partial_u +\frac{1}{2}\left((x^-)^2 -
1\right)\partial_- -\frac{u^2}{2}\partial_+ \right)\sp g\to
\mathrm{e}^{-\frac{\lambda}{2}\sigma^1}g,
\nonumber \\
L_2 &=& -\frac{1}{2}\left(\zeta_{02} - \zeta_{31}\right) =-i\left(
\frac{1}{2}u\partial_u +x^-\partial_- \right)\sp g\to
\mathrm{e}^{-\frac{\lambda}{2}\sigma^3}g\nonumber \\
L_3 &=& \frac{1}{2}\left(\zeta_{03} - \zeta_{12}\right) =i\left(
\frac{x^-}{2}u\partial_u +\frac{1}{2}\left((x^-)^2 +
1\right)\partial_- -\frac{u^2}{2}\partial_+ \right)\sp g\to
\mathrm{e}^{i\frac{\lambda}{2}\sigma^2}g,
\nonumber \\
R_1 &=& \frac{1}{2}\left( \zeta_{01} + \zeta_{32}\right)=i\left(
\frac{x^+}{2}u\partial_u +\frac{1}{2}\left((x^+)^2 -
1\right)\partial_+ -\frac{u^2}{2}\partial_-\right)\sp g\to
g\mathrm{e}^{\frac{\lambda}{2}\sigma^1},
\nonumber \\
R_2 &=& \frac{1}{2}\left(\zeta_{31} - \zeta_{02}\right) =-i\left(
\frac{1}{2}u\partial_u +x^+\partial_+ \right)\sp g\to
g\mathrm{e}^{-\frac{\lambda}{2}\sigma^3}\nonumber \\
R_3 &=& \frac{1}{2}\left(\zeta_{03} + \zeta_{12}\right) =-i\left(
\frac{x^+}{2}u\partial_u +\frac{1}{2}\left((x^+)^2 +
1\right)\partial_+ -\frac{u^2}{2}\partial_- \right)\sp g\to
g\mathrm{e}^{i\frac{\lambda}{2}\sigma^2}. \nonumber
\end{eqnarray}
Both sets satisfy the algebra (\ref{comm}). Notice also that in
terms of Euler angles defined by
\begin{equation}
g  =   \mathrm{e}^{i (t+\phi) \sigma_2/2} \mathrm{e}^{r \sigma_1}
\mathrm{e}^{i(t-\phi) \sigma_2/2} , \label{euler}
\end{equation}
$L_3$ and $R_3$ simplify considerably:
\begin{equation}
L_3 +R_3 = -i\partial_t \sp L_3 -R_3 = -i\partial_\phi;
\label{rotran}
\end{equation}
these generate time translations and rotations
around the center.

We will now focus on the WZW model on $SL(2,\mathbb{R})$. The
above isometries turn into symmetries of the action displayed in
Eq. (\ref{wzw}), leading thereby to conserved currents. In writing
Eq. (\ref{sigma}), we have chosen a gauge for the $B$ field:
\begin{equation}
B=-\frac{1}{2u^2}dx^+\wedge dx^-. \nonumber
\end{equation}
The two-form is not invariant under $R_{1,3}$ and $L_{1,3}$, and
the action (\ref{sigma}) leads correspondingly to boundary terms
which must be properly taken into account in order to reach the
conserved currents. The latter can be put in an improved-Noether
form, in which they have only holomorphic (for $L_i$'s) or
anti-holomorphic (for $R_j$'s) components. These are called $J^i
(z)$ and $\bar J^j (\bar z)$ respectively. Their expressions are
the following:
\begin{eqnarray}
J^1(z) \pm J^3(z) = -\frac{k}{8\pi}\mathrm{Tr} \left(\sigma^1\mp i
\sigma^2 \right)
\partial g \, g^{-1} \! \! \! &, & J^2 (z) = - \frac{k}{8\pi}
\mathrm{Tr} \sigma^3
\partial g \,
g^{-1}, \nonumber \\
\bar J^1(\bar z)\pm \bar J^3(\bar z) =  \frac{k}{8\pi}\mathrm{Tr}
\left(\sigma^1\pm i \sigma^2 \right)g^{-1} \bar
\partial g
\! \! \! &, & \bar J^2 (\bar z) = - \frac{k}{8\pi}
\mathrm{Tr}\sigma^3  g^{-1} \bar \partial g. \nonumber
\end{eqnarray}
These yield in Poincar\'e coordinates:
\begin{eqnarray}
J^1 +J^3 &=&-\frac{k}{4\pi}\frac{\partial x^+}{u^2}
=-\frac{k}{4\pi}J \label{curJ} \\
J^1 -J^3 &=&\frac{k}{4\pi}\left( 2 x^- \frac{\partial
u}{u}-\partial x^- + (x^-)^2\frac{\partial x^+}{u^2} \right) \\
J^2 &=& \frac{k}{4\pi}\left( \frac{\partial u}{u} + x^-
\frac{\partial x^+}{u^2}\right), \label{curJ2} \\ \bar J^1 + \bar
J^3 &=&\frac{k}{4\pi}\frac{\bar\partial
x^-}{u^2}=\frac{k}{4\pi}\bar J  \\\bar J^1 - \bar J^3 &=&
\frac{k}{4\pi}\left(- 2 x^+ \frac{\bar\partial u}{u}+ \bar\partial
x^+ -
(x^+)^2\frac{\bar\partial x^-}{u^2} \right) \\
\bar J^2 &=& \frac{k}{4\pi}\left( \frac{\bar\partial u}{u} + x^+
\frac{\bar\partial x^-}{u^2}\right), \label{curJ2b}
\end{eqnarray}
where $J$ and $\bar J$ are  the null currents introduced in Eq.
(\ref{nullcurr}).

At the quantum level, these currents, when properly normalized,
satisfy the following $\widehat{SL}(2,\mathbb{R})_{\mathrm{L}}
\times \widehat{SL}(2,\mathbb{R})_{\mathrm{R}}$ OPA\footnote{In
some conventions the level is $x=-k$. This allows to unify
commutation relations for the affine
$\widehat{SL}(2,\mathbb{R})_x$ and $\widehat{SU}(2)_x$ algebras.
Unitarity demands $x<-2$ for the former and $0<x$ with integer $x$
for the latter.}:
\begin{eqnarray}
J^3(z) J^3(0) & \sim & - \frac{k}{2z^2}, \nonumber \\
J^3(z) J^{\pm}(0)& \sim & \pm \frac{J^{\pm}}{z},  \label{LOPA} \\
J^+(z) J^-(0) & \sim & \frac{2J^3}{z}-\frac{k}{z^2},   \nonumber
\end{eqnarray}
and  similarly for the right movers. Equivalently on the modes of
these currents generate the affine Lie algebra:
\begin{eqnarray}
\left[ J_{n}^3 , J_{m}^3 \right] & = & - \frac{k}{2} n
\delta_{m,-n},
\nonumber \\
\left[ J_{n}^3 , J_{m}^{\pm} \right] & = & \pm J^{\pm}_{n+m},\nonumber \\
\left[ J_{n}^+ , J_{m}^- \right] & = & 2 J^{3}_{n+m} - k n
\delta_{m,-n}. \nonumber
\end{eqnarray}

The Virasoro algebra generators of the conformal field theory are
built out of these currents:
\begin{eqnarray}
L_0 & = & \frac{-1}{k-2} \left[ \frac{1}{2} \left( J_{0}^+ J_{0}^-
+ J_{0}^- J_{0}^+ \right) + (J_{0}^{3})^2 + \sum_{m=1}^{\infty}
\left( J_{-m}^+ J_{m}^- + J_{-m}^- J_{m}^+ +2 J_{-m}^3 J_{m}^3
\right) \right]
\nonumber \\
L_n & = & \frac{-1}{k-2} \sum_{m=1}^{\infty} \left( J_{n-m}^+
J_{m}^- + J_{n-m}^- J_{m}^+ +2 J_{n-m}^3 J_{m}^3 \right).\nonumber
\end{eqnarray}
The central charge is $c= 3+ 6/(k-2)$.

Lowest-weight representation of this CFT can be constructed by
using the standard rule: start with a set of primary states
annihilated by the operators $J^{i}_{n}$ with $n\geq 1$; these
ground states fall into representations of the global algebra
generated by the zero modes $J^{\pm,3}_{0}$. The module is then
constructed by acting with the creation operators $J^{i}_{-n}$ ($n
\geq 1$).


Because the metric of the algebra is indefinite, the
representations of the affine algebra will contain negative norm
states, and the CFT is not unitary. However, by using the Virasoro
constraints it is possible to construct a unitary string theory
containing the $SL(2,\mathbb{R})$ CFT. Since the level 0
generators commute with the Virasoro algebra, the spectrum of the
string theory must be constructed out of unitary representations
of $SL(2,\mathbb{R})$. The unitary representations relevant here
are: the \emph{discrete representations} $\mathcal{D}^{\pm}(j)$
with $j >0$ and the \emph{principal continuous representations}
$\mathcal{C}_p (b,a)$. The second step in the proof of the
unitarity of the spectrum is to show that the negative norm states
obtained with the creation operators are removed at each level by
the Virasoro constraints. For the discrete representations, this
is true only if the spin of the allowed representations is
bounded: $0<j<k/2$. This is not consistent with the general
structure of string theory ; in fact assuming that the internal
CFT contributes positively to $L_0$, this restriction on the spin
puts an \emph{absolute upper bound} on the level of string
excitations: $N \leq 1 + k/4$. The case of continous
representations is worse:  the only allowed states are tachyons.

From the representations given above, it is possible to construct
new ones by acting with an automorphism of the affine algebra
called \emph{spectral flow} ($w \in \mathbb{Z}$):
\begin{eqnarray}
\tilde{J}^{3}_n & = & J_{n}^3 - \frac{k}{2} w \ \delta_{n,0}, \nonumber \\
\tilde{J}^{\pm}_{n} & = & J^{\pm}_{n\pm w}.\nonumber
\end{eqnarray}
This solves the above consistency problem. The eigenvalues of the
states are then shifted according to:
\begin{equation}
\tilde{m}= m - \frac{k}{2} w \ , \ \ \tilde{\bar{m}} = \bar{m}-
\frac{k}{2} w,\nonumber
\end{equation}
and the Virasoro generators as:
\begin{equation}
\tilde{L}_n = L_n + w J_{n}^3 - \frac{k}{4} w^2 \delta_{n,0} = L_n
+ w \tilde{J}_{n}^3 + \frac{k}{4} w^2 \delta_{n,0}.\nonumber
\end{equation}
The flowed representations obtained from the lowest-weight
representations constructed above are generically not bounded from
below~; after imposing the Virasoro constraints, one can show that
the physical spectrum of the string theory still contains only
positive norm states.


\section{Free-boson conformal blocks}

The generic conformal blocks for a free compactified boson are the
$U(1)$ characters,
\begin{equation}
\zeta\oao{\omega}{\mu}\left(R^2\right) = \frac{R}{\sqrt{\tau_2}}
\exp -\frac{\pi R^2}{\tau_2}\left|\omega \tau - \mu\right|^2,
\label{coblofre}
\end{equation}
where $R$ is the compactification radius (imaginary for a
time-like boson), and $\omega$, $\mu$ need not be integers.

The partition function for an ordinary, free compactified boson
reads (in the Lagrangian representation):
\begin{equation}
Z(R)= \frac{\Gamma_{1,1}(R)}{\eta \bar\eta} =\frac{1}{\eta
\bar\eta} \sum_{m,w \in \mathbb{Z}}
\zeta\oao{w}{m}\left(R^2\right). \label{combo}
\end{equation}
Notice that for $(\omega, \mu) \in \mathbb{R}^2$ in
(\ref{coblofre}), a modular-invariant combination is provided by
\begin{eqnarray}
\tilde Z(R)&=& \frac{1}{\eta \bar\eta} \int_{\mathbb{R}^2}
d\omega\, d\mu\, R^2\, \zeta\oao{\omega}{\mu}  \left(R^2\right)
\label{coblofrecon}\\
&=& \frac{R}{\sqrt{\tau_2}\eta\bar\eta},
\end{eqnarray}
which is the partition function of a decompactified free boson
(the measure $d\omega\,d\mu\, R^2$ ensures the correct scaling of
with $R$).

Bosons can also be twisted or shifted. This corresponds to
ordinary or freely acting orbifolds. We will first focus on
$\mathbb{Z}_N$ shifts, whose spectra are also captured in Eq.
(\ref{coblofre}). The shifted-partition-function sectors read in
this case:
\begin{equation}
Z\oao{h}{g}(R)= \frac{\Gamma_{1,1}\oao{h}{g}(R)}{\eta \bar\eta}
=\frac{1}{\eta \bar\eta} \sum_{m,w \in \mathbb{Z}}
\zeta\oao{w+h/N}{m+g/N}\left(R^2\right) \ ,\ \ h,g \in \{0,\ldots,
N-1\}, \label{shbl}
\end{equation}
and satisfy the periodicity conditions
\begin{equation}
Z\oao{h}{g}(R)=Z\oao{h+N}{g}(R)=Z\oao{h}{g+N}(R). \label{shblper}
\end{equation}
The basic properties of the quantities introduced so far are
summarized as follows:
\begin{eqnarray}
\tau \to \tau +1  \ : \ \ \zeta \oao{\omega}{\mu}\left(R^2\right)&
\to &\zeta\oao{\omega}{\mu - \omega}\left(R^2\right), \nonumber  \\
\tau \to -\frac{1}{\tau}  \ : \ \
\zeta\oao{\omega}{\mu}\left(R^2\right)&\to& |\tau|
\zeta\oao{\mu}{-\omega}\left(R^2\right), \nonumber
\end{eqnarray}
and
\begin{equation}
\frac{1}{N} \sum_{h,g=0}^{N-1} Z\oao{h}{g}(R)  =
Z\left(\frac{R}{N}\right). \label{RN}
\end{equation}
Notice finally that the duality symmetry of the partition function
(\ref{combo}), namely $Z(R)=Z\left(R^{-1}\right)$, does not
survive the $\mathbb{Z}_N$ shift. This becomes clear in the
following identity, obtained by double Poisson resummation:
\begin{equation}
\sum_{m,w \in \mathbb{Z}}
\zeta\oao{w+h/N}{m+g/N}\left(R^{-2}\right) =\sum_{y,n \in
\mathbb{Z}}\mathrm{e}^{\frac{2i\pi}{N}(ng-yh)}
\zeta\oao{n}{y}\left(R^{2}\right).\nonumber
\end{equation}
As a consequence, the two limits ($R\to 0$ or $\infty$) of
(\ref{shbl}) are distinct:
\begin{equation}
Z\oao{h}{g}(R) \underarrow{R\to\infty} \left\{ \begin{array}{cc}
R/\sqrt{\tau_2}{\eta \bar\eta} & {\rm for} \ h=g=0 ,
\\
0 & {\rm otherwise},
\end{array}  \right.
\label{lolil}
\end{equation}
whereas
\begin{equation}
Z\oao{h}{g}(R) {\underarrow{R\to 0}}\frac{1}{R\sqrt{\tau_2}{\eta
\bar\eta}} \ \ \forall h,g, \label{l0lis}
\end{equation}
up to exponentially suppressed terms\footnote{When $h/N$ and $g/N$
become continuous variables, $\delta$-functions appear, which must
be carefully normalized.}.

We now consider $\mathbb{Z}_N$ twists of a two-torus, for $N \leq 4$.
The corresponding sums read:
\begin{eqnarray}
Z_{2,2}\oao{2h/N}{2g/N}& = & \frac{\Gamma_{2,2}(T,U)}{\eta^2
\bar\eta^2} \ \ \mathrm{for} \ \ h=g=0, \nonumber \\ & = &4
\frac{\eta \bar\eta}{\left\vert \vartheta\oao{1+2h/N}{1-2g/N}(0|
\tau)\right\vert^2} \sin^2 \pi \frac{\Lambda(h,g)}{N} \
\mathrm{otherwise}, \label{twbl}
\end{eqnarray}
where $T,U$ are the usual $T^2$-compactification moduli. Here
$\Lambda(h,g)$ is an integer which is correlated to the number of
fixed points of the torus, depending of the twisted sector under
consideration. In the case of a $T^4$, the
$\mathbb{Z}_N$ twists give rise to twisted sectors which are the
square of the those given in Eq. (\ref{twbl}). The
$T^2/\mathbb{Z}_N$ twisted partition function reads:
\begin{equation}
Z_{T^2/\mathbb{Z}_N}^{\rm twisted}(T,U)= \frac{1}{N}
\sum_{h,g=0}^{N-1} Z_{2,2}\oao{2h/N}{2g/N}. \label{twpf}
\end{equation}

Shifts and twists can be combined. We will consider here two
cases, which happen to play a role in the analysis of the
$SL(2,\mathbb{R})$ and of its cosets. The first is a
$\mathbb{Z}_N$ orbifold of a compact boson of radius $R$ times a
two-torus. The $\mathbb{Z}_N$ acts as a twist on the $T^2$ and as
a shift on the orthogonal $S^1$. The order of the orbifold is
restricted to $N \leq 4$ by the symmetries of the lattice of the
two-torus. The partition function for this model reads:
\begin{eqnarray}
Z_{\mathbb{Z}_N} & = & \frac{1}{N}\sum_{h,g=0}^{N-1}
\frac{\Gamma_{1,1}\oao{h}{g}(R)}{\eta \bar\eta}
Z_{2,2}\oao{2h/N}{2g/N} \nonumber \\
&=&\frac{1}{N}\frac{\Gamma_{1,1}(R)\, \Gamma_{2,2}(T,U)}{\eta^3
\bar\eta^3} \nonumber \\
& +& \frac{1}{N}\sum_{(h,g)\neq (0,0)} 4 \sin^2 \pi \frac{\Lambda
(h,g)}{N} \frac{\mathrm{e}^{2\pi \tau_2 h^2/N^2 }}{\left\vert
\vartheta_1 \left(\frac{h\tau - g}{N}| \tau\right)\right\vert^2}
\sum_{m,w \in \mathbb{Z}}\zeta\oao{w+h/N}{m+g/N}\left(R^2\right) .
\label{ZN}
\end{eqnarray}

Similarly, we can consider a $\mathbb{Z}_N\times \mathbb{Z}_N$
orbifold of four free bosons. The first $\mathbb{Z}_N$ acts as a
shift on a compact boson of radius $R_1$ and as a twist on a
$T^2$; the second $\mathbb{Z}_N$ acts as a shift on the compact
boson of radius $R_1$ and similarly on another compact boson of
radius $R_2$. The partition function now reads:
\begin{eqnarray}
Z_{\mathbb{Z}_N\times \mathbb{Z}_N} & = &
\frac{1}{N^2}\sum_{h_1,g_1=0}^{N-1}\sum_{h_2,g_2=0}^{N-1}
\frac{\Gamma_{1,1}\oao{h_1-h_2}{g_1-g_2}(R_1)}{\eta \bar\eta}
\frac{\Gamma_{1,1}\oao{h_2}{g_2}(R_2)}{\eta \bar\eta}
Z_{2,2}\oao{2h_1/N}{2g_1/N} \nonumber \\
&=&\frac{1}{N^2}\frac{\Gamma_{1,1}(R_1)\,\Gamma_{1,1}(R_2)\,
\Gamma_{2,2}(T,U)}{\eta^4
\bar\eta^4} \nonumber \\
& +& \frac{1}{N^2}\sum_{(h_1,g_1)\neq (0,0)}
\frac{\Gamma_{1,1}\oao{h_1}{g_1}(R_1)\,\Gamma_{1,1}(R_2)}{\eta^2
\bar\eta^2}Z_{2,2}\oao{2h_1/N}{2g_1/N}
 \nonumber \\
& +& \frac{1}{N^2}\sum_{(h_2,g_2)\neq (0,0)}
\frac{\Gamma_{1,1}\oao{-h_2}{-g_2}(R_1)\,
\Gamma_{1,1}\oao{h_2}{g_2}(R_2)\,\Gamma_{2,2}(T,U)}{\eta^4
\bar\eta^4}
\nonumber \\
& +& \frac{1}{N^2}\frac{1}{\eta \bar\eta}\sum_{(h_1,g_1)\neq
(0,0)}\sum_{(h_2,g_2)\neq (0,0)} 4 \sin^2 \pi \frac{\Lambda
(h_1,g_1 )}{N} \frac{\mathrm{e}^{2\pi \tau_2 h_1^2/N^2
}}{\left\vert \vartheta_1 \left(\frac{h_1\tau - g_1}{N}|
\tau\right)\right\vert^2} \times  \nonumber \\ & & \times
\sum_{m_1,w_1,m_2,w_2 \in
\mathbb{Z}}\zeta\oao{w_1+(h_1-h_2)/N}{m_1+(g_1-g_2)/N}\left(R_1^2\right)\,
\zeta\oao{w_2+ h_2/N}{m_1+ g_2/N}\left(R_2^2\right) . \label{ZNZN}
\end{eqnarray}

Models based on freely acting orbifolds exhibit rich
decompactification properties \cite{KK95} \cite{KKPR96}
\cite{KKPR98}, which are due to the breaking of the duality
symmetries. There are two limits of interest\footnote{Similarly,
for $R_1 \to 0$ or $\infty$, we obtain respectively (\ref{l1}) or
(\ref{l2}), with $R_1\leftrightarrow R_2$.} here:
\begin{equation}
Z_{\mathbb{Z}_N\times \mathbb{Z}_N} {\underarrow{R_2\to
0}}\frac{1}{R_2\sqrt{\tau_2}{\eta \bar\eta}} Z\left( \frac{R_1}{N}
\right)Z_{T^2/\mathbb{Z}_N}^{\rm twisted}(T,U) \label{l1}
\end{equation}
and
\begin{equation}
Z_{\mathbb{Z}_N\times \mathbb{Z}_N} {\underarrow{R_2\to
\infty}}\frac{R_2}{N\sqrt{\tau_2}{\eta \bar\eta}}
Z_{\mathbb{Z}_N}(R_1,T,U) \label{l2}
\end{equation}
obtained by using the above equations. In the first limit the two
circles decouple from the $T^2$, and the only reminiscence of the
second $\mathbb{Z}_N$ shift is the rescaling $R_1/N$. In the
second limit, only the decompactifying circle decouples.

\section{Derivation of the spectrum}

\boldmath
\subsection{The spectrum of $SL(2, \mathbb{R})$}
\unboldmath

In this appendix we solve for the constraints $s_{1}$ and $s_2$ to
obtain the spectrum of $SL(2,\mathbb{R})$,  following the
lines of~\cite{Maldacena:2000kv} and~\cite{Hanany:2002ev}.
The total exponential
factor in~(\ref{partsl}) after expanding the $\vartheta_1$
function and integrating $t_2$ is
\begin{eqnarray}
\exp \left\{ \vphantom{\frac{1}{2}} -\pi \tau_2    k (w_{+} +
s_{1} ) (w_- -2t_1 + s_{1} ) - 2i\pi \tau_1  n ( w_{+} + s_{1} )
\right.
\nonumber \\
+ 2i\pi  n s_{2} - 4\pi \tau_2 \frac{1}{4(k-2)}  + 2\pi \tau_2
s_{1}^{2}
-2\pi \tau_2 \left( q+\bar{q} + 1 \right) s_{1} \nonumber \\
\left. + 2i\pi \tau_1 \left( q-\bar{q} \right) s_{1} -2i\pi \left(
q-\bar{q} \right) s_{2} -2\pi\tau_2 \left( N + \bar{N}\right) +
2i\pi \tau_1 \left(N-\bar{N}\right) \vphantom{\frac{1}{2}}
\right\}, \nonumber
\end{eqnarray}
where $q$ is the number of $J^{+}_{n<0}$ minus $J^{-}_{n<0}$
operators acting on the ground state, and similarly for $\bar{q}$.
The integration over $s_{2}$ gives simply the constraint: $ q -
\bar{q} = n $. The remaining integration over $s_{1}$ reads:
\begin{equation}
\int_{0}^{1} ds_{1} \exp \left\{ -2\pi \tau_2  s_{1} \left( k(w -
t_1) +q+\bar{q} + 1 \right)
 - (k-2) \pi \tau_2 s_{1}^{2} \right\}. \nonumber
\end{equation}
As in~\cite{Hanany:2002ev}, we introduce an auxiliary variable $s$
in order to integrate the constraint:
\begin{eqnarray}
\exp \left\{ -2\pi \tau_2  s_{1} \left( k(w -t_1) +q+\bar{q} + 1
\right)
 - (k-2) \pi \tau_2 s_{1}^{2} \right\} \nonumber \\
 =  \ 2 \sqrt{\frac{\tau_2}{k-2}}
\int_{-\infty}^{+ \infty}ds \ \exp \left\{ - 4\pi \tau_2
\frac{s^2}{k-2} - 2\pi \tau_2 \left( 2is + q+ \bar{q} + 1 + k (w -
t_1 )  \right) s_1 \right\}. \nonumber
\end{eqnarray}
We would like to emphasize that $s$ is not just an auxiliary
variable but stands for the momentum of the Liouville field
coupled to the other degrees of freedom of the theory. At this
stage, the interpretation of the partition function is clear in
the free-field representation (see Sec.~\ref{uncover}). In this
case, the BRST constraint relates the Liouville momentum, the
oscillator number and the lattice of the light-cone coordinates.
The integration over $s_{1}$ gives finally
\begin{equation}
\frac{\mathrm{e}^{-4\pi \tau_2 \frac{s^2}{k-2}}}{\pi \sqrt{\tau_2
(k-2)}} \left[ \frac{1}{2is + q + \bar{q} + 1 + k (w-t_1)} -
\frac{\mathrm{e}^{-2\pi\tau_2 (2is+q+\bar{q}+1+k(w-t_1))}}{2is + q
+ \bar{q} + 1 + k (w-t_1)} \right].\nonumber
\end{equation}
We complete the square in the second term by shifting the
integration contour of $s$ in the complex plane: $s \to  s - i
(k-2)/2$. We thus pick up residues corresponding to the discrete
representations, for $$ \mathrm{Im} s =
\frac{q+\bar{q}+1+k(w-t_1)}{2}.$$ These poles are located in the
strip
$$- (k-2) < q+\bar{q} +1 +k(w- t_1)<0,$$ and correspond to the
states of the discrete spectrum, obeying the constraint
\begin{equation}
\tilde{m}+\tilde{\bar{m}} =2j + q + \bar{q} = - k
(w-t_1).\nonumber
\end{equation}
Hence, we obtain the discrete representations in the correct range
given in~\cite{Maldacena:2000hw}:
\begin{equation}
\frac{1}{2} < j < \frac{k-1}{2}.\nonumber
\end{equation}
We note here that due to the continuous shift $t_1$ the spin is
not quantized by the constraint.

Putting all factors together, we obtain the following weights for
the discrete spectrum:
\begin{eqnarray}
L_0 = - \frac{j(j-1)}{k-2} + w_+ \left( -\tilde{m} - \frac{k}{4}
w_+ \right) +N, \nonumber \\
\bar{L}_0 = - \frac{j(j-1)}{k-2} + w_+ \left( -\tilde{\bar{m}}-
\frac{k}{4} w_+ \right) +\bar{N}.\nonumber
\end{eqnarray}
The remaining part of the partition function reads:
\begin{eqnarray}
\sum_{w,w_+} \int ds \ \left[ \frac{1}{2is+q+\bar{q} +1 + k(w-t_1
)}- \frac{\mathrm{e}^{-2\pi \tau_2 (q+\bar{q}+k(w-t_1
+1/2))}}{2is+q+\bar{q} -1 + k(w-t_1+1)} \right] \times
\nonumber \\
\times \exp \left\{ -2\pi \tau_2 \left(2\frac{s^2}{k-2} + k w_+
\left(w - t_1 - \frac{w_+}{2} \right) + N + \bar{N}- 2 \right) +
2i\pi \tau_1 \left(N-\bar{N}-nw_+\right) \right\}. \nonumber
\end{eqnarray}
In order to identify the continuous part of the spectrum, we note
that the exponent of the second term, namely
$$ -2\pi \tau_2 \left(2\frac{s^2}{k-2} +  2 (w_+ + 1) \left( k(w - t_1) -
\frac{k}{2}-\frac{k}{4} (w_+ +1) \right) + N + q+ \bar{N} +
\bar{q} - 2 \right), $$ can be seen as the spectral flow by one
unit of the $w_+$ sector of the theory:  $w_+ \to w_+ +1$, $m \to
m -k/2$, $\bar{m} \to \bar{m} - k/2$, $N \to N + q$, $ \bar{N} \to
\bar{N} + \bar{q}$.  We can then combine the first term from the
$w_+$ sector and the second term from the flowed $w_+ -1$ sector
to obtain:
\begin{eqnarray}
\sum_{w,w_+} \int ds \ \left[ \frac{1}{2is+q+\bar{q} +1 + k(w-t_1
)}-
\frac{1}{2is+q+\bar{q} -1 + k(w-t_1)} \right] \ \times \nonumber \\
\nonumber \\
\times \exp \left\{-2\pi \tau_2 \left(\frac{s^2}{k-2} + k w_+
\left(w - t_1 - \frac{w_+}{2} \right) +
 N + \bar{N}- 2 \right) + 2i\pi \tau_1 \left(N-\bar{N}-nw_+\right) \right\}. \nonumber
\end{eqnarray}
The second line represents the density of long-string states, and
gives a divergence while summing over $q$. By regularizing the sum
as explained in~\cite{Maldacena:2000kv},
$$\sum_{r=0}^{\infty} \frac{1}{A+r} \mathrm{e}^{-r\epsilon} = \log \epsilon - \frac{d}{dA} \log \Gamma (A),$$
we obtain the density of states of the continuous spectrum:
\begin{equation}
\rho (s) = \frac{1}{\pi}\log \epsilon + \frac{1}{4\pi i}
\frac{d}{ds} \log \frac{\Gamma \left(\frac{1}{2}-is-\tilde{m}
\right)\Gamma \left(\frac{1}{2}-is+\tilde{\bar{m}}\right) }
{\Gamma \left(\frac{1}{2}+ is-\tilde{m}\right) \Gamma
\left(\frac{1}{2}+is+\tilde{\bar{m}}\right) }.\nonumber
\end{equation}
The weights of the continuous spectrum are
\begin{eqnarray}
L_0 = \frac{s^2 +1/4}{k-2} + w_+ \left( -\tilde{m} - \frac{k}{4} w_+ \right) +N,  \nonumber \\
\bar{L}_0 = \frac{s^2+1/4}{k-2} + w_+ \left( -\tilde{\bar{m}}-
\frac{k}{4} w_+ \right) +\bar{N}, \nonumber
\end{eqnarray}
with $\tilde{m}+\tilde{\bar{m}} = -k(w-t_1)$ and
$\tilde{m}-\tilde{\bar{m}} = n$. Therefore we have identified both
types of representations, including the sectors obtained by
spectral flow.

\boldmath
\subsection{The spectrum of the null-deformed $SL(2, \mathbb{R})$}
\unboldmath

We now derive the first-order spectrum of the null-deformed
$SL(2,\mathbb{R})$ theory, whose partition function is given
in~(\ref{partnulldef}).We expand, as previously, all oscillator
terms and obtain the overall exponential factor
\begin{eqnarray}
\exp \Bigg\{ -2\pi \tau_2    \alpha k w_{+} (w -t_1 -
w_+ /2 ) -4\pi\tau_2 \frac{s^2 + 1/4}{k-2} - 2i\pi \tau_1  n w_{+}
\nonumber \\
\left. + 2i\pi  s_{2} (n-q+\bar{q} ) -2\pi \tau_2 \left( 2i
\sqrt{\frac{\alpha k-2}{k-2}} s + q+\bar{q} + 1 +
\alpha k (w-t_1) \right) s_{1} \right\},\nonumber
\end{eqnarray}
where $$\alpha = \frac{M^2+1}{M^2}.$$ After integrating over
$s_1$ we are left with
\begin{eqnarray}
& & \frac{\exp \left\{ -4\pi\tau_2 \frac{s^2 + 1/4}{k-2}\right\}}
{2\pi \tau_2 \left( 2i \sqrt{\frac{\alpha k-2}{k-2}} s
+
q+\bar{q} + 1 + \alpha k (w-t_1) \right)}  \nonumber \\
 & - & \ \ \frac{ \exp \left\{-2\pi \tau_2 \left( 2 \frac{s^2 +
1/4}{k-2} + 2i \sqrt{\frac{\alpha k-2}{k-2}} s +
q+\bar{q} + 1 + \alpha k (w-t_1) \right) \right\}}{2\pi
\tau_2 \left( 2i \sqrt{\frac{\alpha k-2}{k-2}} s +
q+\bar{q} + 1 + \alpha k (w-t_1) \right)}.
 \nonumber
\end{eqnarray}
As previously, we complete the square in the second term by
shifting the Liouville momentum:
\begin{equation}
s \to s - \frac{i}{2} \sqrt{(\alpha k  -2)(k-2)}.\nonumber
\end{equation}
The poles corresponding to the discrete representations are now
$$ \mathrm{Im} (s) = \frac{1}{2} \sqrt{\frac{k-2}{\alpha k-2}}
\left[ q+\bar{q}+1+ \alpha k (w-t_1) \right];$$ they are
located in the strip:
$$-\sqrt{(\alpha k -2)(k-2)} < q+\bar{q} +1 +\alpha k(w- t_1)<0.$$
The deformed discrete spectrum, for $\alpha = 1+\varepsilon$,
$\varepsilon \ll 1$, reads:
\begin{eqnarray}
L_0 & = & - \frac{j(j-1)}{k-2} - w_+ \left( \tilde{m} +
\frac{\varepsilon}{2} (\tilde{m}+\tilde{\bar{m}}) \right) - \frac{k}{4} (1+
\varepsilon) w_{+}^{2} + N \nonumber \\
& & + \frac{\varepsilon (1-2j)}{2(k-2)} \left( \tilde{m} + \tilde{\bar{m}} +
\frac{k}{2(k-2)} (1-2j) \right)
, \nonumber \\
\bar{L}_0 & = & - \frac{j(j-1)}{k-2} - w_+ \left( \tilde{\bar{m}} +
\frac{\varepsilon}{2} (\tilde{m}+\tilde{\bar{m}}) \right) - \frac{k}{4} (1+
\varepsilon) w_{+}^{2} + \bar{N}\nonumber \\
& & + \frac{\varepsilon (1-2j)}{2(k-2)} \left( \tilde{m} +
\tilde{\bar{m}} + \frac{k}{2(k-2)} (1-2j)\right).
\end{eqnarray}
The last terms of both equations are due to the displacement of
the poles corresponding to the discrete representations. By using
the parameterization considered in the supersymmetric model,
$\alpha k = p^2 (k-2) +2 $, the expressions are simpler: the poles
of the discrete representations are now located at
$$ \mathrm{Im} (s) =  \frac{1}{2p} \left[ q+\bar{q}+1+\left(2+ (k-2)p^2\right)(w-t_1)\right],$$
inside the strip $$- p(k-2) < q+\bar{q} +1 + \left(2+ (k-2)p^2\right) (w- t_1)<0.$$
If we define, by analogy with the undeformed theory: 
$$2\mathrm{Im} (s) \equiv 1-2j,$$ we find the discrete spectrum:
\begin{eqnarray}
L_0 = - \frac{j(j-1)}{k-2} - w_+ \left( j + q+\frac{1-p}{2} (1-2j)
\right) - \frac{p^2 (k-2)+2}{4} w_{+}^2 
+N, \nonumber \\
\bar{L}_0 = - \frac{j(j-1)}{k-2}  - w_+ \left( j +\bar{q}+ \frac{1-p}{2} (1-2j)
\right) - \frac{p^2 (k-2)+2}{4} w_{+}^2 
 +\bar{N}.\nonumber
\end{eqnarray}
For the continuous spectrum, we use the spectral flow as in the
undeformed case:
\begin{eqnarray}
\left\{ \frac{1}{2i \sqrt{\frac{\alpha k-2}{k-2}} s +
q+\bar{q} + 1 + \alpha k (w-t_1)} - \frac{1}{2i
\sqrt{\frac{\alpha k-2}{k-2}} s +
q+\bar{q} - 1 + \alpha k (w-t_1+1)} \right\} \times \nonumber \\
\exp \left\{ -2\pi \tau_2  \left( 2\frac{s^2 + 1/4}{k-2} - w_+
\alpha k (w- t_1)
 - \frac{\alpha k}{2} w_{+}^{2}
+ N + \bar{N} \right) +2i\pi \tau_1 \left(N-\bar{N} -nw_+\right)
\right\}. \nonumber
\end{eqnarray}
This gives the deformed continuous spectrum at first order:
\begin{eqnarray}
L_0 =  \frac{s^2+1/4}{k-2} - w_+ \left( \tilde{m} +
\frac{\varepsilon}{2} (\tilde{m}+\tilde{\bar{m}}) \right) +
\frac{k}{4} (1+ \varepsilon) w_{+}^{2} + N, \nonumber \\
\bar{L}_0 = \frac{s^2 + 1/4}{k-2} - w_+ \left( \tilde{\bar{m}} +
\frac{\varepsilon}{2} (\tilde{m}+\tilde{\bar{m}}) \right) +
\frac{k}{4} (1+ \varepsilon) w_{+}^{2} + \bar{N},\nonumber
\end{eqnarray}
with the density of long-string states:
\begin{eqnarray}
\rho (s)  = \vphantom{\frac{1}{2}} & \frac{1}{\pi}\log \epsilon +
\frac{1}{4\pi i} \frac{d}{ds} \log &\frac{\Gamma
\left(\frac{1}{2}-i\left(1+
\frac{k\epsilon}{2(k-2)}\right)s-\tilde{m} -
\frac{\tilde{m}+\tilde{\bar{m}}}{2}\epsilon\right)} {\Gamma
\left(\frac{1}{2}+ i\left(1+
\frac{k\epsilon}{2(k-2)}\right)s-\tilde{m} -
\frac{\tilde{m}+\tilde{\bar{m}}}{2} \epsilon \right)} \times
\nonumber \\
& \times &\frac{\Gamma \left(\frac{1}{2}-i\left(1+
\frac{k\epsilon}{2(k-2)}\right)s+\tilde{\bar{m}} +
\frac{\tilde{m}+\tilde{\bar{m}}}{2}\epsilon \right)}{\Gamma
\left(\frac{1}{2}+i\left(1+
\frac{k\epsilon}{2(k-2)}\right)s+\tilde{\bar{m}} +
\frac{\tilde{m}+\tilde{\bar{m}}}{2} \epsilon \right)}. \nonumber
\end{eqnarray}

\section{Theta functions}

We recall here the basic properties of Jacobi functions. Our
conventions are
\begin{equation}
 \vartheta \oao{a}{b} (v | \tau)=
 \sum_{p \in \mathbb{Z}} \mathrm{e}^{\pi i \tau \left(p+{a \over 2}\right)^2 +
2\pi i \left(v + {b \over 2} \right)\left(p + {a \over 2} \right)
}\nonumber
\end{equation}
$a,b \in \mathbb{R}$, so that
\begin{equation} \vartheta_1= \vartheta \oao{1}{1} \sp \vartheta_2= \vartheta
\oao{1}{0} \sp \vartheta_3= \vartheta \oao{0}{0} \sp \vartheta_4=
\vartheta \oao{0}{1} .\nonumber
\end{equation}
We also recall that
\begin{eqnarray*}
\vartheta_1 (v | \tau) & = & -2 q^{1/8} \sin \pi v
\prod_{m=1}^{\infty} \left(1-\mathrm{e}^{2i\pi v}
q^m\right)\left(1-q^m\right)\left(1-\mathrm{e}^{-2i\pi v
}q^m\right), \\
\eta (\tau)& = & q^{1/24} \prod_{m=1}^{\infty} (1-q^m),
\end{eqnarray*}
and
\begin{equation}
 \vartheta_1' = - 2 \pi \eta^3 = - \pi  \vartheta_2 \vartheta_3
\vartheta_4, \nonumber
\end{equation}
where the prime stands for $\partial_v |_{v=0}$. Notice that
\begin{equation}
\left\vert \vartheta_1 (a\tau +b| \tau)\right\vert^2 =
\mathrm{e}^{2\pi \tau_2 a^2} \left\vert \vartheta \oao{1+2a }{1+2
b  } (0| \tau)\right\vert^2, \nonumber
\end{equation}
which leads in particular to the following:
\begin{equation}
\left\vert \vartheta\oao{1+2h/N}{1-2g/N}(0| \tau)\right\vert
 =\mathrm{e}^{-\pi \tau_2 h^2/N^2} \left\vert
\vartheta_1 \left(\frac{h\tau - g}{N}| \tau\right)\right\vert.
\label{thid}
\end{equation}

Finally, the Riemann identity\footnote{We use the short-hand
notation $\vartheta \oao{a}{b} (v)$ for $\vartheta \oao{a}{b}
(v\vert \tau)$.} reads:
\begin{eqnarray}
\frac{1}{2}\sum_{a,b=0}^1 (-1)^{a + b + \mu ab}
 \vartheta \oao{a}{b} \left(v_0\right)\prod_{j=1}^{3}
 \vartheta \oao{a+h_j}{b+g_j}\left(v_j\right) = \\
= \vartheta\oao{1 }{ 1} \left({(-)^{\mu+1}v_0-\sum_j v_j\over
2}\right) \prod_{j=1}^{3}\vartheta \oao {1 -h_j}{ 1-g_j}
\left({(-)^{\mu+1}v_0+\cdots-v_j+\cdots) \over 2}\right),
\nonumber
\end{eqnarray}
where the parameter $\mu =0$ or $1$, and $\sum_j h_j = \sum_j g_j
= 0$.


\begin{thebibliography}{99}

\bibitem{Callan:1991ky}
C.G.~Callan, J.A.~Harvey and A.~Strominger, ``Worldbrane actions
for string solitons'', Nucl.\ Phys.\ {\bf B367} (1991) 60.

\bibitem{Antoniadis:1994sr}
I.~Antoniadis, S.~Ferrara and C.~Kounnas, ``Exact supersymmetric
string solutions in curved gravitational backgrounds'', Nucl.\
Phys.\ {\bf B421} (1994) 343 [arXiv: hep-th/9402073].

\bibitem{Sfetsos:1998xd}
K.~Sfetsos, ``Branes for Higgs phases and exact conformal field
theories'', JHEP {\bf 9901} (1999) 015 [arXiv: hep-th/9811167].

\bibitem{Giveon:1999px}
A.~Giveon and D.~Kutasov, ``Little string theory in a
double-scaling limit'', JHEP {\bf 9910} (1999) 034 [arXiv:
hep-th/9909110].

\bibitem{Kiritsis:2002xr}
E.~Kiritsis, C.~Kounnas, P.M.~Petropoulos and J.~Rizos,
``Five-brane configurations without a strong-coupling regime'',
Nucl.\ Phys.\ {\bf B652} (2003) 165 [arXiv: hep-th/0204201].

\bibitem{Antoniadis:1989mn}
I.~Antoniadis, C.~Bachas and A.~Sagnotti, ``Gauged supergravity
vacua in string theory'', Phys.\ Lett.\  {\bf 235B} (1990) 255.

\bibitem{Boonstra:1998yu}
H.J.~Boonstra, B.~Peeters and K.~Skenderis,
Nucl.\ Phys.\ {\bf B533} (1998) 127
[arXiv: hep-th/9803231].

\bibitem{Maldacena:1998bw}
J.M.~Maldacena and A.~Strominger, ``AdS$_3$ black holes and a
stringy exclusion principle'', JHEP {\bf 9812} (1998) 005 [arXiv:
hep-th/9804085].

\bibitem{Giveon:1998ns}
A.~Giveon, D.~Kutasov and N.~Seiberg, ``Comments on string theory
on AdS$_3$'', Adv.\ Theor.\ Math.\ Phys.\  {\bf 2} (1998) 733
[arXiv: hep-th/9806194].

\bibitem{Balog:1988jb}
J.~Balog, L.~O'Raifeartaigh, P.~Forgacs and A.~Wipf, ``Consistency
of string propagation on curved space--times: an $SU(1,1)$ based
counterexample'', Nucl.\ Phys.\  {\bf B325} (1989) 225.

\bibitem{Petropoulos:1989fc}
P.M. Petropoulos, ``Comments on $SU(1,1)$ string theory'', Phys.
Lett. {\bf 236B} (1990) 151.

\bibitem{Petropoulos:1999nc}
P.M.~Petropoulos, ``String theory on AdS$_3$: some open
questions'', [arXiv: hep-th/9908189].

\bibitem{Maldacena:2000hw}
J.M.~Maldacena and H.~Ooguri, ``Strings in AdS$_3$ and $SL(2,R)$
WZW model. I'', J.\ Math.\ Phys.\  {\bf 42} (2001) 2929 [arXiv:
hep-th/0001053].

\bibitem{Maldacena:2000kv}
J.M.~Maldacena, H.~Ooguri and J.~Son, ``Strings in AdS$_3$ and the
$SL(2,R)$ WZW model. II: Euclidean black hole'', J.\ Math.\ Phys.\
{\bf 42} (2001) 2961 [arXiv: hep-th/0005183].

\bibitem{Maldacena:2001km}
J.M.~Maldacena and H.~Ooguri, ``Strings in AdS$_3$ and the
$SL(2,R)$ WZW model. III: correlation  functions'', Phys.\ Rev.\
{\bf D65} (2002) 106006 [arXiv: hep-th/0111180].

\bibitem{Henningson:1991jc}
M.~Henningson, S.~Hwang, P.~Roberts and B.~Sundborg, ``Modular
invariance of $SU(1,1)$ strings'', Phys.\ Lett.\ {\bf 267B} (1991)
350.

\bibitem{Gawedzki:1991yu}
K.~Gawedzki, ``Non-compact WZW conformal field theories'', [arXiv:
hep-th/9110076].

\bibitem{Hanany:2002ev}
A.~Hanany, N.~Prezas and J.~Troost, ``The partition function of
the two-dimensional black hole conformal  field theory'', JHEP
{\bf 0204} (2002) 014 [arXiv: hep-th/0202129].

\bibitem{Witten:1991yr}
E.~Witten, ``On string theory and black holes'', Phys.\ Rev.\
{\bf D44} (1991) 314.

\bibitem{Forste:1994wp}
S.~F\"orste, ``A Truly marginal deformation of $SL(2, R)$ in a
null direction'', Phys.\ Lett.\  {\bf B338} (1994) 36 [arXiv:
hep-th/9407198].

\bibitem{Klimcik:1994wp}
C.~Klim\v{c}ik and A.A.~Tseytlin, ``Exact four-dimensional string
solutions and Toda like sigma models from ``null-gauged" WZNW
theories'', Nucl.\ Phys.\  {\bf B424} (1994) 71 [arXiv:
hep-th/9402120].

\bibitem{Giveon:1999zm}
A.~Giveon, D.~Kutasov and O.~Pelc, ``Holography for non-critical
superstrings'', JHEP {\bf 9910} (1999) 035 [arXiv:
hep-th/9907178].

\bibitem{Maldacena:1997re}
J.M.~Maldacena, ``The large-$N$ limit of superconformal field
theories and supergravity'', Adv.\ Theor.\ Math.\ Phys.\  {\bf 2}
(1998) 231 [Int.\ J.\ Theor.\ Phys.\  {\bf 38} (1999) 1113]
[arXiv: hep-th/9711200].


\bibitem{Boonstra:1998mp}
H.J.~Boonstra, K.~Skenderis and P.K.~Townsend,
JHEP {\bf 9901} (1999) 003
[arXiv: hep-th/9807137].



\bibitem{Aharony:1998ub}
O.~Aharony, M.~Berkooz, D.~Kutasov and N.~Seiberg, ``Linear
dilatons, NS5-branes and holography'', JHEP {\bf 9810} (1998) 004
[arXiv: hep-th/9808149].

\bibitem{Seiberg:1997zk}
N.~Seiberg, ``New theories in six dimensions and matrix
description of M-theory on  $T^5$ and $T^5/Z_2$'', Phys.\ Lett.\
{\bf B408} (1997) 98 [arXiv: hep-th/9705221].

\bibitem{David:2002wn}
J.R.~David, G.~Mandal and S.R.~Wadia, ``Microscopic formulation of
black holes in string theory'', Phys.\ Rept.\  {\bf 369} (2002)
549 [arXiv: hep-th/0203048].




\bibitem{deBoer:1998ip}
J.~de Boer, ``Six-dimensional supergravity on $S^3 \times$ AdS$_3$
and 2d conformal field  theory'', Nucl.\ Phys.\ {\bf B548} (1999)
139 [arXiv: hep-th/9806104].


\bibitem{Hassan:1992gi}
S.F.~Hassan and A.~Sen, ``Marginal deformations of WZNW and coset
models from $O(d,d)$ transformation'', Nucl.\ Phys.\ {\bf B405}
(1993) 143 [arXiv: hep-th/9210121].

\bibitem{Giveon:1993ph}
A.~Giveon and E.~Kiritsis, ``Axial vector duality as a gauge
symmetry and topology change in string theory'', Nucl.\ Phys.\
{\bf B411} (1994) 487 [arXiv: hep-th/9303016].

\bibitem{Giveon:1994fu}
A.~Giveon, M.~Porrati and E.~Rabinovici, ``Target-space duality in
string theory'', Phys.\ Rept.\  {\bf 244} (1994) 77 [arXiv:
hep-th/9401139].

\bibitem{Horowitz:1994ei}
G.T.~Horowitz and A.A.~Tseytlin, ``On exact solutions and
singularities in string theory'', Phys.\ Rev.\  {\bf D50} (1994)
5204 [arXiv: hep-th/9406067].


\bibitem{Gepner:1986}
D. Gepner and Z. Qiu, ``Modular invariant partition functions for
parafermionic field theories'', Nucl.\ Phys.\ {\bf B285} (1987)
423.


\bibitem{Dixon:1989}
L.J. Dixon, M.E. Peskin and J. Lykken, ``$N=2$ superconformal
symmetry and $SO(2,1)$ current algebra'', Nucl.\ Phys.\ {\bf B325}
(1989) 329.

\bibitem{Mohammedi:1990}
N. Mohammedi, ``On the unitarity of string propagation on
$SU(1,1)$'', Int.\ J.\ Mod.\ Phys.\ Lett.\  {\bf A5} (1990) 3201.

\bibitem{Hwang:1990aq}
S.~Hwang, ``No ghost theorem for $SU(1,1)$ string theories'',
Nucl.\ Phys.\ {\bf B354} (1991) 100.

\bibitem{Evans:1998qu}
J.M.~Evans, M.R.~Gaberdiel and M.J.~Perry, ``The no-ghost theorem
for AdS$_3$ and the stringy exclusion principle'', Nucl.\ Phys.\
{\bf B535} (1998) 152 [arXiv: hep-th/9806024].


\bibitem{Kiritsis:1993ju}
E.~Kiritsis, ``Exact duality symmetries in CFT and string
theory'', Nucl.\ Phys.\ {\bf B405} (1993) 109 [arXiv:
hep-th/9302033].

\bibitem{Hikida:2000ry}
Y.~Hikida, K.~Hosomichi and Y.~Sugawara, ``String theory on
AdS$_3$ as discrete light-cone Liouville theory'', Nucl.\ Phys.\
{\bf B589} (2000) 134 [arXiv: hep-th/0005065].


\bibitem{Griffin:1990zt}
P.~A.~Griffin and O.F.~Hernandez, ``Structure of irreducible
$SU(2)$ parafermion modules derived via the Feigin--Fuchs
construction'', Int.\ J.\ Mod.\ Phys.\ A {\bf 7} (1992) 1233.


\bibitem{Satoh:1997xe}
Y.~Satoh, ``Ghost-free and modular invariant spectra of a string
in $SL(2,R)$ and  three-dimensional black hole geometry'', Nucl.\
Phys.\ {\bf B513} (1998) 213 [arXiv: hep-th/9705208].

\bibitem{Kiritsis:1994ij}
E.~Kiritsis, C.~Kounnas and D.~L\"ust, ``Superstring gravitational
wave backgrounds with space--time supersymmetry'', Phys.\ Lett.\
{\bf B331} (1994) 321 [arXiv: hep-th/9404114].


\bibitem{Banks:1988yz}
T.~Banks and L.J.~Dixon, ``Constraints on string vacua with
space--time supersymmetry'', Nucl.\ Phys.\ {\bf B307} (1988) 93.

\bibitem{Ademollo:1976wv}
M.~Ademollo {\it et al.}, ``Dual string models with non-abelian
color and flavor symmetries'', Nucl.\ Phys.\ {\bf B114} (1976)
297.

\bibitem{Sevrin:1988ew}
A.~Sevrin, W.~Troost and A.~Van Proeyen, ``Superconformal algebras
in two-dimensions with $N=4$'', Phys.\ Lett.\ {\bf 208B} (1988)
447.

\bibitem{Kounnas:1990ud}
C.~Kounnas, M.~Porrati and B.~Rostand, ``On $N=4$ extended
superliouville theory'', Phys.\ Lett.\  {\bf 258B} (1991) 61.

\bibitem{Kounnas:1993ix}
C.~Kounnas, ``Four-dimensional gravitational backgrounds based on
$N=4$, $c = 4$ superconformal systems'', Phys.\ Lett.\  {\bf B321}
(1994) 26 [arXiv: hep-th/9304102].

\bibitem{Ivanov:ec}
I.T.~Ivanov, B.~Kim and M.~Ro\v{c}ek, ``Complex structures,
duality and WZW models in extended superspace'', Phys.\ Lett.\
{\bf B343} (1995) 133 [arXiv: hep-th/9406063].

\bibitem{Gepner:1987qi}
D.~Gepner, ``Space--time supersymmetry in compactified string
theory and superconformal models'', Nucl.\ Phys.\ {\bf B296}
(1988) 757.

\bibitem{Berenstein:1999gj}
D.~Berenstein and R.G.~Leigh, ``Space--time supersymmetry in
AdS$_3$ backgrounds'', Phys.\ Lett.\ {\bf B458} (1999) 297 [arXiv:
hep-th/9904040].

\bibitem{Friedan:1985ge}
D.~Friedan, E.J.~Martinec and S.H.~Shenker, ``Conformal
invariance, supersymmetry and string theory'', Nucl.\ Phys.\ {\bf
B271} (1986) 93.



\bibitem{Pakman:2003cu}
A.~Pakman, ``Unitarity of supersymmetric $SL(2,R)/U(1)$ and
no-ghost theorem for  fermionic strings in AdS$_3 \times  N$'',
JHEP {\bf 0301} (2003) 077 [arXiv: hep-th/0301110].


\bibitem{Bershadsky:1995qy}
M.~Bershadsky, C.~Vafa and V.~Sadov, ``D-branes and topological
field theories'', Nucl.\ Phys.\  {\bf B463} (1996) 420 [arXiv:
hep-th/9511222].

\bibitem{Cappelli:xt}
A.~Cappelli, C.~Itzykson and J.-B.~Zuber, ``The ADE classification
of minimal and $A_1(1)$ conformal invariant theories'', Commun.\
Math.\ Phys.\  {\bf 113} (1987) 1.

\bibitem{Elitzur:cb}
S.~Elitzur, A.~Forge and E.~Rabinovici, ``Some global aspects of
string compactifications'', Nucl.\ Phys.\  {\bf B359} (1991) 581.

\bibitem{Forste:2003km}
S.~Forste and D.~Roggenkamp,
``Current current deformations of conformal field theories, and WZW  models,''
JHEP {\bf 0305} (2003) 071
[arXiv:hep-th/0304234].


\bibitem{Kutasov:1998zh}
D.~Kutasov, F.~Larsen and R.G.~Leigh, ``String theory in magnetic
monopole backgrounds'', Nucl.\ Phys.\  {\bf B550} (1999) 183
[arXiv: hep-th/9812027].



\bibitem{Buscher:sk}
T.H.~Buscher, ``A symmetry of the string background field
equations'', Phys.\ Lett.\  {\bf 194B} (1987) 59.

\bibitem{Dijkgraaf:1991ba}
R.~Dijkgraaf, H.~Verlinde and E.~Verlinde, ``String propagation in
a black hole geometry'', Nucl.\ Phys.\ {\bf  B371} (1992) 269.



\bibitem{Bakas:1991fs}
I.~Bakas and E.~Kiritsis, ``Beyond the large-$N$ limit: non-linear
$W_\infty$ as symmetry of the $SL(2,R) / U(1)$ coset model'',
Int.\ J.\ Mod.\ Phys.\  {\bf A7} (1992) 55 [arXiv:
hep-th/9109029].




\bibitem{Balasubramanian:1998sn}
V.~Balasubramanian, P.~Kraus and A.E.~Lawrence, ``Bulk vs.
boundary dynamics in anti-de Sitter space--time'', Phys.\ Rev.\
{\bf D59} (1999) 046003 [arXiv: hep-th/9805171].

\bibitem{Johnson:1999qt}
C.V.~Johnson, A.W.~Peet and J.~Polchinski, ``Gauge theory and the
excision of repulson singularities'', Phys.\ Rev.\ {\bf D61}
(2000) 086001 [arXiv: hep-th/9911161].


\bibitem{Kutasov:2001uf}
D.~Kutasov, ``Introduction to little string theory'',
{\it ICTP Spring School on Superstrings and Related Matters,}
Trieste, Italy, April 2--10, 2001.

\bibitem{Maldacena:1997cg}
J.M.~Maldacena and A.~Strominger, ``Semiclassical decay of
near-extremal fivebranes'', JHEP {\bf 9712} (1997) 008 [arXiv:
hep-th/9710014].





\bibitem{Hori:2001ax}
K.~Hori and A.~Kapustin, ``Duality of the fermionic 2d black hole
and $N = 2$ Liouville theory as  mirror symmetry'', JHEP {\bf
0108} (2001) 045 [arXiv: hep-th/0104202].



\bibitem{groups}{B.G. Wybourne, ``Classical groups for
physicists", John Wiley \& Sons, New York, 1974.}

\bibitem{KK95}
E.~Kiritsis and C.~Kounnas,
``Perturbative and non-perturbative partial supersymmetry breaking:
$N = 4 \ \to \ N = 2 \ \to \ N = 1$'',
Nucl.\ Phys.\ {\bf B503} (1997) 117
[arXiv: hep-th/9703059].

\bibitem{KKPR96}
E.~Kiritsis, C.~Kounnas, P.M.~Petropoulos and J.~Rizos,
``Solving the decompactification problem in string theory'',
Phys.\ Lett.\ {\bf B385} (1996) 87
[arXiv: hep-th/9606087].

\bibitem{KKPR98}
E.~Kiritsis, C.~Kounnas, P.M.~Petropoulos and J.~Rizos,
``String threshold corrections in models with spontaneously broken  supersymmetry'',
Nucl.\ Phys.\ {\bf B540} (1999) 87
[arXiv: hep-th/9807067].


\end{thebibliography}
\end{document}